\documentclass[prb,twocolumn,superscriptaddress,float,aps]{revtex4-2}
\usepackage{bm,color,amsmath,amssymb,mathrsfs,latexsym,graphicx,psfrag,tabularx,booktabs}

\usepackage[hidelinks,colorlinks,linkcolor=blue,
citecolor=blue,urlcolor=blue]{hyperref}

\newcommand{\bs}[1]{\boldsymbol{#1}}

\newcommand{\bra}[1]{\left\langle#1\right|}
\newcommand{\ket}[1]{\left|#1\right\rangle}

\newcommand{\beq}{\begin{equation}}
\newcommand{\eneq}{\end{equation}}

\def\kk{\mathbf{k}}

\def\RR{\mathbf{R}}

\def\rr{\mathbf{r}}
\def\GG{\mathbf{G}}

\def\aa{\mathbf{a}}
\def\bb{\mathbf{b}}

\def\GG{\mathbf{G}}

\def\RR{\mathbf{R}}

\def\aa{\mathbf{a}}
\def\bb{\mathbf{b}}

\def\gs{{\bs g}}
\def\rr{\mathbf{r}}

\usepackage{etoolbox}
\makeatletter
\patchcmd{\@maketitle}{\@author}{\@author\show\@thanks}{}{}
\makeatother

\usepackage{dsfont}
\begin{document}

\title{Ideal Topological Flat Bands in Two-dimensional Moir\'e Heterostructures with Type-II Band Alignment}

\author{Yunzhe Liu}
\affiliation{Department of Physics, The Pennsylvania State University, University Park, Pennsylvania 16802, USA}
\author{Anoj Aryal}
\affiliation{Department of Physics, Northeastern University, MA 02115} 
\author{Kaijie Yang}
\affiliation{Department of Materials Science and Engineering, University of Washington, Seattle, Washington 98195, USA}
\author{Dumitru Calugaru}
\affiliation{Department of Physics, Princeton University, Princeton, NJ 08544}
\author{Zhenyao Fang}
\affiliation{Department of Physics, Northeastern University, MA 02115} 
\author{Haoyu Hu}
\affiliation{Donostia International Physics Center, P. Manuel de Lardizabal 4, 20018 Donostia-San Sebastian, Spain}
\author{Qimin Yan}
\affiliation{Department of Physics, Northeastern University, MA 02115} 
\author{B. Andrei Bernevig}
\affiliation{Department of Physics, Princeton University, Princeton, NJ 08544} 
\affiliation{Donostia International Physics Center, P. Manuel de Lardizabal 4, 20018 Donostia-San Sebastian, Spain}
\affiliation{IKERBASQUE, Basque Foundation for Science, Bilbao, Spain}
\author{Chao-xing Liu$^*$}
\affiliation{Department of Physics, The Pennsylvania State University, University Park,  Pennsylvania 16802, USA}

\begin{abstract} 
Topological flat bands play an essential role in inducing exotic interacting physics, ranging from fractional Chern insulators to superconductivity, in moir\'e materials. In this work, we propose a design principle for realizing topological flat bands with "ideal quantum geometry", namely the trace of Fubini-Study metric equals to the Berry curvature, in a class of two-dimensional moir\'e heterostructures with type-II band alignment. We first introduce a moir\'e Chern-band model to describe this system and show that topological flat bands can be realized in this model when the moir\'e superlattice potential is stronger than the type-II atomic band gap of the heterostructure. Next, we map this model into a topological heavy fermion model that consists of a localized orbital for "f-electron" and a conducting band for "c-electron". We find that both the flatness and quantum geometry of the flat band in the topological heavy fermion model depend on the energy gap between c-electron and f-electron bands at {$\Gamma$} which is experimentally controllable via external gate voltages. This tunability will allow us to realize an ideal topological flat band with zero band-width and ideal quantum geometry. Our design strategy of topological flat bands is insensitive of twist angle. We also discuss possible material candidates for moir\'e heterostructures with type-II band alignment. 
\end{abstract}

\date{\today}


\maketitle

\section{Introduction}
The discovery of flat bands in two-dimensional (2D) moir\'e materials, including twisted bilayer graphene (tBG)     \cite{bistritzer2011moire,cao2018unconventional,lu2019superconductors,xie2021fractional,cao2018correlated,lisi2021observation,pons2020flat,tarnopolsky2019origin,serlin2020intrinsic,zhang2019twisted,bultinck2020mechanism,song2022magic,ledwith2020fractional,repellin2020chern,oh2021evidence,andrei2020graphene,ohta2012evidence,xie2019spectroscopic}
and twisted transition metal dichalcogenides (TMDs)     \cite{wu2018hubbard,pan2020band,xu2022tunable,mak2022semiconductor,wu2019topological,regan2022emerging,zhang2020flat,devakul2021magic,cai2023signatures, zeng2023thermodynamic,xu2023observation,xia2025superconductivity,morales2024magic,tang2021recent,liu2014evolution,zhang2020twist,tang2020simulation,xu2020correlated,marcellina2021evidence}, has sparked intense research interest due to their potential to host exotic correlated phases \cite{adak2024tunable,nuckolls2024microscopic,mak2022semiconductor,andrei2021marvels,nimbalkar2020opportunities,carr2020electronic,torma2022superconductivity}, such as superconductivity and fractional Chern insulator (FCI).  
Unlike flat bands associated with localized atomic orbitals in a trivial insulator, the flat bands in these moir\'e materials possess nontrivial quantum geometry     \cite{torma2022superconductivity,song2019all,song2021twisted,po2019faithful,ahn2019failure,adak2024tunable,xie2021fractional,serlin2020intrinsic,bultinck2020mechanism,zhang2019twisted,song2022magic,cai2023signatures,zeng2023thermodynamic,xu2023observation,xia2025superconductivity,oh2021evidence,ledwith2020fractional,repellin2020chern,yu2024quantum,liu2025quantum,torma2023essay}
characterized by Berry curvature and Fubini-Study metric, which plays a crucial role in the emergence of exotic correlated states 
like the FCI phase \cite{ju2024fractional,sun2011nearly,regnault2011fractional,tang2011high,ledwith2020fractional,wang2021exact,wang2023origin,ledwith2023vortexability}
. Quantum geometry generally obeys the trace condition, where the trace of Fubini-Study metric has a lower bound set by the magnitude of Berry curvature. When the "ideal quantum geometry" condition \cite{roy2014band}, namely, the trace of Fubini-Study metric equals the Berry curvature, is satisfied, it has been shown \cite{ledwith2023vortexability,wang2021exact,wang2023origin,wang2022hierarchy,ledwith2020fractional} that the FCI phase emerges as the ground state under short-ranged electron-electron interaction for topological flat bands. It was theoretically proposed that the ideal quantum geometry condition is satisfied in a tBG model under the "chiral" limit \cite{tarnopolsky2019origin,ledwith2020fractional} with a vanishing tunneling between the AA sites, which \textit{cannot} can be directly applied to the realistic tBG systems \cite{koshino2018maximally,carr2019exact}. Theoretical models for ideal topological flat bands have also been proposed in other "chiral" model systems \cite{khalaf2019magic, wang2022hierarchy, li2022magic,eugenio2023twisted, wan2023topological, guerci2024chern,becker2026degenerate, sarkar2025ideal,ledwith2022family,le2024double, gao2023untwisting, cui2024classification, sarkar2025symmetry} and electron gas or lattice systems with periodic magnetic fields\cite{aharonov1979ground,dubrovin1980ground,kapit2010exact, dong2022dirac, shi2025effects}, while their experimental realization remains unexplored. An experimentally feasible and tunable platform to realize topological flat bands with ideal quantum geometry is desirable for exploring interacting topological states and other correlated states in moir\'e materials. 

In certain moir\'e systems like tBG, where the nontrivial quantum geometry is localized around specific regions in the moir\'e Brillouin zone (mBZ), a topological heavy fermion (THF) model   \cite{song2022magic, cualuguaru2023twisted,shi2022heavy,herzog2024topological,lau2025topological} provides a compelling theoretical framework for understanding the coexistence of local moment behavior from strongly interacting localized states and metallic/superconducting behaviors from conducting states, and thus can serve as a starting point to understand a variety of interacting physics   \cite{song2022magic,herzog2024topological,cualuguaru2024thermoelectric,liu2024electron,wang2025electron,herzog2024heavy,hu2025projected,yu2023magic}. Despite these theoretical advancements, a significant challenge lies in identifying new two-dimensional (2D) moir\'e materials exhibiting THF physics, 
i.e. heavy fermion simulators \cite{herzog2024topological}.

\begin{figure*}
    \centering
\includegraphics[width=2\columnwidth]{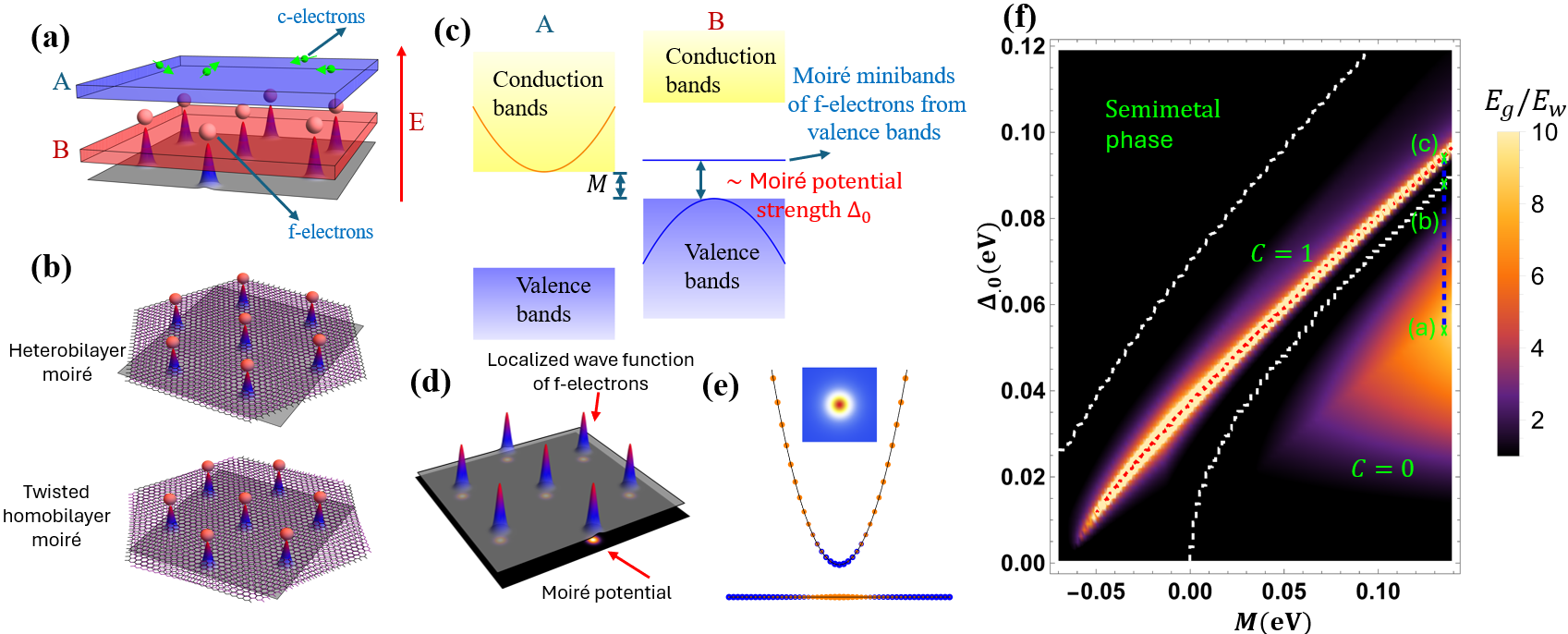}
\caption{\label{fig;1} (a) Illustration of the setup, which consists of two parts, A and B. In 
part A, electrons with a small effective mass behave as itinerant “c-electrons”, depicted by green spheres. In 
part B, the electron wavefunctions are periodic and localized due to moir\'e potential, and can be regarded as “f electrons”, depicted by pink spheres. An electric field can be applied to tuning band alignment between the parts A and B. (b) Part B can be formed by either twisted homobilayers or heterobilayers. The moir\'e potential arises from twisting or lattice mismatch. (c) Illustration of type-II band alignment. The conduction and valence bands originate from the parts A and B, respectively. The atomic band gap of the heterostructure is labeled as $M$. Due to the moir\'e potential, the moir\'e mini-band from the top valence band can become flat and shift upward in energy. When the strength of the moir\'e potential $\Delta_0$ is larger than $M$, band inversion can occur, leading to a topological flat band. (d) Schematic figure of localized wavefunctions at the peak positions of moir\'e potential. (e) The formation of topological flat band after band inversion in the THF model. The sizes of orange and blue dots represent the contribution from “c-electrons” and “f-electrons”. Inset shows the concentration of Berry curvature around $\mathbf{\Gamma}$. 
(f)  Phase diagram for $\Delta_0>0$. $E_g$ and  $E_w$ denote the band gap between VB1 and the other minibands and the bandwidth of VB1, respectively. The background color shows $E_g$/$E_w$. White dashed lines separate different phases and the red line guides the peak for $E_g/E_w$ of VB1. 
}
\end{figure*}

In this work, we tackle these challenges by proposing a design paradigm for realizing an "ideal" THF model, in which the topological flat bands with ideal quantum geometry can be achieved by tuning external gate voltages, using moir\'e heterostructures composed of  2D materials with type-II band alignment. The design principle of the THF model in moir\'e heterostructures is illustrated in Fig.\ref{fig;1}. The THF model consists of the conducting electrons, called "c-electron", and the localized electrons, dubbed "f-electron". Our setup in Fig.\ref{fig;1}(a) includes two parts,  part A, which is made of 2D material with high mobility conduction bands to provide c-electrons, and  part B for f-electron, which is realized via 2D moir\'e materials either by forming a superlattice potential in a moir\'e hetero-bilayer or twisting the 2D material homo-bilayer, as shown in Fig.\ref{fig;1}(b). Due to the moir\'e potential, the highest-energy moir\'e miniband from the valence band in part B gets localized and forms periodic bound states for f-electrons, as schematically shown in Fig.\ref{fig;1}(d). Since conduction band electrons normally have a smaller effective mass compared to valence band electrons in many semiconducting compounds, we below assume the c-electron is from the conduction bands of part A while the f-electron is from the valence bands of part B. But it should be pointed out that it is also possible to interchange the roles of the conduction and valence bands, namely the c-electron is from the valence band of part B while the f-electron is from the conduction band of part A, when the moir\'e potential is imposed on part A. To be concrete, we here consider all the low-energy bands around ${\bf \Gamma}$ in the atomic Brillouin zone (ABZ). In order to form topological flat bands with c- and f-electrons, two criteria are required: (1) the energy difference between the conduction band minimum (CBM) in part A and the valence band maximum (VBM) in part B at ${\bf \Gamma}$ in ABZ is small (a few hundreds of meV) to form a type-II band alignment of the hetero-structure \cite{belabbes2012electronic,ozccelik2016band, ng2022tunable,lo2011emergent,hong2014ultrafast}, so that applying moir\'e potential can reverse the band ordering, as shown in Fig.\ref{fig;1}(c); and (2) the electronic states at ${\bf \Gamma}$ for the c-electrons in part A and f-electrons in part B should belong to different irreducible representations (irreps) of the crystal symmetry group for the heterostructure. For a small moir\'e potential in part B, we would expect the periodic bound states for f-electrons to form a trivial flat band. With increasing moir\'e potential strength, denoted as $\Delta_0$ below, the trivial flat bands are pushed up in energy, and when the moir\'e potential strength $\Delta_0$ is stronger than the type-II atomic band gap $M$ of the heterostructure, we would expect the band ordering between the CBM of c-electrons and the moir\'e miniband of f-electrons from valence bands reverses at ${\bf \Gamma}$, potentially leading to topological flat bands when the irreps of c- and f-electrons at ${\bf \Gamma}$ are different, as shown in Fig.\ref{fig;1}(e). 

Below we will introduce a moir\'e Chern-band model to illustrate this general scheme for designing topological flat bands in Fig.\ref{fig;1}(a)-(e) and numerically demonstrate the emergence of topological flat bands in this model for two limits of the moir\'e superlattice potential: (1) the potential within the first-shell approximation and (2) the $\delta$-function-like potential. Next, we will map the moir\'e Chern-band model to the THF model and demonstrate that the exact topological flat band with "ideal quantum geometry" can be realized in this THF model by tuning the type-II atomic band gap $M$ in Fig.\ref{fig;1}(c). We emphasize that this exact topological flat band is insensitive to the twist angle and can be achievable by tuning external gate voltages in experiments. Finally, we will discuss the potential material realization of this model in 2D moir\'e material heterojunctions and semiconductor heterostructures with superlattice potentials from patterned dielectric substrates. We exemplify 2D moir\'e material heterojunction with the type-II band alignment by Tl$_2$Se$_2$/Zn$_2$Te$_2$ with the moir\'e potential on the Tl$_2$Se$_2$ layer and show the existence of topological flat bands in this moir\'e heterojunction. We also provide a list of possible choices of two different 2D materials in this class of materials. More detailed calculations about realistic materials can be found in Supplementary Materials (SM) Sec.D\cite{SM}.

\begin{figure*}
    \centering
\includegraphics[width=2\columnwidth]{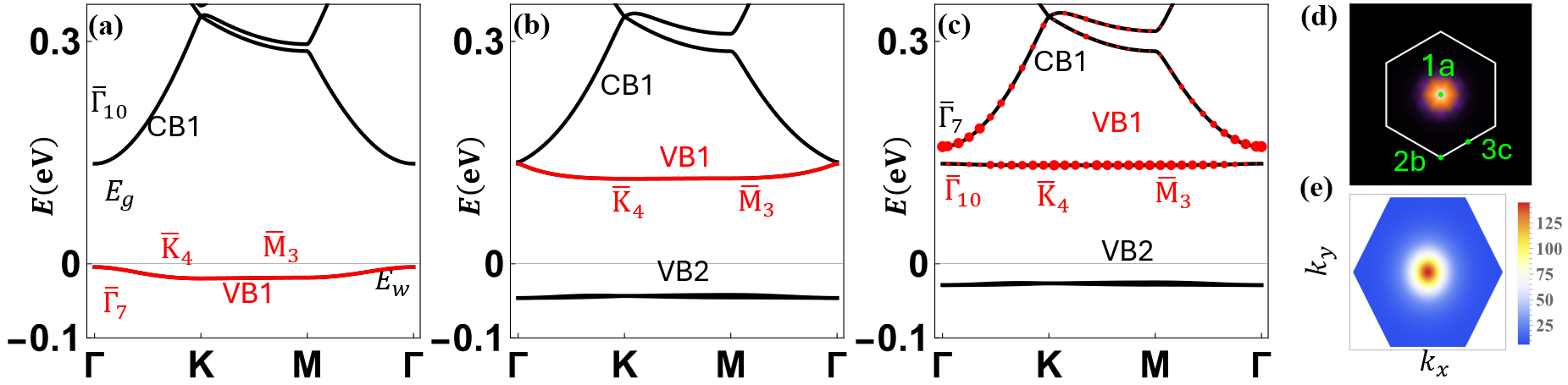}
\caption{\label{fig;2}(a)-(c) show the band dispersion for the parameters $M=0.135$eV and $\Delta_0 = 0.054$ eV, $0.089$ eV and $0.094$ eV, respectively, as denoted by the points (a)-(c) in the phase diagram of Fig.\ref{fig;1}(f).  The band irreps of VB1 and CB1 are labeled in (a)-(c). In (c), the size of red dots in the VB1 and CB1 represents the overlap between the Bloch states and the Wannier orbital that is localized at the Wyckoff positions $1a$, as shown in (d). Three different Wyckoff positions are labeled by green dots in (d). (e) shows the Berry curvature distribution of VB1 in (c). }
\end{figure*}

\section{Moir\'e Chern-band model}
We first introduce the Chern-band model under moir\'e superlattice potential, and its physical realization will be discussed in Sec.\ref{sec_material}. The model Hamiltonian is given by   \cite{tan2024designing,miao2025artificial}
\begin{eqnarray} 
\hat{H}_{mC}=\hat{H}_{C}+\hat{H}_{\text m}.
\label{eq_main:HammC}
\end{eqnarray}

$\hat{H}_{C}$ describes a two-band effective model that allows for a non-zero Chern number and is given by     \cite{qi2006topological,qi2011topological,chang2023colloquium}
\begin{eqnarray}
&& \hat{H}_{C} = \sum_{\kk \alpha \beta}a_{\alpha \kk }^\dagger 
 [H_{C}(\kk)]_{\alpha \beta} a_{\beta \kk} \nonumber \\
&& H_C(\kk) =  \left( \begin{array}{cc} \epsilon(\kk) + \mathcal{M}(\kk)&Ak_+\\ 
  A k_-& \epsilon(\kk) - \mathcal{M}(\kk)
 \end{array} \right), \label{eq:Ham_Chernband} 
\end{eqnarray}
where $a_{\alpha \kk}$ ($a^\dagger_{\alpha \kk}$) is the fermion annihilation (creation) operator with $\alpha = 1, 2$ indexing orbital basis, $\kk$ is the momentum expanded around $\Gamma$ in the ABZ, 
$k_\pm=k_x\pm i k_y $, $\mathcal{M}(k)=M+B k^2$, and $\epsilon(k)=Dk^2$.  
$M, B, D, A$ are material dependent parameters, in which $M$ and $B$ govern the system's topology. Specifically, the system 
has Chern number $C=\pm 1$ when $MB<0$, and $C=0$ when $MB>0$. $H_C$ has continuous translation and rotation symmetries in the x-y plane. For our heterostructure setup in Fig.\ref{fig;1}(a), we would expect the $\alpha=1$ orbital mainly contributes to the conduction band in part A while the $\alpha = 2$ orbital mainly gives the valence band in part B,  forming the type-II band alignment in Fig.\ref{fig;1}(c). In the calculation below, we always choose $B, M> 0$ so that the Hamiltonian $H_C$ is in the trivial insulator regime and furthermore choose $B, D > 0$ so that the valence band has a larger effective mass compared to the conduction band. The linear term $A k_{\pm}$ describes the interlayer coupling between the conduction band in part A and the valence band in part B. For the material systems discussed in Sec.\ref{sec_material}, we find its value  around $\sim 0.72 eV\cdot \AA$ for InAs/GaSb quantum wells and $\sim 1 eV\cdot \AA$ for Tl$_2$Se$_2$/Zn$_2$Te$_2$ heterojunctions, which is several times smaller than the value $\sim 3.9 eV\cdot \AA$ in HgTe quantum wells \cite{LIU201359}. 
It should be noted that the above Hamiltonian $\hat{H}_C$ breaks time reversal (TR) and thus it serves as the spin-up block in the full Hamiltonian for the  nonmagnetic materials that  respect TR  symmetry, as discussed in Sec.\ref{sec_material}, while the spin-down block can be related to $\hat{H}_C$ by TR. 

$\hat{H}_{\text m}$ describes the moir\'e superlattice potential. To be concrete, we choose a hexagonal superlattice form described by the space group $P6/mmm$, leading to the $P6$ symmetry for $\hat{H}_{mC}$, which can be generated by six-fold rotation $C_{6z}$. Furthermore, we consider the diagonal term for the moir\'e potential, 
\begin{eqnarray}
&& \hat{H} _ {\text{m}}  = \sum_{\alpha = 1, 2 }\int d^2\rr  a_{\alpha \rr }^\dagger H ^{\alpha \alpha}_ {\text{m}} (\rr) a_{\alpha \rr },  \label{eq_main:moirepotential}
\end{eqnarray}
with $a_{\alpha \rr }=\frac{1}{\sqrt{V}}\int d^2k a_{\alpha \kk } e ^{i \kk \cdot \rr}$, and $H^{\alpha\alpha}_ {\text{m}}(\rr+\RR) = H^{\alpha\alpha}_ {\text{m}}(\rr) $, where $\RR$ is the moir\'e superlattice vectors, given by $\RR= n_1 \aa^M_1 + n_2 \aa^M_2$ with $n_1, n_2 \in \mathcal{Z}$ and $\aa^M_1=(1,0)a^M_0$, $\aa^M_2=(\frac{1}{2},\frac{\sqrt{3}}{2})a^M_0$, where $a^M_0$ is the moir\'e primitive lattice constant. We can expand the moir\'e potential in the moir\'e reciprocal lattice space as 
\begin{eqnarray}\label{eq_main:moirepotential_1}
    H_m^{\alpha\alpha}(\rr)=\sum_{\gs \in \GG_M} \Delta_\alpha(\gs) e^{i \gs \cdot \rr}
\end{eqnarray}
and $\GG_M = \{ n_1 \bb^M_1 + n_2 \bb^M_2 | n_1, n_2 \in \mathcal{Z}  \}$. Here, $\bb^M_1=(2\pi,-\frac{2\pi}{\sqrt{3}})\frac{1}{a^M_0}$ and $\bb^M_2=(0,\frac{4\pi}{\sqrt{3}})\frac{1}{a^M_0}$ label the primitive moir\'e reciprocal lattice vectors. We will consider two limits for $\Delta_\alpha(\gs)$ below. 

\subsection{First-Shell Potential Approximation}
The first limit is to consider the first-shell potential approximation for the moir\'e potential, which has been successfully applied to the Bistritzer-MacDonald model of tBG \cite{bistritzer2011moire,bernevig2021twisted}. For our model, the first momentum shell includes six momenta, $\GG^{\mathds{1}}_M=\left\{\bb^M_2,\bb^M_1,-\bb^M_2,-\bb^M_1,\bb^M_2+\bb^M_1,-\bb^M_2-\bb^M_1\right\}$, which can be connected by six-fold rotation symmetry $C_{6z}$ around the z axis, and thus one parameter $\Delta_\alpha$ for one $\alpha$ can characterize the moir\'e potential strength within the first-shell potential approximation. We further choose $\Delta_1=0$ and $\Delta_2(\gs)=\Delta_0$ for $\gs \in \GG^{\mathds{1}}_M$ for the calculations below to describe the moir\'e potential only existing in the valence bands of part B, where $\Delta_0$ represents the moir\'e potential strength. We note that the zero shell $\gs = 0$ just corresponds to a constant energy shift, so it can be absorbed into the Fermi energy and atomic gap $M$. We choose the parameters for the valence band to possess a heavy effective mass compared to the conduction band, so we would expect the valence band will get localized by the moir\'e potential and thus focus on the highest valence mini-bands, denoted as VB1 below (see Fig.\ref{fig;2}(a)-(c)). We also choose $M>0$ for which the atomic Hamiltonian $\hat{H}_C$ is in the normal insulator regime.

Fig.\ref{fig;1}(f) reveals the mini-band flatness and topological phase diagram for the VB1 as a function of atomic band gap  $M$ and moir\'e potential strength $\Delta_0$. The color represents the ratio $E_g/E_w$, where $E_g$ denotes the minimal mini-band gap between the VB1 and other mini-bands (mini-bands VB2 or CB1 labelled in Fig.\ref{fig;2}(b)), while $E_w$ denotes the bandwidth of VB1. The bright color is for a large $E_g/E_w$, implying the occurrence of the flat VB1 mini-bands that are well separated in energy from VB2 and CB1. The white dashed lines label the mini-gap closing between the VB1 and other mini-bands, potentially corresponding to the topological phase transition of the VB1. The Chern number $C$ of the VB1 is calculated and labeled in each insulating regime in Fig.\ref{fig;1}(f). We notice a striking bright line, as depicted by the red dashed line in  Fig.\ref{fig;1}(f), in the $C = 1$ regime, which corresponds to topologically nontrivial flat mini-bands. To understand the formation of topological flat bands, we examine the band dispersion of the VB1 and the corresponding irreps of VB1 at high symmetry momenta (${\bf \Gamma}$, ${\bf K}$ and ${\bf M}$) in Fig.\ref{fig;2}(a)-(c) for the parameter values marked by green crosses along the blue dashed line in Fig.\ref{fig;1}(f). The irreps of VB1 are calculated from the rotation eigenvalues of the eigen-wavefunctions of $\hat{H}_{mC}$ (see SM Sec.B.1 \cite{SM}). 
For a small $\Delta_0$, the VB1 in Fig.\ref{fig;2}(c) is described by periodic localized orbitals from the valence band, which are confined by the maxima of moir\'e potential that correspond to the $1a$ Wyckoff position in the moir\'e unit cell, since our atomic model is in the normal insulator regime ($M>0$). The irreps at high symmetry momenta of VB1, $(\bar{\mathbf{\Gamma}}_{7},\bar{\textbf{K}}_4,\bar{\textbf{M}}_3)$, coincide with the elementary band representation (EBR) $^2\Bar{E}_{1}\uparrow G(1)$ of the $P6$ group, which corresponds to the $J_z=3/2$ orbital located at the $1a$ Wyckoff position, where $J_z$ labels the out-of-plane total angular momentum (See Sec.B.2 in SM\cite{SM} for more details). The localized orbital plays the role of f-electron in the THF model. With increasing $\Delta_0$, the VB1 is getting more localized and pushed up in energy, leading to a band inversion between the VB1 and CB1 at $\mathbf{\Gamma}$ point, as shown in Fig.\ref{fig;2}(b). After band inversion, the irreps at high symmetry momenta of VB1 are changed to $(\Bar{\mathbf{\Gamma}}_{10},\Bar{\textbf{K}}_4,\Bar{\textbf{M}}_3)$, which is no longer an EBR. We directly evaluate the rotation eigenvalues of VB1 at high-symmetry momenta: $C_{6z}$ eigenvalue $\eta(\mathbf{\Gamma})=e^{-i \frac{\pi}{6}}$ at $\mathbf{\Gamma}$, three-fold rotation $C_{3z}$ eigenvalue $\theta(\textbf{K})=-1$ at $\mathbf{K}$, and two-fold rotation $C_{2z}$ eigenvalue $\epsilon(\textbf{M})=i$ at $\mathbf{M}$, respectively. Using the connection between the Chern number and rotation eigenvalues, $e^{i C \pi/3}=- \eta(\mathbf{\Gamma}) \theta(\textbf{K})\epsilon(\textbf{M})$    \cite{fang2012bulk}, 
we find $C = 1$ mod 6, which is consistent with the value from direct calculations in Fig.\ref{fig;1}(f). The red dots in the energy dispersion of Fig.\ref{fig;2}(c) represent the projection of localized Wannier functions for $J_z=3/2$ orbitals, as shown in Fig.\ref{fig;2}(d). This Wannier orbital dominates the flat band VB1 away from $\mathbf{\Gamma}$ but forms the conduction band bottom of CB1 around $\mathbf{\Gamma}$ due to band inversion, and thus serves as the f-electron in the THF model discussed below. Fig.\ref{fig;2}(e) shows the Berry curvature distribution of VB1, which concentrates around $\mathbf{\Gamma}$. Fig.\ref{fig;1}(f) and Fig.\ref{fig;2} demonstrate the existence of topological flat bands of VB1 with concentrated Berry curvature for the moir\'e potential within the first-shell approximation. 

We note that the topological flat band of VB1 is not limited to one point (c) in Fig.\ref{fig;1}(f), but exists along the whole red dashed line, as demonstrated in SM Sec.B.1\cite{SM}. 
Below we will provide a brief overview of the evolution of the VB1 along the red dashed line in Fig.\ref{fig;1}(f).
As shown in Fig.S1 of SM, we find that the VB1 moves from the conduction band bottom for a large $\Delta_0$ and $M>0$ to the valence band top for a small $\Delta_0$ and $M<0$, while the flatness and topological nature ($C=1$) of VB1 remain along the red line in Fig.\ref{fig;1}(g). Along this line, the Berry curvature distribution of VB1 is changed. For a large $\Delta_0$, the Berry curvature is concentrated around $\mathbf{\Gamma}$ in Fig.\ref{fig;2}(e). When reducing $\Delta_0$ along the red line in Fig.\ref{fig;1}(f), the Berry curvature becomes distributed across the entire mBZ (See Fig.S1 of SM\cite{SM}).


We note that the parameter regime for the topological flat bands with relatively uniform Berry curvature has been studied for the Hamiltonian $\hat{H}_{mC}$ in Eq.(\ref{eq_main:HammC}) within the first-shell potential approximation in Refs.\cite{tan2024designing,miao2025artificial}. 
Our work here focuses on different parameter regimes as compared to Refs.\cite{tan2024designing,miao2025artificial}. There are two main differences: (1) the conduction and valence bands have the same absolute value of the effective mass in Ref.\cite{tan2024designing}, which corresponds to $D=0$ in Eq.(\ref{eq:Ham_Chernband}); and (2) moir\'e potential exists for both the conduction and valence bands in Refs.\cite{tan2024designing,miao2025artificial}, which corresponds to $\Delta_1=\Delta_2$ in Eq.(\ref{eq_main:moirepotential_1}). Consequently, both conduction and valence bands feel strong moir\'e potential and form moir\'e minibands, leading to a relatively uniform Berry curvature distribution in mBZ for topological flat bands. In our case, the small effective mass and the absence of moir\'e potential $\Delta_1=0$ for the conduction bands make the electrons in part A itinerant, thus providing c-electrons, while the large effective mass and strong moir\'e potential in part B  induce periodic localized states from the valence bands, naturally leading to f-electrons. Thus, the topological flat band possesses concentrated Berry curvature distribution as shown in Fig.\ref{fig;2}
(e), which can be described by the THF model in Sec.\ref{sec:THFmodel} and are thus in different parameter regimes from those with a relatively uniform Berry curvature distribution discussed in Refs.\cite{tan2024designing,miao2025artificial}. However, we emphasize that the form of moir\'e potential only existing on the valence bands is not substantial. As discussed in SM Sec.B.4\cite{SM}, even with equal moir\'e potential strength on both conduction and valence bands ($\Delta_1=\Delta_2$), the asymmetry of effective mass between valence and conduction bands can still lead to topological flat bands with concentrated Berry curvature.

\subsection{Periodic $\delta$-function-like Potential}
The second limit is to consider the periodic $\delta$-function-like moir\'e potential, which can more naturally illustrate our physical picture of topological flat bands with concentrated Berry curvature and allow for analytical expressions for the parameters of the THF model discussed in Sec.\ref{sec.THF map}. The $\delta$-function potential, described by 
\begin{eqnarray}\label{eq_main:delta-potential}
H^{\alpha\alpha}_m(\rr) = \Delta_0 \Omega_0 \sum_{\RR}  \delta (\rr - \RR)
\end{eqnarray}
with $\alpha = 1, 2$, $\Delta_0$ representing moir\'e potential and $\Omega_0$ the moir\'e unit cell area, introduces uniform coupling between momentum states separated by moir\'e reciprocal lattice vectors $\gs \in \GG_M$ as its Fourier components retain equal amplitude for all $\gs$s. It should be noted that we introduce $\delta$-function potential into both the conduction and valence bands in Eq.(\ref{eq_main:delta-potential}) for the convenience of deriving the Green's function formalism below. When $\Delta_0 > 0$, we find the $\delta$-function potential has little influence on the energy states around the conduction band bottom, as discussed in SM Sec.C.1\cite{SM}.  
Thus, we would expect introducing $\delta$-potential with $\Delta_0 > 0$ for both conduction and valence bands in Eq.(\ref{eq_main:delta-potential}) will mainly give rise to the bound states of the valence band and have little influence on low-energy conduction bands. 
In the momentum space, the $\delta$-function potential can be written as
\begin{eqnarray}\label{eq_main_H_delta_momentumspace}
  \hat{H} _ \text m = \Delta_0\sum_{\kk  \alpha } \sum_{\gs, \gs'}  a_{\alpha, \kk + \gs}^\dagger a_{\alpha, \kk+\gs'},
\end{eqnarray}
in the second quantization form, where $\kk$ is chosen within the first mBZ, $\gs, \gs'$ are for any moir\'e reciprocal lattice vector, $\gs, \gs' \in \GG_M$. We note that the $\delta$-function potential couples a state at the momentum $\kk + \gs$ to a state at $\kk+\gs'$ with equal strength $\Delta_0$ for any values of $\gs, \gs'$, and this long-range coupling in momentum space will lead to the divergence of the bound state energy under the ideal 2D $\delta$-function potential (See SM Sec.C.1\cite{SM}). Therefore, we introduce a momentum cutoff $G_\Lambda$, $|\gs|, |\gs'|\leq|G_\Lambda|$, for the $\gs,\gs'$ summations in the moir\'e reciprocal space for Eq.(\ref{eq_main_H_delta_momentumspace}). We also test different choices of approximate forms of the $\delta$-function-like potential in SM Sec.C.1 \cite{SM} and find similar results.


This problem for the $\delta$-function-like potential can be solved using the Green's function formalism\cite{mahan2013many}, which is discussed in SM Sec.B.2\cite{SM}. In momentum space, the full Hamiltonian with the $\delta$-function-like mori\'e potential can be written as 
\begin{eqnarray}
    H_{mC}(\kk) = H_0(\kk) +N_g\Delta_0 P_0 \label{eq_main:Hmc_deltapotential}
\end{eqnarray}
where $H_0(\kk)$ is a block diagonal matrix given by 
\begin{eqnarray}
    H_0(\kk) = \begin{pmatrix}
H_{C} (\kk + \gs_1) & 0 & \dots& 0\\
0 &  H_{C} (\kk + \gs_2) & \dots& 0\\
\vdots & \vdots & \ddots&  \vdots\\
0 & 0 & \dots & H_{C} (\kk + \gs_{N_g})
\end{pmatrix}. \nonumber
\end{eqnarray}
$\gs_j$ is the moir\'e reciprocal lattice vectors within the momentum cut-off $G_\Lambda$ ($j = 1, 2,..., N_g$), and $P_0 = \ket{\phi_0}\bra{\phi_0} \otimes \sigma_0$ is the projection operator with $\ket{\phi_0}=\frac{1}{\sqrt{N_g}} [1, 1, ..., 1]^T$ and $\sigma_0$ is the 2-by-2 identity matrix. For this Hamiltonian, the trace of the Green's function, $G = (E 1_{2N_g\times 2N_g}-H_{mC}(\kk))^{-1}$, can be derived as (See SM Sec.C.2\cite{SM})
\begin{eqnarray}
&& Tr(G) = Tr_\sigma(\mathbf{g}(\kk,E))+Tr_{\sigma}\left(\Delta_0 \sum_j \mathbf{g}^0(\kk+\gs_j, E)\right. \nonumber \\
&& \left.\frac{1}{\sigma_0-\Delta_0\mathbf{g}(\kk,E)} \mathbf{g}^0(\kk+\gs_j, E)\right),\label{eq_main:trace BHZ delta}
\end{eqnarray}
where $Tr_\sigma$ means the trace over 2-by-2 $\sigma$ matrix on the orbital basis $\alpha=1,2$, $\mathbf{g}^0(\kk+\gs_j, E) = (E \sigma_0-H_{C}(\kk+\gs_j))^{-1}$ is the free Green's function of $H_C$ and $\mathbf{g}(\kk, E)=\sum_j \mathbf{g}^0(\kk+\gs_j, E)$. The poles of $Tr(G)$ determine the energy spectrum of $H_{mC}$ (denoted as $\mathcal{E}_n(\kk)$ below). 
For any eigen-energy $E_{\eta}(\kk+\gs_j)$ of $H_{C}(\kk+\gs_j)$, one can prove the limit
$\lim_{E\rightarrow E_{\eta}(\kk+\gs_j)} Tr(G) \approx -\frac{1}{\Delta_0}$
(See SM Sec.C.2 \cite{SM}). 
Therefore, the eigen-energy of $H_{C}(\kk+\gs_j)$ for any $j$ is {\it not} the pole of $Tr(G)$. From Eq.(\ref{eq_main:trace BHZ delta}), the poles of $Tr(G)$ can only come from $\frac{1}{\sigma_0-\Delta_0\mathbf{g}(\kk,E)}$ and thus the eigen-energy spectrum $\mathcal{E}_n(\kk)$ of $H_{mC}$ is determined by 
\begin{eqnarray} \label{eq_main:delta_potential_eqn_energy}
Det(\sigma_0-\Delta_0 \mathbf{g}(\kk,\mathcal{E}_n(\kk)))=0.
\end{eqnarray}

\begin{figure}
\centering
\includegraphics[width=1\columnwidth]{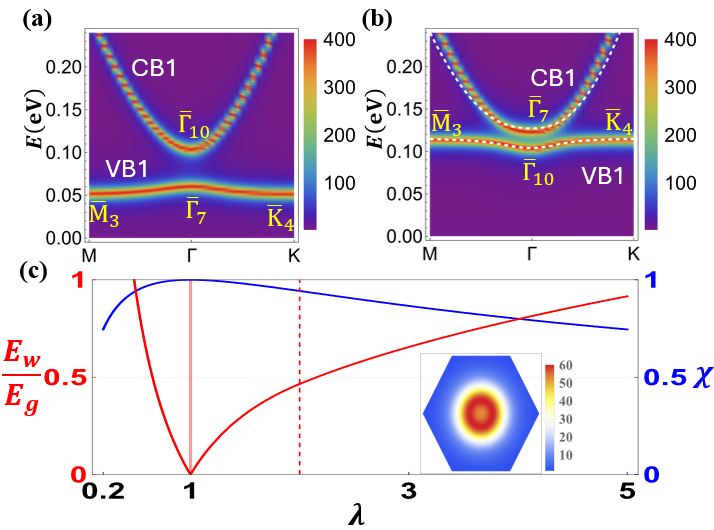}
\caption{\label{fig;3}  (a) and (b) show the energy spectra before and after band inversion for moir\'e Chern-band model with periodic $\delta$-function-like potential. The potential strength $\Delta_0$ is $3.85$meV and $4.05$meV, respectively. In panel (b), the white dashed line 
represents the dispersion obtained from the THF model. (c) $E_w/E_g$ and $\chi=\frac{\int \Omega(\kk) d\kk}{\int  Tr (g(\kk))d\kk }$ 
as a function of $\lambda$. 
When $\lambda=\frac{\Delta E\alpha}{\beta^2}=1$, $\frac{\Omega_{xy}(\kk)}{Tr (g(\kk))}=1$ for every $\kk$ and $E_w/E_g=0$. For the  parameters of the THF fitting to the band dispersion in (b), we find $\lambda=2.01$ indicated by the red dashed line. The insert shows the Berry curvature distribution of topological flat band. }
\end{figure}

The spectrum function $\mathcal{A}(\kk,E)=-\Im(Tr(G))/\pi$ is numerically calculated from Eq.(\ref{eq_main:trace BHZ delta}) and depicted in Fig.\ref{fig;3}(a) and (b), where $\Im$ denotes the imaginary part. In Fig.\ref{fig;3}(a) for a small $\Delta_0$, we find a flat band, denoted as VB1, which is originated from the valence band and is extremely localized at the $1a$ Wyckoff position by the $\delta$-function-like potential (see Fig.S11 in SM Sec.C.2\cite{SM}). Using the perturbation theory (SM Sec.C.2\cite{SM}), we can analytically solve the binding energy of VB1, given by \begin{eqnarray}\label{eq_main:bindingenergy}
    \mathcal{E}_0 = -M-\frac{A^2}{2B}+\frac{ (B-D)G^2_{\Lambda}}{e^\frac{\Omega_B(B-D)}{\Delta_0 \pi}-1}
\end{eqnarray} 
where $\Omega_B = \frac{8 \pi^2}{\sqrt{3}(a^M_0)^2}$ is the area of mBZ. The binding energy in Eq.(\ref{eq_main:bindingenergy}) diverges with $G_\Lambda^2$ and fits well with the full numerical calculations for a large $G_\Lambda$ (See Fig.S12 of SM\cite{SM}). With further increasing the moir\'e potential $\Delta_0$ in Fig.\ref{fig;3}(b), the band inversion between VB1 and CB1 around $\mathbf{\Gamma}$ leads to the emergence of a topological flat mini-band. We also directly calculate the irreps for VB1 at high symmetry momenta, e.g. $\mathbf{\Gamma}$, $\textbf{K}$ and $\textbf{M}$, in Fig.\ref{fig;3}(a) and (b), which coincide with those in Fig.\ref{fig;2}(a) and (c). This is because the potential maximum appears at the $1a$ Wyckoff position in the moir\'e unit cell, and the trivial flat band from the valence band can be regarded as localized $J_z=3/2$ orbitals at the $1a$ Wyckoff position for both cases of moir\'e potential.  
Based on the studies of both the first-shell approximation and $\delta$-function-like potential, we conclude that the physical picture shown in Fig.\ref{fig;1} can be generally applied and that the topological flat band of VB1 can emerge (Fig.\ref{fig;2}(c) and Fig.\ref{fig;3}(b)) once the moir\'e potential strength $\Delta_0$ is larger than the atomic band gap $M$, irregardless of the detailed moir\'e potential form and the twist angle, which only determines the length scale of the moir\'e unit cell. 

\section{The THF model}
\label{sec:THFmodel}
\subsection{Mapping to THF model} \label{sec.THF map}
The band inversion for moir\'e Chern-band model with either the first-shell approximation or periodic $\delta$-function-like potential can be captured by a THF model with the form
\begin{eqnarray}\label{Eq_main:HAM_THF}
&&\hat{H}_{HF}^0 = \sum_{|\kk|<\Lambda_c} \left[\alpha \kk^2 c^{\dagger}_{\kk}c_{\kk}+\Delta E f^{\dagger}_{\kk}f_{\kk}+(\beta k_+c^{\dagger}_{\kk}f_{\kk}+h.c.) \right],\nonumber \\
\label{Eq_main:HAM_THF0}
\end{eqnarray}
where $k_\pm = k_x \pm i k_y$, the annihilation operator $c_{\kk}$ is for c-electrons of CB1 while $f_{\kk}$ is for f-electrons of VB1. 
The parameters $\alpha, \Delta E, \beta$ can be extracted from the moir\'e Chern-band model in Eq.(\ref{eq_main:HammC}) for both types of moir\'e potentials. Particularly, the expressions for the relation between $\alpha, \Delta E, \beta$ in the THF model and Chern-band model parameters such as $M$ and $\Delta_0$ can be derived analytically for the periodic $\delta$-function-like potential, as shown below in Eq.(\ref{eq_main:THFparameter_analytical}). The $\alpha$ term describes the dispersion of c-electrons, the $\Delta E$ term describes the energy difference between the c-electron and f-electron at $\mathbf{\Gamma}$, and the $\beta$ term describes the hybridization between c-electron and f-electron. We have set the energy of the conduction band bottom to be zero. The full THF model should also include the interaction terms $\hat{H}_{HF}=\hat{H}_{HF}^0+\hat{H}_{int}$ with $\hat{H}_{int}$ describing the screened Coulomb interaction previously derived in Ref.    \cite{herzog2024topological}. We focus on the derivation of $\hat{H}_{HF}^0$ in this section. 

For the first-shell potential approximation of moir\'e potential, the mapping to the THF     model can be constructed via the maximally localized Wannier function method    \cite{marzari2012maximally,souza2001maximally}, as detailed in SM Sec.B.2\cite{SM}. To do that, we first construct the maximally localized Wannier function $|W^{(1)}_{\RR}\rangle$ for $J_z=3/2$ orbital at the $1a$ position, as shown in Fig.\ref{fig;2}(d), and its projection to the VB1 and CB1 is shown by the sizes of red  dots in Fig.\ref{fig;2}(c). 
We also consider the wavefunction of the conduction band bottom at $\mathbf{\Gamma}$, denoted as $|\phi_{1,\kk=0}\rangle$, and  project the full Hamiltonian $\hat{H}_{mC}(\kk)$ into the subspace spanned by $|W^{(1)}_{\RR}\rangle$ and $|\phi_{1,\kk=0}\rangle$, leading to the THF model $\hat{H}_{HF}^0$ in Eq.(\ref{Eq_main:HAM_THF0}) with the parameters $\alpha, \beta, \Delta E$ extracted numerically in Eq.(S.B17) 
of SM Sec.B.2 \cite{SM}. 

A similar procedure can be applied to the periodic $\delta$-function-like moir\'e potential to map the moir\'e Chern-band model to the THF model. Due to the simplicity of periodic $\delta$-function-like potential, we can solve all the parameters in the THF model via the L\"owdin perturbation theory    \cite{winkler2003spin}
, given by 
\begin{eqnarray}\label{eq_main:THFparameter_analytical}
\alpha=B+D;\Delta E=\mathcal{E}_0-M-\Delta_0;\beta=\frac{A N_v}{\mathcal{E}_0+M},
\end{eqnarray}
where $N_v$ is a normalization factor of the localized f-electron wavefunction $f^\dagger_0\ket{0}$ 
and $\mathcal{E}_0$ is the binding energy in Eq.(\ref{eq_main:bindingenergy}). More details for the derivations of these parameters can be found in SM Sec.C.3\cite{SM}. We emphasize that $\Delta E$ depends on both $M$ and $\Delta_0$, so it can be tuned by an external gate voltage experimentally. In Fig.\ref{fig;3}(b), the white dashed lines are calculated from the THF model in Eq.(\ref{Eq_main:HAM_THF0}) with the parameters directly extracted from the perturbation theory in Eq.(\ref{eq_main:THFparameter_analytical}), and the dispersion fits well with the full numerical calculations for CB1 and VB1. The momentum-space Berry curvature distribution of the topological flat band in Fig.\ref{fig;3}(b) is calculated for the THF model and plotted in the inset of Fig.\ref{fig;3}(c), which is concentrated around $\mathbf{\Gamma}$ and similar to the result shown in Fig.\ref{fig;2}(e). 


\subsection{Topological Flat Band with Ideal Quantum Geometry in THF model}
The THF Hamiltonian $\hat{H}_{HF}^0$ in Eq.(\ref{Eq_main:HAM_THF0}) can possess an exact flat band with "ideal quantum geometry" when the condition $\Delta E = \beta^2/\alpha $ is satisfied. To see that, we introduce 
the mode operator
\begin{eqnarray}
    \gamma^\dagger_{1, \kk} = (\begin{array}{cc}c^\dagger_\kk&f^\dagger_\kk\end{array}) u_{1, \kk}; \quad u_{1, \kk} =  \sqrt{\frac{1}{\alpha E_\kk}} \left(\begin{array}{c}
      \alpha k_+ \\
      \beta 
    \end{array}\right)
\end{eqnarray}
so that the Hamiltonian (\ref{Eq_main:HAM_THF0}) becomes
\begin{eqnarray}\label{eq_main:HAM_THF0_zeromode}
    \hat{H}_{HF}^0 = \sum_{|\kk|<\Lambda_c} \left[ E_{\kk} \gamma_{1,\kk}^\dagger \gamma_{1,\kk} + (\Delta E - \frac{\beta^2}{\alpha} ) f^\dagger_{\kk} f_{\kk} \right],
\end{eqnarray}
where $E_{\kk} = \alpha k^2 + \frac{\beta^2}{\alpha}$. When $\Delta E = \frac{\beta^2}{\alpha}$, the second term in Eq.(\ref{eq_main:HAM_THF0_zeromode}) vanishes so $\hat{H}_{HF}^0$ only depends on a single mode operator $\gamma^\dagger_{1, \kk}$, but not on the other mode
\begin{eqnarray}\label{eq_main:zeromode_operator}
    \gamma^\dagger_{0, \kk} = (\begin{array}{cc}
       c^\dagger_\kk  & f^\dagger_\kk 
    \end{array}) u_{0, \kk}; \quad  u_{0, \kk} = \sqrt{\frac{1}{\alpha E_\kk}} \left(\begin{array}{c}
      -\beta \\
      \alpha k_-
    \end{array}\right)
\end{eqnarray}
which is orthogonal to $\gamma^\dagger_{1, \kk}$. Thus, the solution of Eq.(\ref{eq_main:zeromode_operator}) corresponds to an exact flat band with zero energy (See Fig.\ref{fig;1}(e)). 
Since the eigen-wavefunction in Eq.(\ref{eq_main:zeromode_operator}) only contains $k_-$ except a normalization factor, this exact flat band realizes "ideal quantum geometry" (See SM Sec.B.3\cite{SM})
\begin{eqnarray}
    Tr(g) = \Omega_{xy} = \frac{2\alpha^2\beta^2}{ (\beta^2 + \alpha^2 k^2 )^2 }, 
\end{eqnarray}
where $\Omega_{xy}=-2 Im (\langle \partial_{k_x}u_{0,\kk}| \partial_{k_y} u_{0,\kk}\rangle )$ 
is the Berry curvature and $g_{ij} = Re \left(\langle \partial_{k_i}u_{0,\kk}| \partial_{k_j} u_{0,\kk}\rangle - \langle\partial_{k_i}u_{0,\kk}|u_{0,\kk}\rangle \langle u_{0,\kk}|\partial_{k_j} u_{0,\kk}\rangle\right)$ ($i,j=x, y$) 
is the Fubini-Study metric tensor of the flat band    \cite{brody2001geometric}. 
We notice that $\Delta E$ describes the energy difference between c-electrons and f-electrons at $\mathbf{\Gamma}$ and is related to the atomic band gap $M$ in Eq.(\ref{eq_main:THFparameter_analytical}), which can be controlled by an external gate voltage (See Fig.\ref{fig;1}(a)). When tuning the dimensionless parameter $\lambda = \Delta E  \frac{\alpha}{\beta^2}$ away from 1, the bandwidth of the flat band increases and the ratio $\chi = \frac{\int \Omega_{xy}(\kk) d\kk}{\int Tr(g(\kk)) d\kk }$  departs from the ideal quantum geometry ($\chi = 1$), as shown in Fig. \ref{fig;3}(c). The energy separation between  the exact flat band ($\gamma_{0, \kk}$) and the dispersive band ($\gamma_{1, \kk}$) is $\Delta E = \frac{\beta^2}{\alpha}$ at ${\bf \Gamma}$. We estimate $\Delta E \approx 40$meV with  $\beta\approx$ 0.2 nm$\cdot$eV and $\alpha\approx$ 1 nm$^2$$\cdot$eV. Therefore, depending on the energy order of the Coulomb interaction energy $U$, our system is in the projected regime for $U \ll \beta \sqrt{\Delta E/\alpha}$ 
and in the Kondo regime when $U > \beta \sqrt{\Delta E/\alpha}$\cite{herzog2024topological}. 


\subsection{FCI phase in the projected regime}

\begin{figure}
\centering
\includegraphics[width=1\columnwidth]{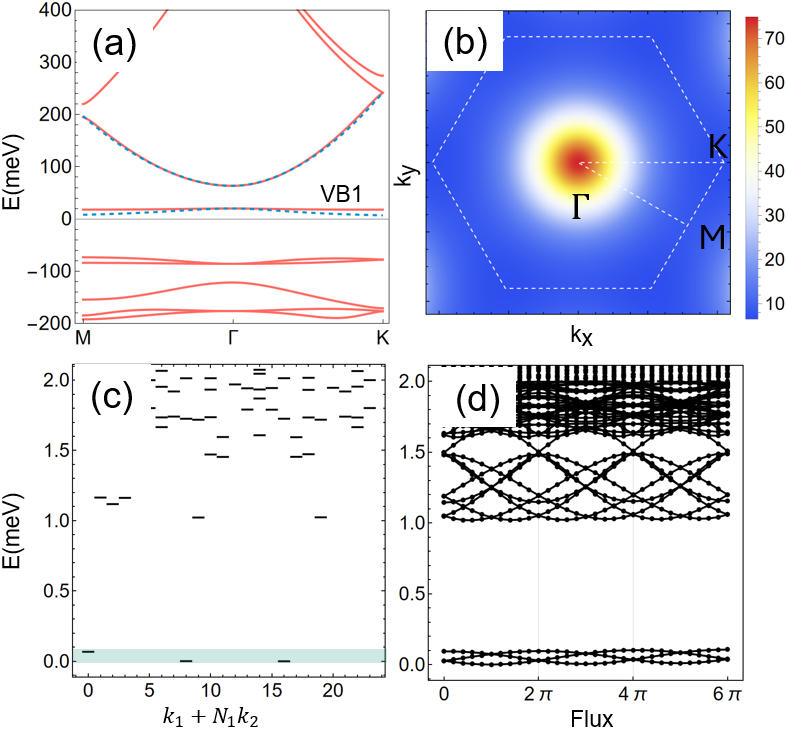}
\caption{\label{fig: fci}  
(a) The single particle spectrum. Red lines give the band structure of the moiré Chern band model with the parameter $M=0.02$eV and $\Delta_0=0.046$eV lying at the peak line of $E_g/E_w$. The blue dashed line is the band structure calculated from the THF model with the parameters $\alpha=1.238$nm$^2\cdot$eV, $\Delta_E=0.043$eV, $\beta=0.27$nm$\cdot$eV. 
(b) The Berry curvature distribution of the band VB1 in the moir\'e BZ indicated by the dashed white lines. 
(c) The many-body spectrum by the ED calculation for the hole filling $-2/3$ of the flat band VB1. The total crystal momentum is $k_1 \mathbf b_1^\text{M}/N_1+k_2 \mathbf{b}_2^\text{M}/N_2$ with $N_1=6,N_2=4$. The blue region includes the nearly degenerate three-fold ground states of FCI. 
(d) The flow of the many-body spectrum under the flux insertion in the $\mathbf b_1^\text{M}$ direction.}
\end{figure}

Based on the exact diagonalization (ED) calculations, we below demonstrate the existence of the FCI phase in the projected regime of the moir\'e Chern band model, which is well captured by the THF model. Fig.~\ref{fig: fci}(a) shows the quantitative agreement between the moir\'e Chern band model and the THF model for $M=0.02$eV and $\Delta_0=0.046$eV, while the other parameters are chosen to be the same as the BHZ Hamiltonian of the HgTe/CdTe quantum wells, as discussed in SM Sec.A.1. Fig.~\ref{fig: fci}(b) depicts the Berry curvature distribution in the mBZ, which is strongly concentrated near the $\Gamma$ point, consistent with the prediction of THF model.

We perform the ED calculations of this model at the hole filling -2/3 of the VB1 labelled in Fig.~\ref{fig: fci}(a) (or equivalently the electron filling 1/3 of the VB1), using a dual gated Coulomb interaction with the potential form 
\begin{equation}
    V(\mathbf q)=\frac{2\pi e^2}{\epsilon_0 \epsilon_r S}\frac{\tanh \vert \mathbf q\vert d}{\vert \mathbf q \vert},   
\end{equation}        
where $e$ is the electron charge, $\epsilon_0$ is the vacuum permittivity, $\epsilon_r=10$ is the relative permittivity, $S$ is the moiré unit cell area, and $d$=30nm is the gate distance. 
The energy scale of the Coulomb interaction is estimated as $U=e^2/(\epsilon_0 \epsilon_r a_0^M )\approx 12$meV, which is a few times smaller than the direct single particle gap 42meV between VB1 and conduction bands. Thus, the system is in the projected regime and we can project the Coulomb interaction into the VB1. We choose the system size to be $6\times 4$ moir\'e unit cells on a torus. The ED spectrum shows three nearly degenerate ground states, which are well separated from other excited states with a gap around 0.95meV, as shown Fig.~\ref{fig: fci}(c). The flux insertion in the $b_1^M$ direction exchanges the three ground states and exhibits $6\pi$ periodicity as shown in Fig~\ref{fig: fci}(d), which is a clear demonstration of the FCI phase, distinguished from the charge density wave phase\cite{regnault2011fractional}. We note that the many-body gap around $0.91$meV persists in the whole spectrum flow. We emphasize that this FCI gap in the spectrum flow remains robust when the Berry curvature becomes more concentrated and the single particle gap is reduced correspondingly. Only when we increase the moir\'e potential parameter to $\Delta_0=0.12$eV and the atomic band gap parameter to $M=0.20$eV, the single particle gap is reduced to 22meV, which approaches $U$, so that the inter-band mixing between the VB1 and the dispersive conduction band becomes significant and the 1-band ED calculations are unreliable. These calculations demonstrate that the moir\'e Chern band with concentrated Berry curvature can still support an FCI ground state, but with a reduced many-body gap, when realistic long-range Coulomb interactions are included in the projected regime and the topological flat band is close to ideal quantum geometry condition. These results are consistent with the early studies of FCIs in flat bands that are close to ideal quantum geometry but possess nonuniform Berry curvature\cite{wang2023origin, shi2025effects, parameswaran2013fractional,bergholtz2013topological,bernevig2025fractional}. 



\section{Material Realization}
\label{sec_material}

\subsection{Semiconductor hetero-structures}
The model Hamiltonian (\ref{eq:Ham_Chernband}) can appear as one spin block in the Bernevig-Hughes-Zhang (BHZ) model \cite{bernevig2006quantum} when there is spin $U(1)$ symmetry or out-of-plane mirror symmetry, 
as discussed in SM Sec.A.1 \cite{SM}, which was originally applied to HgTe/CdTe \cite{bernevig2006quantum,konig2007quantum,konig2008quantum}, InAs/GaSb quantum wells \cite{liu2008quantum,knez2011evidence},  topological insulator films of (Bi,Sb)$_2$(Te,Se)$_3$ compounds \cite{liu2010model,liu2010oscillatory} and Dirac semimetal Cd$_2$As$_3$ thin films \cite{miao2025artificial,rashidi2024tuning}. 
We would expect two degenerate topological flat bands with opposite spin appearing in the BHZ model with moir\'e potential, and these two bands are related to each other by TR symmetry and carry non-trivial $Z_2$ topological invariant. We note that the inverted band ordering at the atomic level is {\it not} necessary, since topological flat bands can occur in the regime of $M > 0$. Therefore, the narrow gap semiconductor heterostructures, such as InAs/AlSb, InSb/InSe or PbTe/PbSnTe    \cite{kroemer20046,wang2020electric,buczko2012pbte},
are also potential candidates. The superlattice potential in semiconductor heterostructures can be achieved by patterning dielectric substrates or via a nearby moir\'e superlattice layer, which has been discussed in the literature    \cite{forsythe2018band, hinnefeld2018graphene,jamalzadeh2025synthetic,zhan2025designing}. In SM Sec.B.1 and C.2 \cite{SM}, we use the moir\'e BHZ model with the material parameters for the HgTe/CdTe quantum wells as a model system for demonstration.

\subsection{2D material hetero-structure with $\mathbf{\Gamma}$-valley}
The existing 2D material databases  \cite{mounet2018two,campi2023expansion,haastrup2018computational,petralanda2024two,jiang20242d} allow for a systematic search of 2D materials for type-II heterostructures with a narrow atomic band gap. We search for two 2D materials, denoted as Material 1 and Material 2, to form a heterostructure based on the following criteria: (1) type-II band alignment between two 2D materials; (2) the location of the CBM of Material 1 and VBM of Material 2 at $\mathbf{\Gamma}$ in the ABZ; (3) different irreps for CBM in Material 1 and VBM in Material 2 at $\mathbf{\Gamma}$  \cite{gao2021irvsp,iraola2022irrep}; (4) the band gap between CBM in Material 1 and VBM in Material 2 smaller than 0.5 eV; and (5) the lattice mismatch smaller than $15\%$. 
Based on these criteria, a list of possible 2D material heterostructures with the symmetry groups $p\bar{3}m1$ is shown in Table SIII of SM Sec.D   \cite{SM}. 

\begin{figure*}
\centering
\includegraphics[width=2\columnwidth]{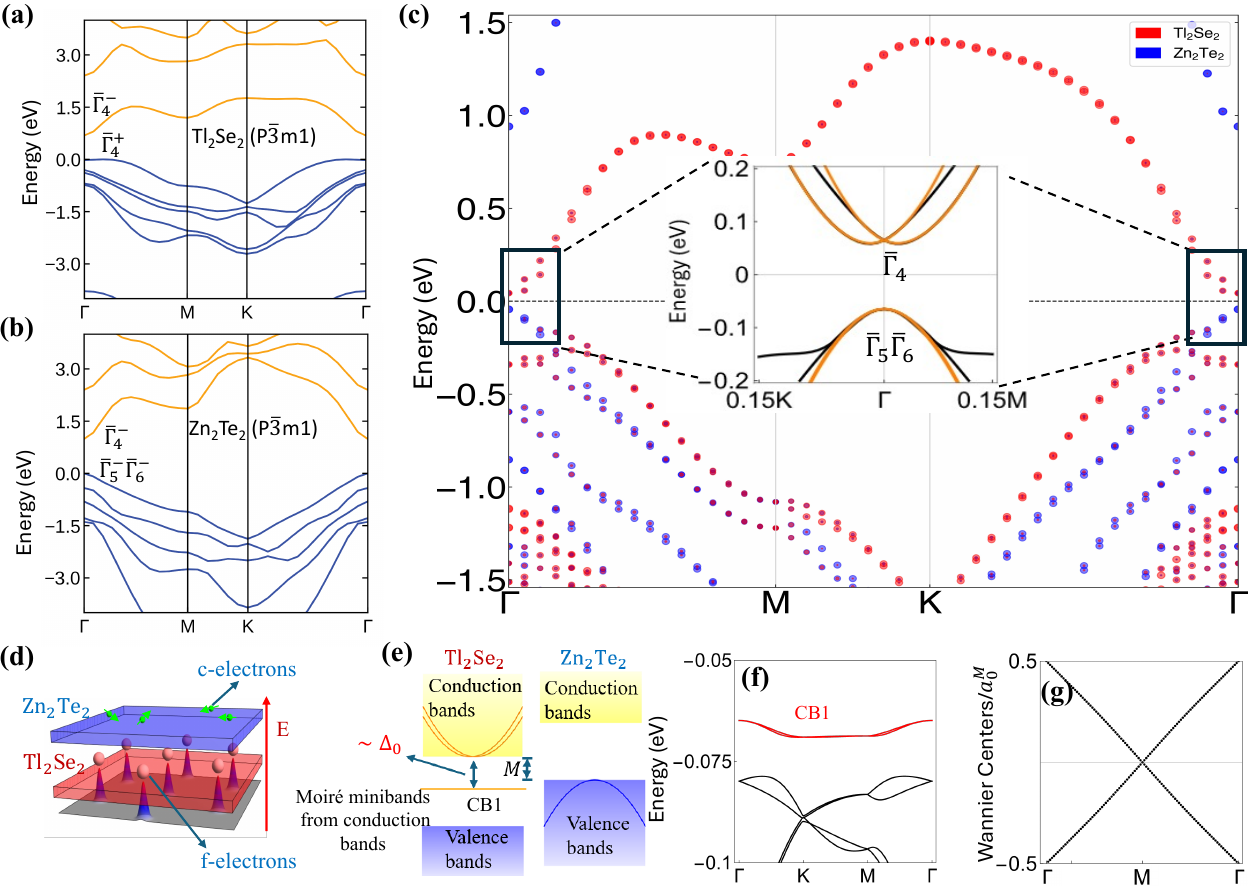}
\caption{\label{fig;4} (a) and (b) show band dispersions near Fermi energy for monolayer Tl$_{2}$Se$_{2}$ and Zn$_{2}$Te$_{2}$, individually. The irreps of conduction band bottom and valence band top at $\mathbf{\Gamma}$ are labeled in the plot for both compounds. (c) The band dispersion of Tl$_2$Se$_2$/Zn$_2$Te$_2$ heterostructure without moir\'e forms a type-II band alignment. The sizes of red and blue dots represent the projections to the Tl$_2$Se$_2$ and Zn$_2$Te$_2$ layers. 
The inset of (c) presents a zoomed-in view of the band dispersion near $\mathbf{\Gamma}$ point from DFT calculation (black curves) and the corresponding effective model fitting (orange curves). 
(d) Schematics for Tl$_2$Se$_2$/Zn$_2$Te$_2$ heterostructure with moir\'e potential applied to Tl$_2$Se$_2$ layer.
(e) Schematics of type-II band alignment. The conduction and valence bands originate from Tl$_2$Se$_2$ and Zn$_2$Te$_2$, respectively. (f) The band dispersion of Tl$_2$Se$_2$/Zn$_2$Te$_2$ heterostructure with moir\'e potential on Tl$_2$Se$_2$ layer. Topological flat bands, labelled by CB1, are highlighted in red color.
(g) The Wannier centers flow of CB1 with a winding
feature demonstrate non-trivial $Z_2$ number of CB1.}
\end{figure*}

Next we will show that the low-energy physics of the heterostructures in Table SIII of SM \cite{SM} can be related to the BHZ model  under moir\'e potentials with additional off-diagonal terms. Here we take Tl$_2$Se$_2$/Zn$_2$Te$_2$ heterostructure as an example with both material compounds belonging to the space group $p\bar{3}m1$ with the wave-vector group of $D_{3d}$ at $\mathbf{\Gamma}$. We will study this heterostructure in three steps. In the first step, we will calculate the electronic band structures and band irreps at ${\bf \Gamma}$ for individual Tl$_2$Se$_2$ and Zn$_2$Te$_2$, as well as the Tl$_2$Se$_2$/Zn$_2$Te$_2$ heterostructure with a matched lattice constant, using the density function theory (DFT) calculations\cite{gao2021irvsp,iraola2022irrep} (Fig.\ref{fig;4}(a), (b) and (c)). Here we relax the lattice constant of Tl$_2$Se$_2$ to match that of Zn$_2$Te$_2$, so that there is no moir\'e pattern formed between Tl$_2$Se$_2$ and Zn$_2$Te$_2$ in this step for the DFT calculations. In the second step, we will build a four-band effective model to reproduce the electronic band structures of Tl$_2$Se$_2$/Zn$_2$Te$_2$ heterostructure within the energy range between $-0.1$eV and $0.1$eV (inset in Fig.\ref{fig;4}(c)). In the last step, we introduce a moir\'e potential into the Tl$_2$Se$_2$ layer using the four-band effective model within the first-shell potential approximation and treat the parameters of moir\'e potential as tuning parameters to identify the regime for the existence of topological flat bands in this system (Fig.\ref{fig;4}(f) and (g)). 

We start from the band structures for individual Tl$_2$Se$_2$ and Zn$_2$Te$_2$ without spin-orbit coupling (SOC). 
The conduction band of Tl$_2$Se$_2$ has a lower energy than that of Zn$_2$Te$_2$, while the valence band of Zn$_2$Te$_2$ has a higher energy than that of Tl$_2$Se$_2$, thus forming a type-II band alignment for Tl$_2$Se$_2$/Zn$_2$Te$_2$ heterostructure.  
According to Table SIII of SM \cite{SM}, the conduction band of Tl$_2$Se$_2$ belongs to 1D irrep $\mathbf{\Gamma}_2^-$ with the z-directional orbital angular momentum $L_z=0$, while the valence band of Zn$_2$Te$_2$ belongs to 2D irrep $\mathbf{\Gamma}_3^-$ with the  $L_z = \pm 1$. The difference in orbital angular momentum  $\Delta L_z = \pm 1$  implies the linear-$k$ coupling between the conduction and valence bands. The atomic band gap between the CBM of Tl$_2$Se$_2$ and the VBM of Zn$_2$Te$_2$ is around $0.2$ eV without SOC, which is in the suitable energy range to form a narrow gap heterostructure.  It should be noted that the inversion symmetry in $D_{3d}$ is broken when Tl$_2$Se$_2$ and Zn$_2$Te$_2$ form a heterostructure, so that the wave-vector group $D_{3d}$ is reduced to $C_{3v}$, and the corresponding irreps $\mathbf{\Gamma}_2^-$ for CBM and $\mathbf{\Gamma}_3^-$ for VBM at $\Gamma$ are reduced to $\mathbf{\Gamma}_1$ with orbital angular momentum $L_z=0$ and $\mathbf{\Gamma}_3$ with $L_z=\pm 1$, respectively, according to the character tables for the $D_{3d}$ and $C_{3v}$ groups shown in Table SIV of Sec.D.2 in SM\cite{SM}. 
Based on the $\mathbf{\Gamma}_1$ and $\mathbf{\Gamma}_3$ irreps for the conduction and valence bands, we can construct the effective Hamiltonian without SOC as
\begin{eqnarray} \label{eq:H0k effecf}
H_0(\kk) =
\begin{pmatrix}
E_1 + A_1 k^2 & i B_0 k_+ & i B_0 k_- \\
-i B_0 k_- & E_2 + A_2 k^2 & 0 \\
-i B_0 k_+ & 0 & E_2 + A_2 k^2
\end{pmatrix}
\end{eqnarray}
on the basis $\{|\mathbf{\Gamma}_1,L=0\rangle, |\mathbf{\Gamma}_3,L=+1\rangle, |\mathbf{\Gamma}_3,L=-1\rangle \}$, where $E_1, E_2, A_1, A_2, B_0$ are material-dependent parameters. $H_0(\kk)$ takes the same form for spin up and spin down basis. 

We next consider the influence of SOC on the electronic band structure using DFT calculations. With SOC, the spin degeneracy for the $L_z=0$ orbital remains and forms the $J_z=\pm \frac{1}{2}$ state, where $J_z$ is the z-directional total angular momentum including orbital angular momentum and spin, while the $L_z=\pm 1$ orbitals are split into the $J_{z}=\pm \frac{3}{2}$ and $J_{z}=\pm \frac{1}{2}$ states. 
According to the character table of the $D_{3d}$ group (Table SIV in SM \cite{SM}), the 1D irrep $\mathbf{\Gamma}_2^-$ becomes the 2D irrep $\bar{\mathbf{\Gamma}}_4^-$, while the 2D irrep $\mathbf{\Gamma}_3^-$ splits into two 2D irreps, $\bar{\mathbf{\Gamma}}_4^-$ and $\bar{\mathbf{\Gamma}}_5^-\bar{\mathbf{\Gamma}}_6^-$. Thus, we expect the CBM of Tl$_2$Se$_2$ to belong to $\bar{\mathbf{\Gamma}}_4^-$ ($J_z = \pm \tfrac{1}{2}$), while the VBM of Zn$_2$Te$_2$ belongs to either $\bar{\mathbf{\Gamma}}_4^-$ ($J_z = \pm \tfrac{1}{2}$) or $\bar{\mathbf{\Gamma}}_5^- \bar{\mathbf{\Gamma}}_6^-$ ($J_z = \pm \tfrac{3}{2}$), depending on the sign of SOC in Zn$_2$Te$_2$. Fig.\ref{fig;4}(a) and (b) show the band structure of Tl$_2$Se$_2$ and Zn$_2$Te$_2$ with SOC, respectively, from the DFT calculations. One finds that the VBM of Zn$_2$Te$_2$ belongs to $\bar{\mathbf{\Gamma}}_5^-\bar{\mathbf{\Gamma}}^-_6$ irrep with $J_{z}=\pm \frac{3}{2}$. The electronic band structure of Tl$_2$Se$_2$/Zn$_2$Te$_2$ heterostructure, in which the lattice constant of Tl$_2$Se$_2$ has been relaxed to match that of Zn$_2$Te$_2$ (no moir\'e pattern), is calculated in Fig.\ref{fig;4}(c) with red and blue colors representing the projections of the bands into the Tl$_2$Se$_2$ and Zn$_2$Te$_2$ layers, respectively. One can see the conduction band is from the Tl$_2$Se$_2$ layer and the topmost valence band around $\mathbf{\Gamma}$  is from the Zn$_2$Te$_2$ layer. Furthermore, the inset of Fig.\ref{fig;4}(c) shows the zoom-in of the band dispersion around $\mathbf{\Gamma}$ (within the black rectangle box) within the energy range between $-0.2$eV and $0.2$eV with the band irreps at $\mathbf{\Gamma}$. We find the CBM at ${\bf \Gamma}$ belongs to $\bar{\mathbf{\Gamma}}_4$ irrep 
($J_{z}=\pm\frac{1}{2}$) while the VBM belongs to $\bar{\mathbf{\Gamma}}_5\bar{\mathbf{\Gamma}}_6$ irrep ($J_{z}=\pm\frac{3}{2}$), consistent with the above symmetry analysis for the positive $\Delta_{SOC}$ case (see SM Sec.D2 \cite{SM}). Based on the symmetry property, the $\kk$-independent SOC term has the form 
\begin{eqnarray}
    &&H^{s}_{SOC} =
\begin{pmatrix}
0 & 0 & 0 \\
0 & s \Delta_{SOC} & 0 \\
0 & 0 & -s \Delta_{SOC}
\end{pmatrix}, 
\end{eqnarray}
within one layer, where we choose $s=+$ and $-$ for spin up $\uparrow$ and spin down $\downarrow$, respectively, and $\Delta_{SOC}$ labels the SOC strength. 

Since both the Hamiltonian $H_0$ and $H_{SOC}$ are block diagonal in the spin basis, we may only consider the spin-up block with a 3-by-3 Hamiltonian
\begin{eqnarray}
    H_{3b,\uparrow} = H_0(\kk) + H^{+}_{SOC}
\end{eqnarray}
on the basis wavefunctions $|\mathbf{\Gamma}_1,J_{z}=\frac{1}{2},\uparrow\rangle |\mathbf{\Gamma}_3,J_{z}=\frac{3}{2},\uparrow\rangle$ and $|\mathbf{\Gamma}_3,J_{z}=-\frac{1}{2},\uparrow\rangle$, which provide one conduction band and two valence bands.
The topmost valence band has the $|\mathbf{\Gamma}_3,J_{z}=\frac{3}{2},\uparrow\rangle$ character for $\Delta_{SOC}>0$ and the $|\mathbf{\Gamma}_3,J_{z}=-\frac{1}{2},\uparrow\rangle$ character for $\Delta_{SOC}<0$. As shown in SM Sec.D.2 \cite{SM}, projecting the Hamiltonian $H_{3b,\uparrow}$ into the subspace of the conduction and topmost valence bands always leads to the Chern-band model $H_C$ in Eq.(\ref{eq:Ham_Chernband}), no matter the sign of $\Delta_{SOC}$, although the basis wavefunctions depend on the sign of $\Delta_{SOC}$. 
In Tl$_2$Se$_2$/Zn$_2$Te$_2$ hetro-structure, our DFT calculations show $\Delta_{SOC}>0$, so that the topmost valence band is from $|\mathbf{\Gamma}_3,J_{z}=\frac{3}{2},\uparrow\rangle$
and the $|\mathbf{\Gamma}_3,J_{z}=-\frac{1}{2},\uparrow\rangle$ state should be projected out. After projecting out this state by perturbation theory, we find $H_{3b,\uparrow}$ is reduced to $H^{\uparrow\uparrow}_{hetero}(\kk)$, given by
\begin{eqnarray} \label{eq_main:Hheteroupup}
    &&H^{\uparrow\uparrow}_{hetero}(\kk) =
\begin{pmatrix}
E_1 + \tilde{A}_1 k^2 & i B_0 k_+ \\
-i B_0 k_- & \tilde{E}_2 + A_2 k^2 \\
\end{pmatrix},
\end{eqnarray}
on the basis wavefunctions $|\mathbf{\Gamma}_1,J_{z}=\frac{1}{2},\uparrow\rangle,$ and $ |\mathbf{\Gamma}_3,J_{z}=\frac{3}{2},\uparrow\rangle$, where $\tilde{A}_1=A_1 + \frac{B_0^2}{E_1 -E_2 + \Delta_{SOC}}$, $\tilde{E}_2=E_2+\Delta_{SOC}$ and other parameters remain the same as those in Eq.(\ref{eq:H0k effecf}).
The TR  partner of $H^{\uparrow\uparrow}_{hetero}(\kk)$ is denoted as $H^{\downarrow\downarrow}_{hetero}(\kk)$, given by

\begin{eqnarray}\label{eq_main:Hheterodowndown}
H^{\downarrow\downarrow}_{hetero}(\kk)=(H^{\uparrow\uparrow}_{hetero}(-\kk))^*,
\end{eqnarray}
on the basis wavefunctions $|\mathbf{\Gamma}_1,J_{z}=-\frac{1}{2},\downarrow\rangle,$ and $ |\mathbf{\Gamma}_3,J_{z}=-\frac{3}{2},\downarrow\rangle$.

For the Tl$_2$Se$_2$/Zn$_2$Te$_2$ heterostructure, the conduction band from the Tl$_2$Se$_2$ layer shows a strong Rashba-type of spin splitting (see the inset of Fig.\ref{fig;4}(c)). Thus, an additional $\kk$-dependent SOC term should be included in the effective model, and the full Hamiltonian reads
\begin{eqnarray} \label{eq:hetero}
    &&H_{hetero}(\kk)=\begin{pmatrix}
H^{\uparrow\uparrow}_{hetero}(\kk) & H^{\uparrow\downarrow}_{hetero}(\kk)  \\
(H^{\uparrow\downarrow}_{hetero}(\kk))^\dagger & H^{\downarrow\downarrow}_{hetero}(\kk) 
\end{pmatrix},
\end{eqnarray}
where $H^{\uparrow\uparrow}_{hetero}(\kk)$ and $H^{\downarrow\downarrow}_{hetero}(\kk)$ are the spin up and down blocks given in Eqs.(\ref{eq_main:Hheteroupup}) and (\ref{eq_main:Hheterodowndown}), respectively.
$H^{\uparrow\downarrow}_{hetero}(\kk)$ takes the form of the Rashba-type of SOC for the conduction band and is given by 
\begin{eqnarray}\label{eq:Rashba}
    &&H^{\uparrow\downarrow}_{hetero}(\kk) =
\begin{pmatrix}
-i R k_- &0  \\
0 & 0 \\
\end{pmatrix}.
\end{eqnarray}
By choosing the parameters $E_1=-\tilde{E}_2=0.0649$ eV, $\tilde{A}_1=16.874$ $\AA^2\cdot$eV, $A_2=-16.80$ $\AA^2\cdot$eV, $B_0=0.909$ $\AA\cdot$eV and $R=0.727$ $\AA\cdot$eV,
we find the band dispersion depicted by the orange lines from the Hamiltonian in Eq.(\ref{eq:hetero}) fits well with that depicted by the black lines from the DFT calculations within the energy range between $-0.1$eV and $0.1$eV in the inset of Fig.\ref{fig;4}(c). It should be noted that the atomic band gap of the Tl$_2$Se$_2$/Zn$_2$Te$_2$ heterostructure (without moir\'e potential) is around $\sim 0.13$ eV, which is smaller than the energy difference $\sim 0.2$eV between the CBM of Tl$_2$Se$_2$ and the VBM of Zn$_2$Te$_2$ (without SOC), shown in Table SIII of SM Sec.D.2\cite{SM}, because the DFT 
calculation of  band structure of the Tl$_2$Se$_2$/Zn$_2$Te$_2$ heterostructure include SOC and changer transfer between two layers and tend to reduce the energy gap.

After establishing the atomic Hamiltonian in Eq.(\ref{eq:hetero}) for the heterostructure, we next consider the moir\'e potential, which can be introduced into either Zn$_2$Te$_2$ layer or Tl$_2$Se$_2$ layer. As the Rashba SOC exists only for the conduction band in the Tl$_2$Se$_2$ layer, but not in the Zn$_2$Te$_2$ layer, it is different for the moir\'e potential occuring in the Tl$_2$Se$_2$ or Zn$_2$Te$_2$ layers. We consider two different configurations for moir\'e potential. The first configuration is to introduce moir\'e potential to the Zn$_2$Te$_2$ layer, which localizes the valence band and pushes the topmost valence mini-band, denoted as VB1, to higher energy and leads to band inversion with the conduction band in the Tl$_2$Se$_2$ layer. However, we find that due to the strong Rashba SOC in the conduction band of the Tl$_2$Se$_2$ layer, the VB1 becomes highly dispersive in the topologically non-trivial regime, as discussed in detail in SM Sec.D.3 \cite{SM}. 


The second configuration is to introduce moir\'e superlattice potential into the conduction band in the Tl$_2$Se$_2$ layer instead, as shown in Fig.\ref{fig;4}(d).  This configuration can be described by the Hamiltonian 
\begin{eqnarray} \label{eq:hetero moire}
    &&\hat{H}=\hat{H}_{hetero}+\hat{H}_{\text{m}},
\end{eqnarray}
where 
\begin{eqnarray} \label{eq: hetero second}
&& \hat{H}_{hetero} = \sum_{\kk \alpha \beta}a_{\alpha \kk }^\dagger 
 [H_{hetero}(\kk)]_{\alpha \beta} a_{\beta \kk},
\end{eqnarray}
with $H_{hetero}(\kk)$ given in Eq.(\ref{eq:hetero}) and the basis index $\alpha, \beta=1,2,3,4$, and the moir\'e potential at the Tl$_2$Se$_2$ layer is given by 
\begin{eqnarray}\label{eq: hetero potential}
    \hat{H} _ \text{m}  = \sum_{\alpha  }\int d^2r  a_{\alpha \rr }^\dagger H ^{\alpha \alpha}_ \text m (\rr) a_{\alpha \rr },
\end{eqnarray} 
with the first shell approximation,  $H_\text{m}^{\alpha\alpha}(\rr)=\sum_{\gs \in \GG^{\mathds{1}}_M} \Delta_0 e^{i \gs \cdot \rr}(\delta_{\alpha,1}+\delta_{\alpha,3})$.  One should note that $\alpha=1, 3$ is for the conduction bands in Tl$_2$Se$_2$ layer and $\alpha=2, 4$ for the valence bands in Zn$_2$Te$_2$ layer, and thus the moir\'e potential only exists at the Tl$_2$Se$_2$ layer.  The moir\'e potential can confine the conduction band into the localized states and decrease the energy of the lowest conduction mini-bands, denoted as CB1 in Fig.\ref{fig;4}(e).  We note that the confining potential quenches the influence of SOC on  CB1 and the spin splitting of  CB1 is reduced for a strong moir\'e potential strength $\Delta_0$. When a strong $\Delta_0$ leads to a band inversion between CB1 and the valence band in the Zn$_2$Te$_2$ layer, the CB1 becomes topologically non-trivial. As shown in Fig.\ref{fig;4}(f) and (g), the CB1 is found to be quite flat with the band width smaller than 10 meV and has negligible spin splitting, while the winding in the Wannier center flow reveals its non-trivial band topology. In addition, for a given moir\'e potential strength $\Delta_0$, the phase transition can be tuned by the atomic gap $M$ via external gate voltages (See Fig.S15 in Sec.D.3 of SM \cite{SM}). It should be noted that the moir\'e potential can be achieved by interfacing Tl$_2$Se$_2$ with another large gap insulating layer to form a hetero-bilayer or by twisting two Tl$_2$Se$_2$ layers. The heterostructure formed by twisted Tl$_2$Se$_2$ homobilayer and Zn$_2$Te$_2$ monolayer is studied in SM Sec.D.4 \cite{SM}.

\section{Conclusion and Discussion}
In this work, we propose a general scheme to realize the nearly ideal topological flat bands using 2D moir\'e heterostructures with type-II band alignment. The ability to continuously tune the atomic band gap $M$ via external gate voltages will allow us to systematically explore different regimes of the THF model discussed in Ref.\cite{herzog2024topological}. The exact flat band with "ideal quantum geometry" was first theoretically discussed for the tractable "chiral" limit of the tBG model   \cite{tarnopolsky2019origin,ledwith2020fractional}, in analog to the Laughlin wavefunctions of the fractional quantum Hall effect, for the realization of FCI phases. However, the chiral limit requires zero tunneling of the AA sites in the tBG model, while the ratio of the tunneling between the AA sites and AB/BA sites in realistic tBG is around $0.7\sim 0.8$   \cite{koshino2018maximally,carr2019exact}, far away from the chiral limit. 
Our design principle of the "ideal" THF model is achievable for a variety of 2D material heterojunctions and semiconductor hetero-structures, as exemplified by our model studies of Tl$_2$Se$_2$/Zn$_2$Te$_2$ moir\'e heterojunction, and HgTe/CdTe quantum wells with patterned dielectric substrates (Sec.B, C and D in SM \cite{SM}). A systematic search of 2D material hetero-structures with type-II band alignment is beyond the scope of this work. Although we focused on the materials with their low-energy bands around ${\bf \Gamma}$ in the ABZ, this scheme of forming heterostructures with type-II band alignment is also possible for other 2D materials with low-energy bands around other high symmetry momenta, such as TMD \cite{tong2017topological,zhu2019gate,zhang2024electronic}. We note that topological heavy fermion physics has recently been revealed in TMD moir\'e systems, MoTe$_2$/WSe$_2$ \cite{zhao2023gate,han2025evidence,guerci2024topological, guerci2023chiral}. We also note that the metallic toy model described in Ref.\cite{soejima2025jellium} possesses ideal quantum geometry as our THF model, but with additional kinetic-energy terms for the study of anomalous Hall crystals.
Our ED calculations in Sec.III.C have demonstrated the existence of the FCI phase in the projected regime, while in the Kondo regime, we estimate the Kondo temperature around $T_K\approx5.8$ meV (see more discussions in SM Sec.C.3 \cite{SM}), which is comparable to the Kondo temperature estimated for twisted bilayer graphene, $T_K\approx1.5$ meV, according to Ref.\cite{zhou2024kondo}.
 Therefore, the ideal topological flat band found in our THF model makes our proposed platform appealing for systematically exploring interacting topological physics such as FCI phase\cite{wang2021exact}, topological Kondo physics   \cite{dzero2016topological,dzero2010topological,checkelsky2024flat,guerci2024topological,zhao2023gate,neupane2013surface}, as well as superconductivity   \cite{song2022magic,herzog2024topological,stewart1984heavy,liu2024electron,liu2025nodal,wang2024molecular}. 
Narrow gap heterostructures with type-II band alignment also provide a platform to explore the physics of exciton insulators and exciton condensate \cite{ma2021strongly,xu2020correlated,gu2022dipolar,nguyen2025quantum}.

\section{Acknowledgment}
We would like to acknowledge Xi Dai, Zhida Song, Jie Wang, Shuolong Yang, Jiabin Yu and Jun Zhu for the helpful discussion. 
Y.L. and C.L. acknowledge the support from the Penn State Materials Research Science and Engineering Center for Nanoscale Science under National Science Foundation award DMR-2011839. B. A. B was supported by DOE Grant No. DESC0016239. H.H. acknowledges the support by the Gordon and Betty Moore Foundation through Grant No. GBMF8685 towards the Princeton theory program, the Gordon and Betty Moore Foundation’s EPiQS Initiative (Grant No. GBMF11070), the Office of Naval Research (ONR Grant No. N00014-20-1-2303), the Global Collaborative Network Grant at Princeton University, the Simons Investigator (Grant No. 404513), the NSF-MERSEC (Grant No. MERSEC DMR 2011750), the Simons Collaboration on New Frontiers in Superconductivity, and the Schmidt Foundation  and the Princeton Catalysis Initiative at the Princeton University. A.A., Z.F., and Q.Y. are supported by the U.S. National Science Foundation under grant No. DMR-2323469.



\bibliography{ref}

\end{document}



\title{Supplementary Materials for Ideal Topological Flat Bands in Two-dimensional Moir\'e Heterostructures with Type-II Band Alignment}

\author{Yunzhe Liu}
\affiliation{Department of Physics, The Pennsylvania State University, University Park, Pennsylvania 16802, USA}
\author{Anoj Aryal}
\affiliation{Department of Physics, Northeastern University, MA 02115} 
\author{Kaijie Yang}
\affiliation{Department of Materials Science and Engineering, University of Washington, Seattle, Washington 98195, USA}
\author{Dumitru Calugaru}
\affiliation{Department of Physics, Princeton University, Princeton, NJ 08544}
\author{Zhenyao Fang}
\affiliation{Department of Physics, Northeastern University, MA 02115} 
\author{Haoyu Hu}
\affiliation{Donostia International Physics Center, P. Manuel de Lardizabal 4, 20018 Donostia-San Sebastian, Spain}
\author{Qimin Yan}
\affiliation{Department of Physics, Northeastern University, MA 02115} 
\author{B. Andrei Bernevig}
\affiliation{Department of Physics, Princeton University, Princeton, NJ 08544} 
\affiliation{Donostia International Physics Center, P. Manuel de Lardizabal 4, 20018 Donostia-San Sebastian, Spain}
\affiliation{IKERBASQUE, Basque Foundation for Science, Bilbao, Spain}
\author{Chao-xing Liu$^*$}
\affiliation{Department of Physics, The Pennsylvania State University, University Park,  Pennsylvania 16802, USA}

\maketitle


\appendix


\section{Moir\'e Chern-band model}
This section introduces the moir\'e Chern-band model and demonstrates how it captures the essential physics of the Bernevig-Hughes-Zhang (BHZ) model under moir\'e superlattice potentials.

The Chern-band model can be written as 
\begin{eqnarray}
&& \hat{H}_{C} = \sum_{\kk \alpha \beta}a_{\alpha \kk }^\dagger 
 [H_{C}(\kk)]_{\alpha \beta} a_{\beta \kk}, \\
&& H_C(\kk) =  \epsilon(\kk)I_{2\times 2} + \left( \begin{array}{cc} \mathcal{M}(\kk)&Ak_+\\ 
  A k_-& -\mathcal{M}(\kk)
 \end{array} \right), \label{Eq:Ham_Chernband} 
\end{eqnarray}
where $a_{\alpha \kk}$ ($a^\dagger_{\alpha \kk}$) is a fermion annihilation (creation) operator with $\alpha = 1, 2$ indexing orbital basis of $H_C$, $\kk$ is the momentum in the atomic Brillouin zone (ABZ), $k_\pm=k_x\pm i k_y $, $\mathcal{M}(\kk)=M+B \kk^2$, $\epsilon(\kk)=D\kk^2$, and $I_{2\times 2}$ is a 2-by-2 identity matrix \cite{qi2006topological,chang2023colloquium}.
The lattice version of this two-band model has been well studied in  Ref.  \cite{bernevig2006quantum,qi2011topological}. A topological phase transition can be induced by changing the sign of the atomic band gap parameter $M$. This model is in the quantum anomalous Hall state with the Chern number $C=\pm 1$ when $MB<0$ and in the trivial insulator state with $C=0$ when $MB>0$.

In addition to $H_C$, we also consider a superlattice potential, which can be generated either by moir\'e pattern  \cite{andrei2020graphene,mak2022semiconductor,devakul2021magic} or by patterning hole arrays in dielectric materials  \cite{forsythe2018band,wang2023formation}. We denote the superlattice potential as $H_{\text{m}}(\rr)$, which satisfies $H_{\text{m}}(\rr+\RR) = H_{\text{m}}(\rr) $, where $\RR$ is the moir\'e superlattice vector. Here we consider hexagonal moir\'e potential for illustration, so that $\RR= n_1 \aa^M_1 + n_2 \aa^M_2$ with $n_1, n_2 \in \mathcal{Z}$ and $\aa^M_1=(1,0)a^M_0$, $\aa^M_2=(\frac{1}{2},\frac{\sqrt{3}}{2})a^M_0$, where $a^M_0$ is the moir\'e primitive lattice constant and chosen to be $|a^{M}_0|=11.46nm$ in the BHZ model calculation below. The choice of moir\'e lattice constant will only affect the size of moir\'e Brillouin zone (mBZ), but will not change any essential physics discussed below. The moir\'e potential Hamiltonian $H_{\text{m}}(\rr)$ is generally given by 
\begin{eqnarray}
&& \hat{H}_{\text{m}}  = \int d^2r  a_{\alpha \rr }^\dagger H ^{\alpha \alpha}_ \text m (\rr) a_{\alpha \rr },   \label{Eq:moirepotential}
\end{eqnarray}
and can be expanded in the moir\'e reciprocal space as
\begin{eqnarray}
&& H^{\alpha\alpha} _ \text m (\rr) = \sum_{\gs \in \GG_M} \Delta_{\alpha}(\gs) e^{i \gs \cdot \rr},\label{Eq:moirepotential_1}
\end{eqnarray}
where $a_{\alpha \rr }=\frac{1}{\sqrt{\Omega_V}}\int d^2k a_{\alpha \kk } e ^{i \kk \cdot \rr}$ with the integral over the momentum $\kk$ in the ABZ and the system area $\Omega_V$. $\GG_M = \{ n_1 \bb^M_1 + n_2 \bb^M_2 | n_1, n_2 \in \mathcal{Z}  \}$ labels the group of the moir\'e reciprocal lattice vectors. The primitive moir\'e reciprocal lattice vectors are $\bb^M_1=(2\pi,-\frac{2\pi}{\sqrt{3}})\frac{1}{a^M_0}$, $\bb^M_2=(0,\frac{4\pi}{\sqrt{3}})\frac{1}{a^M_0}$. 

The full Hamiltonian of our moir\'e Chern-band model is given by
\begin{eqnarray} 
\hat{H}_{mC}=\hat{H}_{C}+\hat{H}_{\text{m}}. \label{Eq:HammC}
\end{eqnarray}
This model Hamiltonian serves as our starting point for the construction of the topological heavy fermion model.  
The symmetry group $G$ of $H_{mC}$ is determined by $G = G^\text M \cap G^0_{C}$, where $G^\text M $ is the space symmetry group of moir\'e superlattice potential ($\hat{H}_{\text{m}}$), and $G^0_{C}$ is the space symmetry group of the atomic model ($H_{C})$. 
The Chern-band model (\ref{Eq:Ham_Chernband}) possesses continuous translation symmetry and full z-directional rotation symmetry ($C_{\theta}$) 
with the representation matrix 
\begin{equation}
D(C_{\theta})= \left( \begin{array}{cc} e^{-i \frac{\theta}{2}} &0  \\
     0&e^{-i \frac{3\theta}{2}}
\end{array}\right)
\end{equation}
on the basis of Chern-band model, where $\theta$ is the rotation angle. This basis corresponds to the angular momentum $\frac{1}{2}$ and $\frac{3}{2}$ for $\alpha=1$ and 2 orbitals. The moir\'e potential $\hat{H}_{\text{m}}$ in Eq.(\ref{Eq:moirepotential_1}) breaks the continuous translation symmetry down to the discrete translation symmetry with the lattice vectors set by $\RR$, and lowers the full rotation symmetry down to the six-fold rotation symmetry $C_{6z}$.

It should be noted that the moir\'e Chern-band model breaks time reversal (TR) symmetry and thus in a TR-invariant system, it should only describe one block of the full Hamiltonian. As we will show below, the moir\'e Chern-band model can describe the spin-up block of the BHZ model under moir\'e superlattice potentials while the spin-down block is its TR partner. 


\subsection{Bernevig-Hughes-Zhang model with moir\'e superlattice potential}
\label{Sec:IA_BHZmoire}
We first consider the BHZ model  with Moir\'e superlattice potential. The BHZ model was first discussed to describe a class of semiconductor quantum wells (QWs), e.g. HgTe/CdTe QWs \cite{bernevig2006quantum} and InAs/GaSb QWs \cite{liu2008quantum}. 
In Sec.\ref{Sec_SM:2Dmaterial}, we will also demonstrate that the BHZ model with additional off-diagonal terms is also applicable to 2D material heterostructures featuring low-energy bands around the $\mathbf{\Gamma}$ point and type-II band alignment, where the conduction and valence bands spatially reside in distinct material layers. 

The BHZ model with the spin U(1) symmetry or out-of-plane mirror symmetry is two copies of the Chern-band models that are related by time reversal (TR) symmetry and can be written as
\begin{eqnarray}
&& \hat{H}_{BHZ} = \sum_{\kk \alpha \beta}a_{\alpha \kk }^\dagger 
 H^{\alpha \beta}_{BHZ}(\kk) a_{\beta \kk} \label{Eq:HBHZ_kspace},\\
&&H_{BHZ}(\kk) = \left( \begin{array}{cc} H_{C}(\kk) & 0 \\ 
0 & H^*_{C}(-\kk)
 \end{array} \right) \label{Eq:H BHZ}, 
\end{eqnarray}
where $a_{\alpha \kk}$ ($a^\dagger_{\alpha \kk}$) is a fermion annihilation (creation) operator with $\alpha, \beta = 1, ..., 4$ indexing four basis of $H_{BHZ}$ and $H_C(\kk)$ is the Chern-band model Hamiltonian (\ref{Eq:Ham_Chernband}). The basis indices $\alpha, \beta = 1, ..., 4$ of the BHZ Hamiltonian correspond to the wavefunctions $\ket{S,\uparrow}$,$\ket{P_+,\uparrow}$, $\ket{S,\downarrow}$ and $\ket{-P_-,\downarrow}$, where $P_\pm$ and $S $ represent $P_x\pm i P_y$ (with orbital angular momentum $\pm 1$) and $S$ orbitals (with orbital angular momentum $0$) while the arrows $\uparrow, \downarrow$ represent spin up and spin down, respectively. The Hamiltonian $H_{BHZ}(\kk)$ has a block diagonal form due to spin $U(1)$ symmetry or the out-of-plane mirror symmetry $M_z$. The basis wave functions $\ket{S,\uparrow}$ and $\ket{P_+,\uparrow}$ of the upper block $H_{C}(\kk)$ are for the basis with $\alpha, \beta = 1, 2$ and possess the $M_z$ eigenvalue $+i$, while the wave functions $\ket{S,\downarrow}$ and $\ket{-P_-,\downarrow}$ of the lower block $H^*_C(\kk)$ are for the basis with $\alpha, \beta = 3, 4$ and possess the $M_z$ eigenvalue $-i$. These two blocks are connected to each other by TR symmetry. In the numerical calculations below, we adopt the material dependent parameters for the HgTe/CdTe quantum wells as $A=0.365$ $nm\cdot$eV, $B=0.706$ $nm^2 \cdot$eV, and $D=0.532$ $nm^2\cdot$eV \cite{rothe2010fingerprint}. 


The moir\'e potential also has the diagonal form
\begin{eqnarray}
    \hat{H}_{\text{m}}  = \int d^2r  a_{\alpha \rr }^\dagger H ^{\alpha\alpha}_ \text m (\rr) a_{\alpha \rr } ,  \label{Eq:moirepotential_BHZ}
\end{eqnarray}
with $H ^{\alpha\alpha}_ \text m (\rr)$ given by Eq.(\ref{Eq:moirepotential_1}) and $\alpha = 1, 2, 3, 4$. The full Hamiltonian of  moir\'e BHZ model is given by
\begin{eqnarray} 
\hat{H}_{mBHZ}=\hat{H}_{BHZ}+\hat{H}_{\text{m}}. \label{Eq:HamMBHZ}
\end{eqnarray}
We notice that the $H_{mBHZ}$ is block-diagonal with its upper block for spin up state ($M_z=+i$) equivalent to the moir\'e Chern-band Hamiltonian $\hat{H}_{mC}$ in Eq.(\ref{Eq:HammC}), while its lower block for spin down state ($M_z=-i$) is the TR partner of $\hat{H}_{mC}$. 

The crystalline symmetry group $G$ of $H_{mBHZ}$ is determined by $G = G^\text M \cap G^0$, where $G^\text M $ is the space symmetry group of moir\'e superlattice potential ($H_{\text{m}}$), and $G^0$ is the space symmetry group of the atomic model ($H_{BHZ}$). For the effective Hamiltonian $H_{BHZ}$ of the BHZ model, we denote $G^0 = D_{\infty h} \otimes G^\text T$, where $G^\text T$ is the continuous translation group and $D_{\infty h}$ includes the rotation group about the z-axis, in-plane mirrors, and inversion. To be concrete, we consider a hexagonal moir\'e potential and,  for the illustrative purpose, choose the symmetry group $G^\text M$ of $\hat{H}_{\text m}$ to be $P6/mmm$, giving rise to $G = G^\text M \cap G^0 = P6/mmm$ for the whole system. It should be noted that choosing different types of moir\'e potential will not change the main physics discussed below. 

\section{Moir\'e superlattice potential with the first-shell approximation} \label{sec:first shell}
In this section, we will discuss the emergence of the topological flat bands in the Chern-band model and the BHZ model with moir\'e superlattice potential under the first-shell approximation. We here mainly use the parameter set from HgTe/CdTe quantum wells \cite{rothe2010fingerprint}. 


\subsection{Topological flat mini-bands and Phase diagram of moir\'e BHZ model}\label{sec:BHZ1}
In this section, we consider the Moir\'e superlattice potential within the first-shell potential approximation, namely the summation of $\gs$ in Eq.(\ref{Eq:moirepotential_1}) which is limited to the first shell, defined by \begin{eqnarray} \label{Eq:first shell potential}
    \GG^{\mathds{1}}_M=\left\{\bb^M_2,\bb^M_1,-\bb^M_2,-\bb^M_1,\bb^M_2+\bb^M_1,-\bb^M_2-\bb^M_1\right\}.
\end{eqnarray}
 These six momenta in $\GG^{\mathds{1}}_M$ 
can be connected by six-fold rotation symmetry $C_{6z}$ around the z axis in the space group $P6/mmm$. Furthermore, we consider the moir\'e potential existing only for the valence band, as discussed in Fig.1 in the main text, and thus only a single real parameter, $\Delta_2(\gs)=\Delta_4(\gs)=\Delta_0$ for $\gs \in \GG^{\mathds{1}}_M$ and $\Delta_1=\Delta_3=0$, can characterize the moir\'e potential strength within the first shell approximation. 


The full Hamiltonian $H_{mBHZ}$ described by Eqs.(\ref{Eq:HBHZ_kspace}), (\ref{Eq:moirepotential_BHZ}) and (\ref{Eq:HamMBHZ}) can be written in the momentum space as
\begin{eqnarray}\label{Eq:moireBHZmodel K space}
&& \hat{H}_{BHZ} = \sum_{\kk \gs}\sum^4_{ \alpha \beta=1}a_{\alpha \kk \gs}^\dagger 
 H^{\alpha \beta}_{BHZ}(\kk+\gs) a_{\beta \kk \gs} \label{Eq:HBHZ_Cmodel_kspace} \\
&& \hat{H}_{\text{m}} = \Delta_0\sum_{\kk \gs \gs' \alpha } \sum_{\qq \in \GG^{\mathds{1}}_M } \delta(\gs' -\gs -\qq) a_{\alpha \kk \gs}^\dagger a_{\alpha \kk \gs'}(\delta_{\alpha,2}+\delta_{\alpha,4}) \label{Eq:HM_kspace},
\end{eqnarray}
where $a_{\alpha \kk \gs}$ is a fermion annihilation operator defined in Eq.(\ref{Eq:HBHZ_kspace}) with 
$\alpha$ and $\beta = 1,...,4$, 
$\kk$ is the momentum in the first mBZ, $\gs, \gs'\in \GG_M$ ($\GG_M$ is the group of moir\'e reciprocal lattice vectors) and $\qq \in \GG^{\mathds{1}}_M $ are the moir\'e reciprocal vectors within the first momentum shell. Here we only consider the moir\'e potential acting on the $\alpha=2, 4$ componentsfor the valence bands. To diagonalize the above moir\'e Hamiltonian in the momentum space numerically, we set the cut-off $G_{\Lambda} =6 |\bb^M_1|$ for the summation of  $\gs, \gs' \in \GG_M$ in Eq.(\ref{Eq:moireBHZmodel K space}) and Eq.(\ref{Eq:HM_kspace}). 

As discussed in Sec.\ref{Sec:IA_BHZmoire}, the BHZ model is block-diagonal, with each block characterized by the eigenvalue of the $M_z$ symmetry and representing a moir\'e Chern-band model. We may consider the $M_z=+i$ block, and in momentum space, the corresponding moir\'e Chern-band model is written as
\begin{eqnarray}\label{Eq:moirecmodel K space}
&&\hat{H}_{mC}=\sum_{\kk \gs \gs' }\sum^2_{\alpha, \beta=1}a_{\alpha \kk \gs'}^\dagger 
 H^{\alpha \beta,\gs'\gs}_{mC}(\kk) a_{\beta \kk \gs}  =\hat{H}_{C}+\hat{H}_{\text{m}},\nonumber\\
&&\hat{H}_{C} = \sum_{\kk \gs }\sum^2_{\alpha, \beta=1}a_{\alpha \kk \gs}^\dagger 
 H^{\alpha \beta}_{BHZ}(\kk+\gs) a_{\beta \kk \gs}  ,\nonumber\\
&& \hat{H}_{\text{m}} = \Delta_0\sum_{\kk \gs \gs' } \sum_{\qq \in \GG^{\mathds{1}}_M } \delta(\gs' -\gs -\qq) a_{2 \kk \gs}^\dagger a_{2 \kk \gs'}. 
\label{Eq:Hc_kspace}
\end{eqnarray}

\begin{figure}
\centering
\includegraphics[width=1\textwidth]{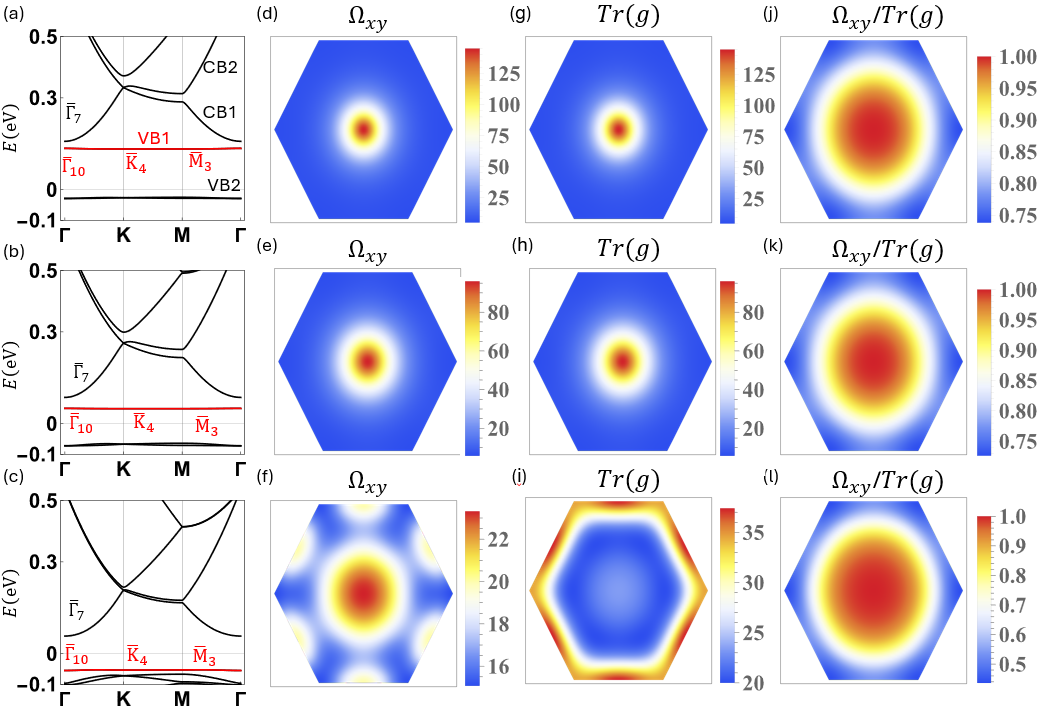}
\caption{\justifying \label{band} (a)-(c) Band dispersions of the moir\'e Chern-band model computed with the parameter sets: 
(a) $M = 0.135$ \text{eV}, $\Delta_0 = 0.094 \,\text{eV}$, 
(b) $M = 0.05$ \text{eV}, $\Delta_0$ = 0.059 \text{eV}, and 
(c) $M = -0.05$ \text{eV}, $ \Delta_0 = 0.011$ \text{eV}. All parameters are along the red dashed lines in the phase diagram of Fig.1(f) in the main text. (d)-(h) The berry curvature $\Omega_{xy}$ of VB1 in the (a)-(c).  (g)-(i) The trace of quantum metric $Tr(g)$ of VB1 in the (a)-(c).  (j)-(l) $\Omega_{xy}/Tr(g)$ of VB1 in the (a)-(c).} 
\end{figure}

We can numerically diagonalize the Hamiltonian of moir\'e Chern-band model Hamiltonian in Eq.(\ref{Eq:Hc_kspace}) and calculate the energy dispersion. The moir\'e BHZ model in Eq.(\ref{Eq:moireBHZmodel K space}) shows exactly the same energy dispersion but with each band doubly degenerate because of the combination of inversion and TR. Therefore, the band dispersion of flat bands and phase diagram are identical to the discussion of the moir\'e Chern-band model in the main text. Here, we focus on the parameter sets for the flat band region  and show the energy dispersions in Fig.\ref{band}(a)-(c) for three different sets of values for $\Delta_0$ and $M$.  
The top valence mini-bands, denoted as the VB1 highlighted by the red color 
in Fig.\ref{band}(a)-(c), can be quite flat with the bandwidth smaller than 10 meV.

We further follow the method in Ref.\cite{song2022magic,song2019all} to  analyze the band irreducible representation (irreps) of the VB1 in Fig.\ref{band} at high symmetry momenta. The symmetry group of $\hat{H}_{mC}$ in Eq.(\ref{Eq:Hc_kspace}) is described by the space group $P6$, with the wave-vector groups at high symmetry momenta $\mathbf{\Gamma}$, $\textbf{K}$, $\textbf{M}$ to be the point groups $C_{6}$, $C_{3}$ and $C_{2}$, respectively. The character tables for these wave-vector groups are shown in  Tab.\ref{tab:P6/mmm charcter}. The six-fold rotation $C_{6z}$ acts on fermion operators as 
\begin{eqnarray}
    C_{6z}a^\dagger_{\alpha, \kk,  \gs}C^{-1}_{6z}=\sum_{\alpha'\gs'}a^\dagger_{\alpha', C_{6z} \kk,  \gs'}D_{\alpha'\gs',\alpha\gs}(C_{6z}).
\end{eqnarray}
with
\begin{eqnarray}\label{Eq: D of C6z}
    D_{\alpha'\gs',\alpha\gs}(C_{6z})=\delta_{\gs',C_{6z}\gs}\delta_{\alpha',\alpha}e^{-i\frac{(2\alpha-1)\pi}{6}}.
\end{eqnarray}
Under $C_{6z}$, $H_{mC}(\kk)$ is invariant
\begin{eqnarray}
   D(C_{6z})H_{mC}(\kk) D^{-1}(C_{6z})=H_{mC}(C_{6z}\kk).
\end{eqnarray}
Based on the $D(C_{6z})$, we can get the representation matrix of two-fold rotation $C_{2z}$ and three-fold rotation $C_{3z}$ given by
\begin{eqnarray}
D(C_{2z})=\left[D(C_{6z})\right]^3;D(C_{3z})=\left[D(C_{6z})\right]^2.
\end{eqnarray} 
For the eigenstates of $H_{mC}(\kk)$, the band irreps at $\mathbf{\Gamma}$, $\textbf{K}$, and $\textbf{M}$ can be determined from the eigenvalues of the  $C_{6z}$, $C_{3z}$, and $C_{2z}$. In Fig.\ref{band}(a)-(c), the $C_{6z}$, $C_{3z}$, and $C_{2z}$ eigenvalues of VB1 at the high symmetry momenta in mBZ can be calculated as $\eta(\mathbf{\Gamma}) = e^{-i \frac{\pi}{6}}$, $\theta(\mathbf{K}) = -1$, and $\epsilon(\mathbf{M}) = i$, respectively, and thus, the band irreps of VB1 is $(\bar{\mathbf{\Gamma}}_{10}, \bar{\textbf{K}}_4,\bar{\textbf{M}}_3)$.
Due to the $C_{6z}$ symmetry, the Chern number of the VB1 can be related to the rotation eigenvalues \cite{fang2012bulk} via
\begin{eqnarray}\label{Eq: cm c6}
    e^{i C \pi/3}=- \eta(\mathbf{\Gamma}) \theta(	\textbf{K})\epsilon(\textbf{M}) 
    =  e^{i  \pi/3} ,
\end{eqnarray}  
leading to $C = 1$ mod $6$, which is consistent with the direct calculations of the Chern number for VB1. 

 Based on the character table in Tab.\ref{tab:P6/mmm charcter}, the band irreps of VB1, $(\bar{\mathbf{\Gamma}}_{10}, \bar{\textbf{K}}_4,\bar{\textbf{M}}_3)$, do not belong to any elementary band representation (EBR) from the
Bilbao Crystallographic Server \cite{bradlyn2017topological,vergniory2017graph,elcoro2017double}, implying the topological nature of the flat mini-band VB1. We notice that if the irreps of VB1 ($\bar{\mathbf{\Gamma}}_{10}$) and CB1  ($\bar{\mathbf{\Gamma}}_{7}$) at $\mathbf{\Gamma}$ point exchange, the VB1 with irreps $(\bar{\mathbf{\Gamma}}_{7}, \bar{\textbf{K}}_4,\bar{\textbf{M}}_3)$ can form an EBR $^2\bar{E}_{1}\uparrow \text{G}(1)@1a$ as shown in Tab.\ref{TAB:EBR}, which means a one-dimensional induced representation of space group $P6$  induced from the 
representation $^2\bar{E}_{1}$ of site-symmetry group $C_{6}$  
at the Wyckoff position $1a$ (the center of moir\'e unit cell).Here the site-symmetry group describes all the symmetry operations in the space group that leave a certain site invariant. 
This EBR corresponds to the orbital with the z-directional total angular momentum $J_z=\frac{3}{2}$ at the $1a$ position, schematically shown in Fig.\ref{wannier orbital}(b).

In addition, the energy of the flat band can be tuned by the values of $\Delta_0$ and $M$. For a small $\Delta_0$ and negative $M$, the VB1 band lies close to the second valence miniband, denoted as VB2 in Fig.~\ref{band}(c).
VB1 moves to the bottom of the conduction mini-bands, denoted as the CB1 in Fig.\ref{band}(a), with a large $\Delta_0$ and positive $M$. The Berry curvature distribution $\Omega_{xy}$ and the trace of the quantum metric $Tr(g)$ of the VB1 in the mBZ are depicted in Fig.\ref{band}(d)-(f) and (g)-(i), respectively. When the VB1 is close to the VB2, $Tr(g)$ is concentrated around the $\textbf{M}$ point, while the $\Omega_{xy}$ is more uniform in mBZ, as shown in Fig.\ref{band}(f) and (i). For a large $\Delta_0$, the VB1 moves close to the bottom of the CB1, and both $\Omega_{xy}$ and $Tr(g)$ are concentrated around the $\mathbf{\Gamma}$ point, as shown in Fig.\ref{band}(d),(e) and (g),(h). Fig.\ref{band}(j)-(l) show the ratio $\Omega_{xy}/Tr(g)$. When the VB1 is close to the CB1 bottom, we find $\Omega_{xy}/Tr(g)$ close to 1 near the $\mathbf{\Gamma}$ point of mBZ, as shown in Fig.\ref{band}(j) and (k). In this work, we focus on the large $\Delta_0$ regime, where both $\mathrm{Tr}(g)$ and $\Omega_{xy}$ of the topological flat band VB1 are concentrated around the $\mathbf{\Gamma}$ point, shown in Fig.\ref{band}(a), (d) and (g). 

\begin{table}[h]
\centering
\renewcommand{\arraystretch}{1.2}
\begin{tabular}{c|cccccc|cccc|ccc}
\hline
 & $\bar{\mathbf{\Gamma}}_7$ & $\bar{\mathbf{\Gamma}}_8$ & $\bar{\mathbf{\Gamma}}_9$ & $\bar{\mathbf{\Gamma}}_{10}$ & $\bar{\mathbf{\Gamma}}_{11}$ & $\bar{\mathbf{\Gamma}}_{12}$ &
 & $\bar{\textbf{K}}_4$ & $\bar{\textbf{K}}_5$ & $\bar{\textbf{K}}_6$ 
& & $\bar{\textbf{M}}_3$ & $\bar{\textbf{M}}_4$ \\
\hline
$E$ & 1 & 1 & 1 & 1 & 1 & 1 & $E$
    & 1 & 1 & 1 
   & $E$ & 1 & 1 \\
$C_{3z}$ & $-1$ & $-1$ & $e^{-i\frac{\pi}{3}}$ & $e^{-i\frac{\pi}{3}}$ & $e^{i\frac{\pi}{3}}$ & $e^{i\frac{\pi}{3}}$ 
        & $C_{3z}$  & $-1$ & $e^{-i\frac{\pi}{3}}$ & $e^{i\frac{\pi}{3}}$ &
        $C_{2z}$  & $i$      &   $-i$    \\
$C_{3z}^{-1}$ & $-1$ & $-1$ & $e^{i\frac{\pi}{3}}$ & $e^{i\frac{\pi}{3}}$ & $e^{-i\frac{\pi}{3}}$ & $e^{-i\frac{\pi}{3}}$ 
          & $C_{3z}^{-1}$  & $-1$ & $e^{-i\frac{\pi}{3}}$ & $e^{-i\frac{\pi}{3}}$ 
             &      &     \\
$C_{2z}$ & $i$ & $-i$ & $i$ & $-i$ & $i$ & $-i$ 
        &     &     &     
        & &  \\
$C_{6z}^{-1}$ & $i$ & $-i$ & $e^{-i\frac{5\pi}{6}}$ & $e^{i\frac{\pi}{6}}$ & $e^{-i\frac{\pi}{6}}$ & $e^{i\frac{5\pi}{6}}$ 
             &     &     &     
             &     &     \\
$C_{6z}$ & $-i$ & $i$ & $e^{i\frac{5\pi}{6}}$ & $e^{-i\frac{\pi}{6}}$ & $e^{i\frac{\pi}{6}}$ & $e^{-i\frac{5\pi}{6}}$ 
        &     &     &     
        &     &     \\
\hline
\end{tabular}
\caption{ \justifying Character tables for the wave-vector group of high symmetry momenta for $P6$ space group. The wave-vector groups at $\mathbf{\Gamma}$, $\textbf{K}$ and $\textbf{M}$ are the point groups $C_{6}$, $C_{3}$ and $C_{2}$, respectively. \label{tab:P6/mmm charcter}}
\end{table}
 
\begin{table}[h!]
\centering
\renewcommand{\arraystretch}{1.5}
\begin{tabular}{|c|c|c|c|c|c|c|}
\hline
\textbf{Wyckoff position} & 1a & 1a & 1a & 1a & 1a & 1a \\
\hline
\textbf{Band representation} & $^2\bar{E}_1 \uparrow G(1)$ & $^1\bar{E}_1 \uparrow G(1)$ & $^2\bar{E}_2 \uparrow G(1)$ & $^1\bar{E}_3 \uparrow G(1)$ & $^2\bar{E}_3 \uparrow G(1)$ & $^1\bar{E}_2 \uparrow G(1)$ \\
\hline
$\mathbf{\Gamma}$ & $\bar{\mathbf{\Gamma}}_7(1)$ & $\bar{\mathbf{\Gamma}}_8(1)$ & $\bar{\mathbf{\Gamma}}_9(1)$ & $\bar{\mathbf{\Gamma}}_{10}(1)$ & $\bar{\mathbf{\Gamma}}_{11}(1)$ & $\bar{\mathbf{\Gamma}}_{12}(1)$ \\\hline
\textbf{K} & $\bar{\textbf{K}}_4(1)$ & $\bar{\textbf{K}}_4(1)$ & $\bar{\textbf{K}}_5(1)$ & $\bar{\textbf{K}}_5(1)$ & $\bar{\textbf{K}}_6(1)$ & $\bar{\textbf{K}}_6(1)$ \\\hline
\textbf{M} & $\bar{\textbf{M}}_3(1)$ & $\bar{\textbf{M}}_4(1)$ & $\bar{\textbf{M}}_3(1)$ & $\bar{\textbf{M}}_4(1)$ & $\bar{\textbf{M}}_3(1)$ & $\bar{\textbf{M}}_4(1)$ \\
\hline
\end{tabular}
\caption{\centering  One dimension spinful EBR table for the $P6$ group without TR}

\label{TAB:EBR}
\end{table}

\begin{figure}
\centering
\includegraphics[width=1\textwidth]{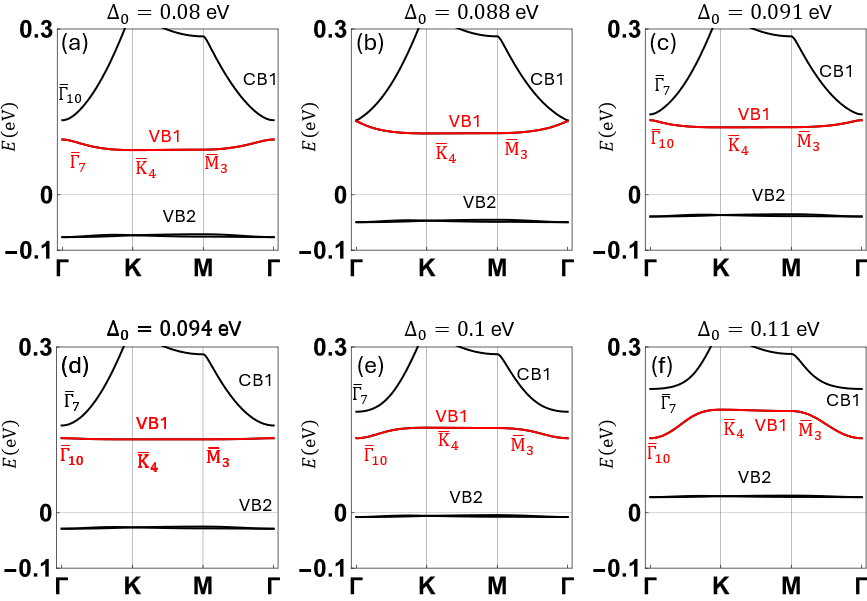}
\caption{\justifying \label{phase} (a)-(f) illustrate band dispersion of the moir\'e Chern-band model for different values of $\Delta_0$ for a fixed $M= 0.135$eV. The VB1 is trivial for $\Delta_0= 0.08$eV in (a). A band touching between the VB1 and CB1 at $\mathbf{\Gamma}$ occurs at $\Delta_0=0.088$ eV in (b). From (c)-(f), the VB1 is topologically nontrivial for $\Delta_0=0.091, 0.094, 0.1, 0.11$eV. The VB1 becomes very flat at $\Delta_0=0.094$eV in (d). }
\end{figure}

\begin{figure}
\centering
\includegraphics[width=1\textwidth]{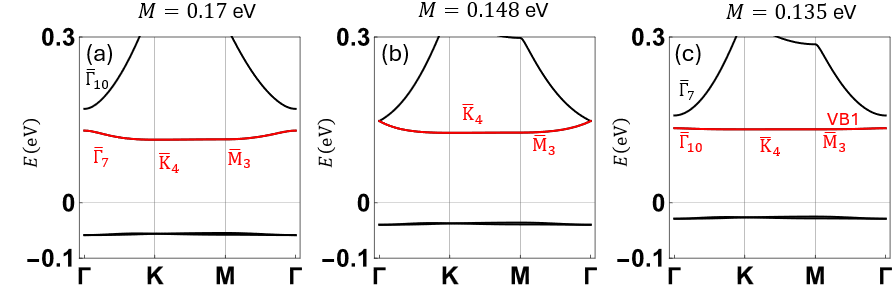}
\caption{\justifying \label{band_changem} (a)-(c) illustrate band dispersion of the moir\'e Chern-band model for different values of $M$ for a fixed $\Delta_0= 0.094$eV. The VB1 is trivial for $M= 0.17$eV in (a). A band touching between the VB1 and CB1 at $\mathbf{\Gamma}$ occurs at $M= 0.148$eV in (b). From (c), the VB1 is topologically nontrivial at $M= 0.135$eV.} 
\end{figure}

The overall topological phase diagram and the band flatness for VB1 are discussed in Fig.1(f) of the main text. Below we will provide more information about the evolution of the band dispersion for VB1 with increasing moir\'e potential strength $\Delta_0$ with a fixed atomic band gap $M$ in Fig.\ref{phase}(a)-(f).
For  $\Delta_0=0.08$eV in Fig.\ref{phase}(a), prior to the band inversion, the VB1 is topologically trivial and forms a localized Wannier orbital according to the band irreps $(\bar{\mathbf{\Gamma}}_{7}, \bar{\textbf{K}}_{4}, \bar{\textbf{M}}_{3} )$ at high symmetry momenta, belonging to the EBR $^2\Bar{E}_{1}\uparrow G(1)$, similar to the discussion of Fig.2(a) in the main text. With increasing $\Delta_0$ to a critical value $\Delta_{0,c1}=0.088$ eV, we find a Dirac type of transition between the VB1 and CB1 at $\mathbf{\Gamma}$ in mBZ in Fig.\ref{phase}(b). 
With further increasing $\Delta_0$, the mini-band VB1 becomes topologically non-trivial due to the exchange of the irreps $\bar{\mathbf{\Gamma}}_7$ and $\bar{\mathbf{\Gamma}}_{10}$ at $\mathbf{\Gamma}$ between the  CB1 and VB1. In this regime (Fig.\ref{phase}(c)-(f)), the dispersion of the VB1 mainly comes from the band repulsion with the CB1 around $\mathbf{\Gamma}$, and thus the VB1 is dispersive only around $\mathbf{\Gamma}$, but flat for the  momenta away from $\mathbf{\Gamma}$ in the mBZ.   For $\Delta_0 = 0.091$eV in Fig.\ref{phase}(c), the dispersion of the VB1 around $\mathbf{\Gamma}$ bends down (like a hole band with negative effective mass around $\mathbf{\Gamma}$), while with increasing $\Delta_0$ to a larger value ($\Delta_0=$0.1eV, 0.11 eV) in Fig.\ref{phase}(e) and (f), the dispersion of the VB1 around $\mathbf{\Gamma}$ bends up (like an electron band with positive effective mass around $\mathbf{\Gamma}$). For a "magic" value of $\Delta_0$ in between ($\Delta_0 =\Delta_{0,c2} \approx 0.094$ eV), we find the VB1 is very flat with bandwidth around 5 meV and remains topological, as shown in Fig.\ref{phase}(d). Although we tune $\Delta_0$ in Fig.\ref{phase} to realize topological flat bands for a fixed $M$, the flatness and topology of VB1 can also be tuned by the atomic band gap $M$ for a fixed $\Delta_0$, as shown in Fig.\ref{band_changem}(a)-(c). This can be achieved via electric gating experimentally.




\subsection{Topological heavy fermion model of moir\'e Chern-band Hamiltonian}\label{sec:Firstshell BHZ}
In this section, we will show that the topological flat mini-bands in the strong moir\'e potential limit, as discussed in Fig.\ref{band}(a) and Fig.\ref{phase}(d), can be well described by a topological heavy fermion (THF) model. We will first illustrate the underlying logic and the physical picture for the emergence of the heavy fermion (HF) picture for the moir\'e Chern-band model under strong moir\'e potential. Let us start from the normal bulk band gap case with $M > 0$ for the moir\'e Chern band model. Based on the symmetry analysis in Sec.\ref{sec:BHZ1}, the valence bands are mainly from the $J_z=\frac{3}{2}$ orbitals, which is composed of the atomic $P_{+}$ orbital with spin $\uparrow$, denoted as $|P_+, \uparrow\rangle$, where $|P_{+}\rangle = \frac{1}{\sqrt{2}} (|P_x\rangle + i |P_y\rangle)$, while the conduction bands are from the $J_z=\frac{1}{2}$ orbitals (without moir\'e potential). When increasing a moir\'e superlattice potential $\Delta_0$, the valence bands with $J_z=\frac{3}{2}$ orbitals in the moir\'e Chern-band model (also called heavy hole band in the semiconductor literature \cite{luttinger1956quantum,winkler2003spin,peter2010fundamentals})
become highly localized due to the large effective mass, leading to the flat bands of VB1 (depicted by the red color in Fig.2(a) in the main text). This type of localized orbitals corresponds to the f-orbitals in HF materials \cite{stewart1984heavy,hewson1993kondo}, and thus we follow this terminology to call them "f-electron" below. In contrast, due to the small effective mass and zero moir\'e potential on conduction bands, the conduction bands remain dispersive as shown by CB1 in Fig.\ref{phase}(d). Therefore, these dispersive conduction bands play the role of metallic conducting electrons in the HF materials and thus are dubbed "c-electron" below. 

\begin{figure}
\centering
\includegraphics[width=1\textwidth]{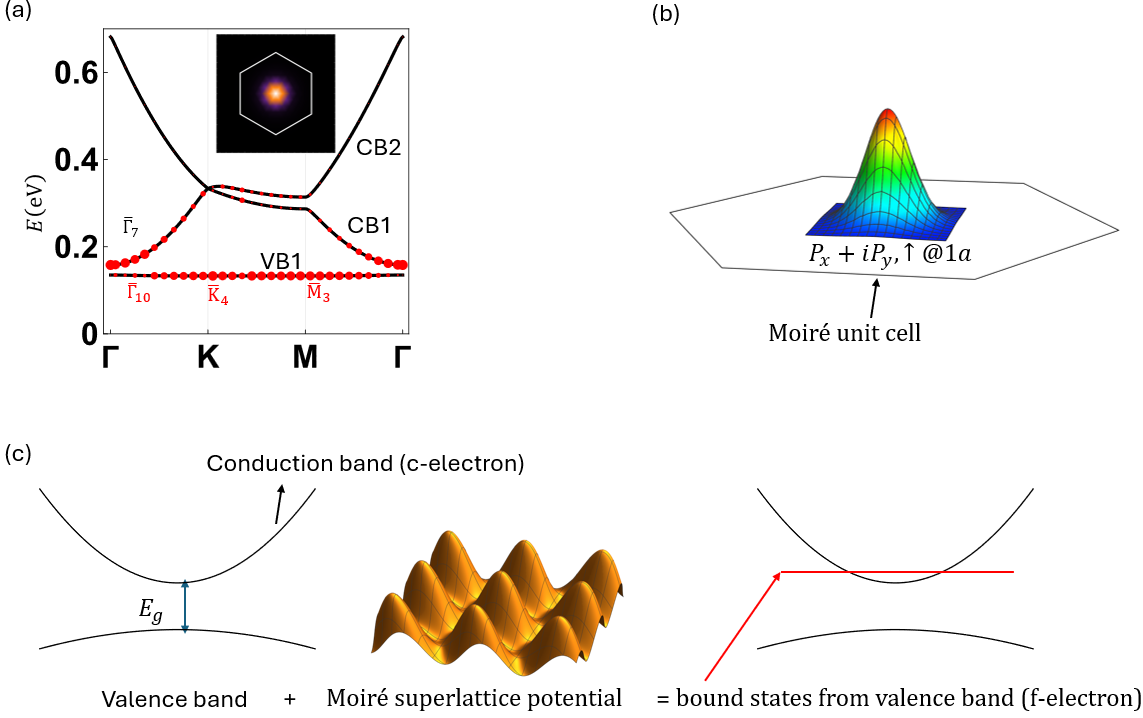}
\caption{\justifying \label{wannier orbital}  (a) Energy dispersion of the moir\'e Chern-band model with topological flat mini-bands. The overlaps between the Bloch states and localized WF are represented by red dots with their sizes representing the overlap magnitude. Inset depicts the localized WF in one moir\'e unit cell. (b) depicts the WF with the $p_x+i p_y,\uparrow $ orbital localized at the Wyckoff position $1a$. (c) Schematics for the mechanism of nontrivial flat mini-band. The band inversion between the periodic localized states that originate from the valence band and conduction band leads to the nontrivial flat mini-band, which can be describe by the THF model. }
\end{figure}

The above physical scenario can be formulated into a HF model based on the Wannier function (WF) methods. To achieve that, we take the same method in Ref.\cite{song2022magic} to resolve the Wannier obstruction originating from the topological property of the VB1.
We consider three mini-bands CB1, CB2 and VB1 for the moir\'e Chern-band model Hamiltonian (\ref{Eq:HammC}) (or the Hamiltonian (\ref{Eq:Hc_kspace})) and construct the corresponding WF which is localized at the Wyckoff position $1a$ based on Wannier 90\cite{marzari2012maximally,mostofi2008wannier90}. We denote the basis wavefunctions as $\ket{\alpha,\kk,\gs}= a_{\alpha \kk \gs}^\dagger\ket{0}$ and $\ket{\alpha,\rr}= a_{\alpha \rr}^\dagger\ket{0}$, where the index $\alpha$ is chosen to be $1, 2$ for the basis of the Hamiltonian (\ref{Eq:Hc_kspace}). The trial Wannier function is defined as
\begin{eqnarray} \label{Eq: try wannier}
&&\ket{W_{0,\RR}}=\frac{1}{\sqrt{N}}\sum_{\kk\in mBZ}\sum_{\gs \in \GG_M} e^{-i \kk \cdot \RR} e^{-i (\kk+\gs) \cdot \tau_a} \ket{2,\kk,\gs},
\end{eqnarray}
where $N$ is the number of moir\'e unit cells, $\RR$ is the moir\'e lattice vector,  $\tau_a=(0,0)$ labels the Wyckoff position $1a$ in the moir\'e unit cell.
According to the EBR $^2\Bar{E}_{1}\uparrow G(1)$, we expect 
the WF from the basis with $\alpha=2$, corresponding to the $J_z=\frac{3}{2}$ orbital. 
We denote the eigen-wave functions of the VB1, CB1 and CB2 as $\ket{\phi_{n,\kk}}$, where $n=$VB1, CB1 or CB2 labels the band index.

We take the overlap $U^{n}_{0,\kk} = \langle \phi_{n,\kk} | W_{0,\RR = 0} \rangle $ 
as an input for the Wannier 90 codes \cite{marzari2012maximally,mostofi2008wannier90}, which will minimize the spread $\bar{\Omega}$ of the Wannier function, defined by $\bar{\Omega}=  (\bra{W_{\RR=0}} {\bf r}^2 \ket{W_{\RR=0}}-|\bra{W_{\RR=0}} {\bf r} \ket{W_{\RR=0}}|^2)$.
The output of the Wannier 90 is a matrix $U^{m}_{\kk}$, from which the WF with the maximal localization can be read out as 
\begin{eqnarray}\label{Eq: wannier function}
    \ket{W_{\RR}}=\frac{1}{\sqrt{N}}\sum_{\kk,m} e^{-i \kk\cdot \RR}U^{m}_{\kk}\ket{\phi_{m,\kk}}.
\end{eqnarray} 
The WF $\ket{W_{\RR}}$ can be projected to the real space by defining $\mathbf{w}_{ \RR}^{\alpha}(\rr)=\langle \rr,\alpha | W_{\RR} \rangle$ where $\rr=(x, y)$ and $\alpha= 1, 2$ label $J_{z}=\frac{1}{2}$ and $\frac{3}{2}$ orbitals, respectively. The real space density distribution of the WF is given by $|\mathbf{w}_{\RR}(\rr)|^2=|\mathbf{w}_{\RR}^{1}(\rr)|^2+|\mathbf{w}_{\RR}^{2}(\rr)|^2$. The insert of  Fig.\ref{wannier orbital}(a) shows $|\mathbf{w}_{\RR=0}(\rr)|^2$ for the WF at $\RR = 0$, 
which is localized at $1a$ position. 
The red dots in the mini-band dispersion of Fig.\ref{wannier orbital}(a) show the overlap amplitude $|U^{m}_{\kk}| = |\langle \phi_{m,\kk} | W_{\RR = 0} \rangle|$ for different mini-bands labeled by the index $m$, from which one can see that a strong overlap between the VB1 and the WF $|W_{\RR=0} \rangle$ occurs for most momenta away from $\mathbf{\Gamma}$, while around $\mathbf{\Gamma}$, the WF $|W_{\RR=0} \rangle$ is mainly mixed into the band bottom of CB1 due to the band inversion. This suggests that the WF $|W_{\RR=0} \rangle$ plays the role of the localized f-electron in the THF model. 
 

\begin{figure}
\centering
\includegraphics[width=1\textwidth]{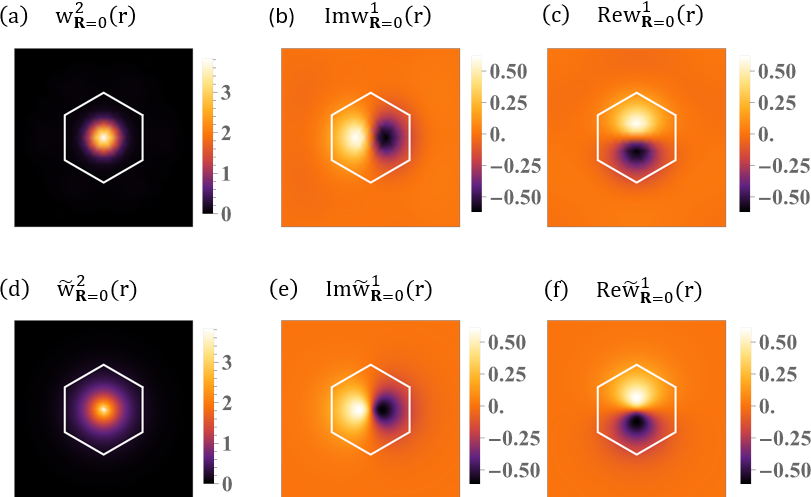}
\caption{\justifying \label{wannier orbital all}
(a) The real-space distribution of $\mathbf{w}^{2}_{\RR=0}(\rr)$ in one moir\'e unit cell. (b) and (c) show the imaginary and real parts of $\mathbf{w}^{1}_{\RR=0}(\rr)$ in one moir\'e unit cell. (d) The real-space distribution of $\tilde{\mathbf{w}}^{2}_{\RR=0}(\rr)$ in moir\'e unit cell. (e) and (f) show the imaginary and real parts of $\tilde{\mathbf{w}}^{1}_{\RR=0}(\rr)$. }
\end{figure}

Given the importance of the f-electron in constructing the THF model, we next verify that the WF $\ket{W_{\RR=0}}$ behaves similarly as the $P_+$ orbital with spin $\uparrow$ under six-fold rotation $C_{6z}$. Fig.\ref{wannier orbital all}(a)-(c) show $\mathbf{w}_{\RR=0}^{2}(\rr)$, the imaginary part and the real part of $\mathbf{w}_{\RR=0}^{1}(\rr)$.
We can approximate them by the Gaussian function form 
\begin{eqnarray} \label{Eq:  wannier Fit}
&&\tilde{\mathbf{w}}_{\RR=0}^{2}(\rr)=\mathcal{W}_2 e^{-r/\lambda_2}; \quad \tilde{\mathbf{w}}_{\RR=0}^{1}(\rr)=\mathcal{W}_1 (y-i x)e^{-r/\lambda_1},
\end{eqnarray}
with $r=|\rr|$. Fig.\ref{wannier orbital all}(d)-(f) show $\tilde{\mathbf{w}}_{\RR}^{2}$, the imaginary part and the real part of $\tilde{\mathbf{w}}_{\RR}^{1}$ with parameters $\mathcal{W}_2=3.84$, $\lambda_2=0.168 a^M_0$, $\mathcal{W}_1=10.85/a^M_0$ and $\lambda_1=0.148 a^M_0$. $\tilde{\mathbf{w}}_{\RR=0}^{2}(\rr)$ and $\tilde{\mathbf{w}}_{\RR=0}^{1}(\rr)$ satisfy the following relations
\begin{eqnarray} \label{Eq: C6z wannier constrain0}
\tilde{\mathbf{w}}_{\RR=0}^{2}(\rr)=
\tilde{\mathbf{w}}_{\RR=0}^{2}( C^{-1}_{6z}\rr); \tilde{\mathbf{w}}_{\RR=0}^{1}(\rr)=e^{i \frac{\pi}{3}}\tilde{\mathbf{w}}_{\RR=0}^{1}( C^{-1}_{6z}\rr).
\end{eqnarray}

From Eq.(\ref{Eq: C6z wannier constrain0}), we see that $\tilde{\mathbf{w}}_{\RR=0}^{1}(\rr)$ has the $P_+$ orbital character, while $\tilde{\mathbf{w}}_{\RR=0}^{2}(\rr)$ possesses $P_z$ orbital character. Under $C_{6z}$, $\ket{\rr,\alpha}$ transforms as 
\begin{eqnarray} \label{Eq: C6z wannier constrain1}
C_{6z}\ket{\rr,1}=e^{-i \frac{\pi}{6}}\ket{C_{6z}\rr,1}; C_{6z}\ket{\rr,2}=e^{-i \frac{\pi}{2}}\ket{C_{6z}\rr,2}.
\end{eqnarray}
Based on Eqs.(\ref{Eq: C6z wannier constrain0}) and (\ref{Eq: C6z wannier constrain1}), we can get 
\begin{eqnarray} \label{Eq: C6z wannier constrain}
C_{6z}\ket{W_{\RR=0}}&=& \sum_\alpha \int d\rr C_{6z} \ket {\rr,\alpha}\langle \rr,\alpha | W_{\RR=0} \rangle \nonumber
\\&\approx&\int d\rr  e^{-i \frac{\pi}{6}} \ket {C_{6z}\rr,1}\tilde{\mathbf{w}}_{\RR=0}^{1}(\rr) +e^{-i \frac{\pi}{2}} \ket {C_{6z}\rr,2}\tilde{\mathbf{w}}_{\RR=0}^{2}(\rr)\nonumber
\\&=&\int d\rr  e^{-i \frac{\pi}{2}} \ket {\rr,1}\tilde{\mathbf{w}}_{\RR=0}^{1}(\rr) +e^{-i \frac{\pi}{2}} \ket {\rr,2}\tilde{\mathbf{w}}_{\RR=0}^{2}(\rr)=e^{-i \frac{\pi}{2}}\ket{W_{\RR=0}}.
\end{eqnarray}

Eq.(\ref{Eq: C6z wannier constrain}) shows the WF $\ket{W_{\RR=0}}$ carries the total angular momentum $J_z = \frac{3}{2}$, corresponding to the $P_+$ orbital with spin $\uparrow$. 


Since $|\mathbf{w}_{ \RR}(\rr)|^2$ is localized near $1a$ position, as shown in the inset of Fig.\ref{wannier orbital}(a), topological flat band VB1 mainly comes from the WF $\ket{W_{\RR}}$ that is composed of the $P_+$ orbital with spin $\uparrow$ at the $1a$ position, and its band character exchanges with that of the CB1, described by the eigen wave function $\ket{\phi_{CB1,\kk}}$, around the $\mathbf{\Gamma}$ point. Thus, we introduce the annihilation operator $f_{\RR}$ for the localized orbital $\ket{W_{\RR}}$, called f-electron below, where $\RR$ denotes the moir\'e superlattice vector, 
and the annihilation operator $c_{\kk}$ for the conduction mini-band CB1, 
called c-electron below, with the moir\'e momentum $\kk$. Here we use the terminology of f-electron and c-electron in analog to the HF materials\cite{stewart1984heavy,fisk1995physics}. 


With the above consideration, the single particle effective Hamiltonian preserving all symmetries in group $P6/mmm $ to the lowest order can be constructed as 
\begin{eqnarray}\label{Eq:HAM_HF}
\hat{H}^0_{HF}=\sum_{|\kk|<\Lambda_c}(E_c+\alpha \kk^2) c^{\dagger}_{\kk}c_{\kk}+\sum_{\RR}E_f f^{\dagger}_{\RR}f_{\RR}+\frac{\beta}{\sqrt{N}}\sum_{|\kk|<\Lambda_c,\RR}[e^{-i \kk \cdot \RR} (k_x +i k_y)c^{\dagger}_{\kk}f_{\RR}+h.c.],
\end{eqnarray}
where $\Lambda_c = |b_{M,1}|$ is the momentum cutoff, and $E_c, \alpha, \beta, E_f$ are the parameters that can be computed from the WF projection as 
\begin{eqnarray} \label{Eq: HF model parameters}
&&E_c = \bra{\phi_{VB1,\kk=0}}H_{mC}(\kk=0)\ket{\phi_{VB1,\kk = 0}}; \alpha = \left.\frac{\partial^2}{\partial k_x^2} \bra{\phi_{VB1,\kk}}H_{mC}(\kk)\ket{\phi_{VB1,\kk}}\right|_{\kk \rightarrow 0}, \nonumber \\
&&E_f = \bra{W_{\RR = 0}}H_{mC}\ket{W_{\RR = 0}};\frac{\beta}{\sqrt{N}} = \left.\frac{\partial\bra{\phi_{VB1,\kk=0}}H_{mC}\ket{W^{(1)}_{\RR =0}}}{\partial k_x}\right|_{\kk\rightarrow 0},
\end{eqnarray}
where the WF $\ket{W_{\RR = 0}}$ is defined in Eq.(\ref{Eq: wannier function}) and  $\ket{\phi_{VB1,\kk}}$ denotes the eigen wavefunction of VB1. 



\begin{figure}
\centering
\includegraphics[width=0.9\textwidth]{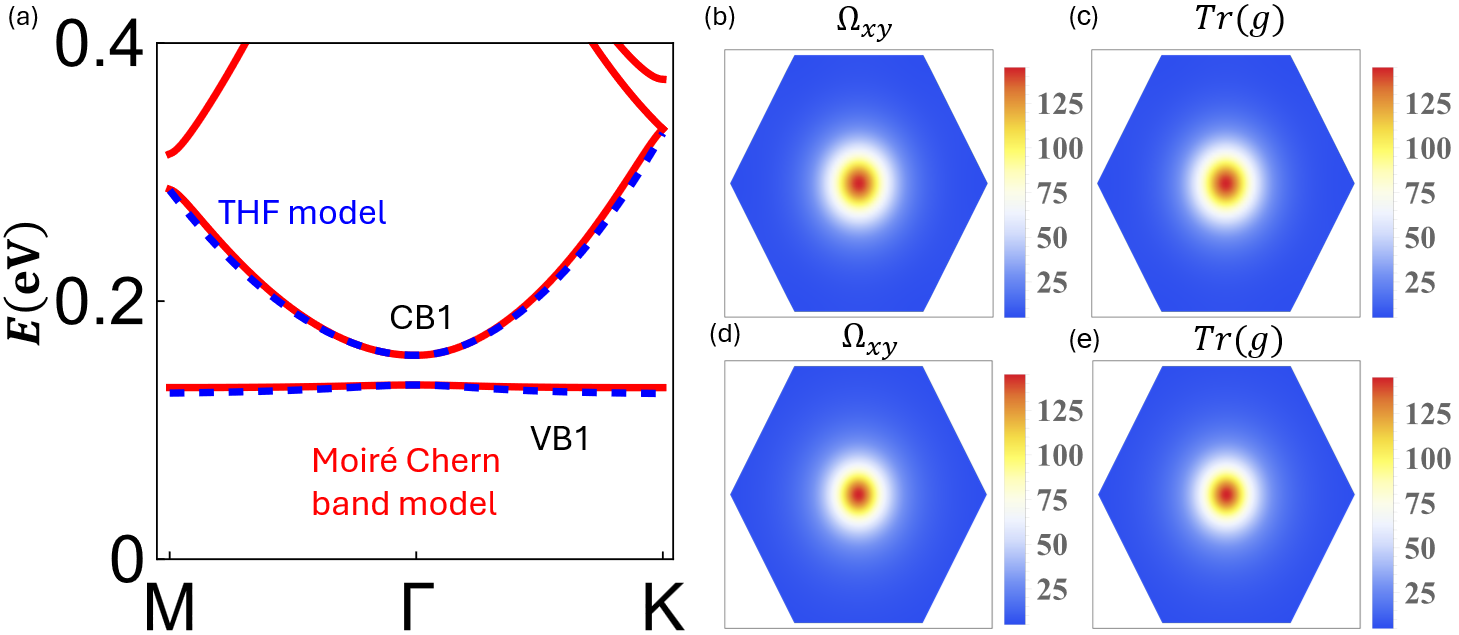}
\caption{\justifying \label{heavy band}  (a)  comparison between band dispersion of effective heavy fermion model and that of continuum model with $M= 0.135$ eV and $\Delta_0=0.094$ eV. (b)-(c) The Berry curvature and the trace of quantum metric of the VB1 in the moir\'e Chern-band model. (d)-(e) shows the distribution of the Berry curvature and the trace of quantum metric of VB1 as a function of the momentum for the THF model.
}
\end{figure}

Fig.\ref{heavy band}(a) shows the comparison of the energy dispersion between the moir\'e Chern-band model $H_{mC}$ in Eq.(\ref{Eq:Hc_kspace}) and the THF model $H_{HF}$ in Eq.(\ref{Eq:HAM_HF}) for the parameters evaluated from Eq.(\ref{Eq: HF model parameters}), given by $E_c=0.135$ eV, $\alpha=1.238$ $nm^2\cdot $ eV,$\beta=0.195$ $nm\cdot$ eV and  $E_f=0.158$ eV, respectively. 
We further show the Berry curvature $\Omega_{xy}$ and the trace of the quantum metric $Tr(g)$ distribution of the VB1 band of $\hat{H}^{0}_{HF}$ in the mBZ, both of which are localized around the $\mathbf{\Gamma}$ point, as shown in  Fig.\ref{heavy band}(d) and (e). They compare well with $\Omega_{xy}$ and $Tr(g)$ of the VB1 band calculated from $H_{mC}$, shown in Fig.\ref{heavy band}(b)-(c).

\subsection{Chern-band model with equal Moir\'e potentials on two layers}

\begin{figure}
\centering
\includegraphics[width=1\textwidth]{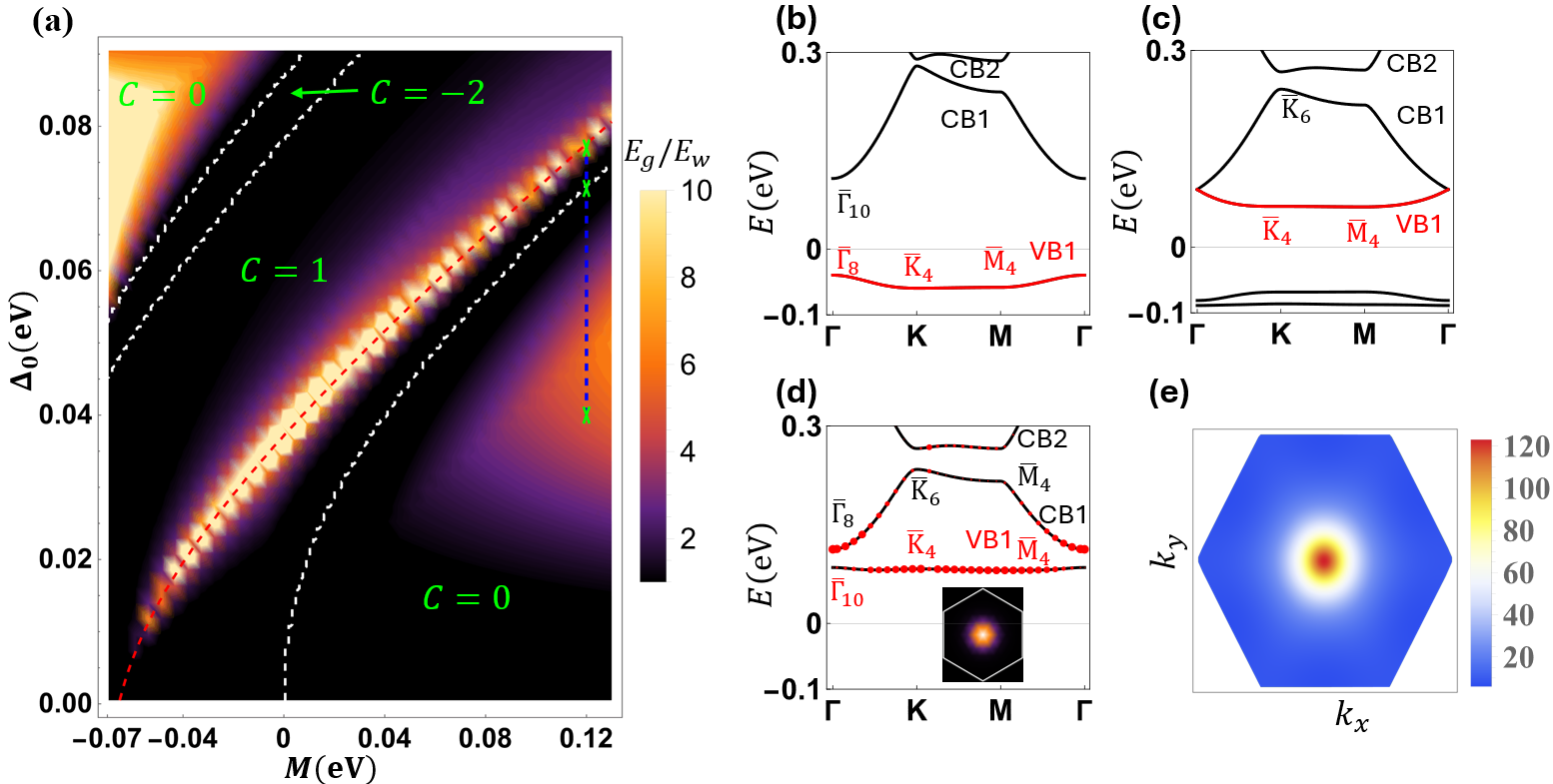}
\caption{\justifying \label{fig:R1} (a) Phase diagram of the moir\'e Chern band model with equal moir\'e potential strength on both layers for $\Delta_0>0$. The background color indicates the ratio $E_g/E_w$. The white dashed lines separate different phases, and the red line traces the peak of $E_g/E_w$ for VB1. (b)-(d) show the band dispersion with the parameters $M=0.12$ eV and $\Delta_0 = 0.04$ eV, $0.0714$ eV and $0.077$ eV, respectively, as denoted by three greeen crosses in (a). In  (d), the sizes of the red dots in VB1, CB1, and CB2 represent the overlap between the Bloch states and a localized Wannier orbital centered at the Wyckoff position 1a (shown in the inset). (e) shows the Berry-curvature distribution of VB1 for the case in (d).} 
\end{figure}

In this section, we will demonstrate that the THF physics discussed above does {\it not} depend on the condition that the moir\'e potential in Sec.II.A of the main text is applied only on the bottom layer. To show that, we explicitly include the moir\'e potential into the top layer with the same strength as the bottom layer, with the potential Hamiltonian given by
\begin{eqnarray} \label{eq: potential equal}
    \hat{H}_{\text{m}} = \Delta_0\sum_{\kk \gs \gs' \alpha } \sum_{\qq \in \GG^{\mathds{1}}_M } \delta(\gs' -\gs -\qq) a_{\alpha \kk \gs}^\dagger a_{\alpha \kk \gs'} \label{Eq:HM_kspace}.
\end{eqnarray}
Since the conduction and valence bands are from two different layers, both bands will feel the same moir\'e potential strength. With this form of Moir\'e potential in Eq.(\ref{eq: potential equal}), we implement the same calculation as that for Fig.1 and Fig.2 in the main text, and the results are summarized in Fig.\ref{fig:R1}. Below we will compare the results from two setups: (i) a moir\'e potential applied only to the valence bands (Figs.1(f) and 2 in the main text), and (ii) a moir\'e potential applied to both the conduction and valence bands (Fig.\ref{fig:R1}).  
 
Fig.\ref{fig:R1}(a) shows the phase diagram for the case where moir\'e potentials with the first shell approximation are applied to both the valence and conduction bands with equal strength. Compared with Fig.1 (f) in the main text, we find the C=1 regime of interest remains essentially unchanged, and the peak line of $E_g/E_w$, as shown by the red dashed line in Fig.\ref{fig:R1}(a), also appears and is almost the same as that in Fig.1(f) of the main text. The main change is that the semimetal phase regime for a large $\Delta_0$ and negative $M$ in Fig.1(f) becomes the insulating phase regimes with C=-2 and C=0 in Fig.\ref{fig:R1}(a). 
 
In Fig.\ref{fig:R1}(b)-(d), we calculate the Moiré band dispersion along the blue dashed line at three green crosses in Fig.\ref{fig:R1}(a). Comparing Fig.\ref{fig:R1}(b)-(d) with Fig.2(a)-(c) in the main text, we do not find much difference on the VB1 band and the bottom of CB1 band for these two cases, and the band inversion between VB1 and CB1 almost remains the same. The main influence of the moiré potential on the conduction bands is to open a gap between CB1 and CB2 at the K point around the energy 0.3 eV in Fig.\ref{fig:R1}(b)-(d), in contrast to Fig.2(a)-(c) in the main text. 
 
\begin{figure}
\centering
\includegraphics[width=1\textwidth]{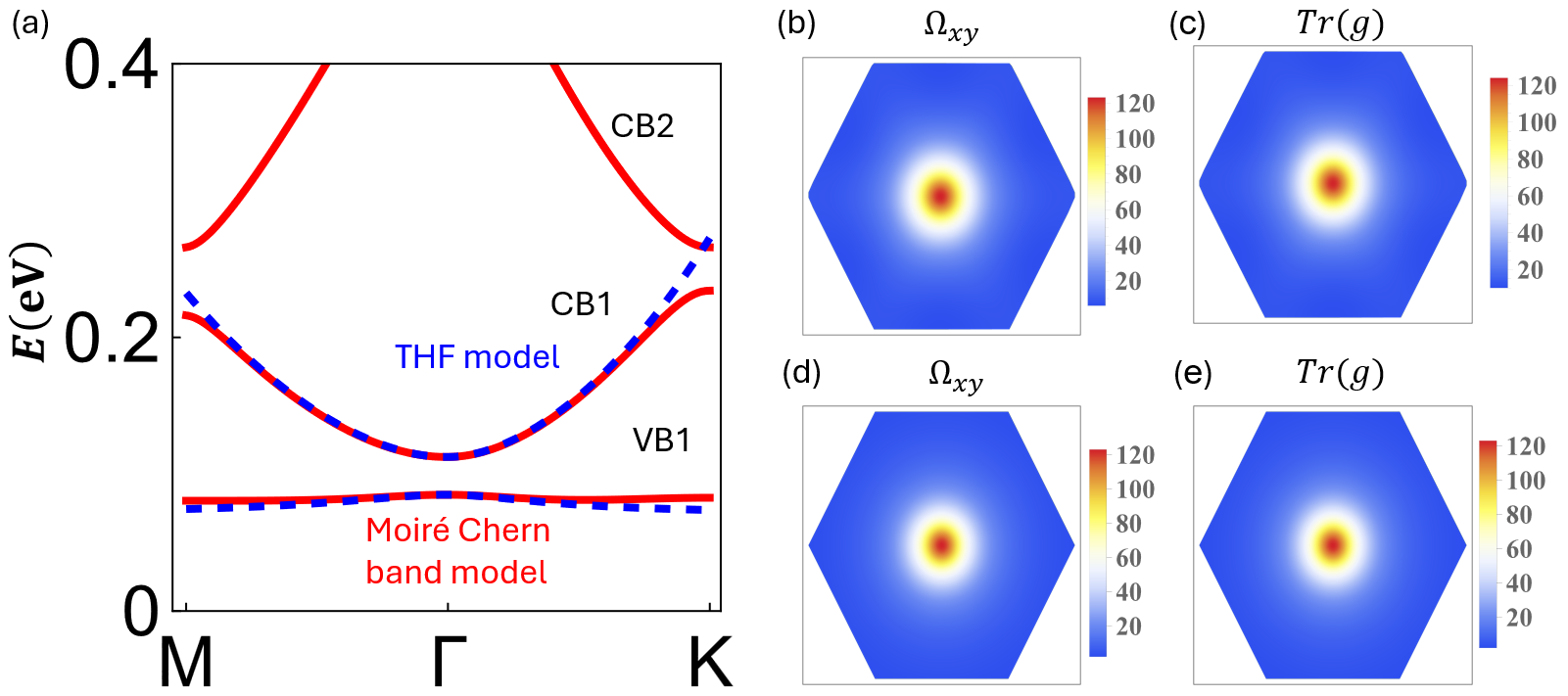}
\caption{\justifying \label{fig:R3} (a) Comparison between the band dispersion from the moir\'e Chern band model and the THF model. For the moir\'e Chern band model, we choose the parameters $M= 0.12$eV and $\Delta_0=0.077$eV, when equal potential strength is applied to both two layers. (b)-(c) The Berry curvature and the trace of quantum metric of the VB1 in the moir\'e Chern-band model. (d)-(e) The Berry curvature and the trace of quantum metric of the VB1 in the THF model.} 
\end{figure}

Fig.\ref{fig:R1}(e) shows the concentrated Berry curvature of VB1 band in Fig.\ref{fig:R1}(d), behaving similarly as that in Fig.2(e) in the main text. We also construct a localized Wannier orbital (inset of Fig.\ref{fig:R1}(d)), and the low-energy bands are well captured by the THF model, as demonstrated in Fig.\ref{fig:R3}, which is almost the same as Fig.\ref{heavy band}. Overall, this comparison indicates that the band inversion is primarily driven by the moir\'e potential in the valence bands. This is because, in our model, the valence band has a much larger effective mass than the conduction band—a feature common to many conventional semiconductors. Consequently, the moir\'e potential mainly leads to the bound states of valence bands, while the conduction band remains dispersive.

\subsection{Ideal Quantum Geometry in topological heavy fermion model} \label{Sec: ideal chern band}
Fig.\ref{band}(d)-(l) show the Berry curvature $\Omega_{xy}$ and the trace of the quantum metric $Tr(g)$, as well as their ratio, defined as $\frac{\Omega_{xy}}{Tr(g)}$, in the momentum space. We see that in the heavy fermion limit, both $\Omega_{xy}$ and $Tr(g)$ reveal peaks around $\mathbf{\Gamma}$, and the ratio $\lambda$ is close to 1 around $\mathbf{\Gamma}$. 
We now study the quantum geometry property of THF model in Eq.(\ref{Eq:HAM_HF}).


First, we rewrite Eq.(\ref{Eq:HAM_HF}) in momentum space as 
\begin{eqnarray}\label{Eq:HAM_HF K}
\hat{H}^0_{HF}
&=& \sum_{|\kk|<\Lambda_c} \left[(E_c+\alpha k^2) c^{\dagger}_{\kk}c_{\kk}+E_f f^{\dagger}_{\kk}f_{\kk}+\beta (k_x +i k_y)c^{\dagger}_{\kk}f_{\kk}+h.c.\right]\nonumber \\
&=& \sum_{|\kk|<\Lambda_c} \left(\begin{array}{cc}
    c^{\dagger}_{\kk} & f^{\dagger}_{\kk}
\end{array}\right) H^0_{HF}(\kk) \left(\begin{array}{c}
    c_{\kk} \\ f_{\kk}
\end{array}\right);\quad 
H^0_{HF}(\kk)=\begin{pmatrix}
        E_c + \alpha k^2&\beta k_+ \\ \beta k_-& E_f
    \end{pmatrix},
\end{eqnarray}
where the Fourier transform of $f_R$ is defined as 
\begin{eqnarray}
    f_{\RR}=\frac{1}{\sqrt{N}}\sum_\kk e^{i \kk \cdot \RR} f_{\kk}.
\end{eqnarray}
We further denote $E_f=E_c+\Delta E$ and choose $\Delta E>0$ which has the same sign as $\alpha>0$, corresponding to the topological band region. 

The eigenstates of $H^{0}_{HF}(\kk)$ are defined as  
\begin{eqnarray} \label{Eq.heavy fermion state}
H^{0}_{HF}(\kk)\ket{u_\pm(\kk)}=\mathcal{E}_{\pm}(\kk) \ket{u_\pm(\kk)},
\end{eqnarray}
where 
\begin{eqnarray}
    \mathcal{E}_{\pm}(\kk)=\frac{1}{2}(2E_c+ \Delta E +\alpha k^2\pm\sqrt{4\beta^2 k^2+(\Delta E-\alpha k^2)^2}),\\
    \ket{u_\pm(\kk)}=\frac{1}{N_\pm}(
-\Delta E + \alpha k^2 \pm \sqrt{4 \beta^2 k^2 + \left(\Delta E - \alpha k^2\right)^2},\ 
2 \beta k_-)^T,
\end{eqnarray}
and $N_\pm= \sqrt{4 \beta^2 k^2 + \left(-\Delta E + \alpha k^2 \pm \sqrt{4 \beta^2 k^2 + (\Delta E - \alpha k^2)^2} \right)^2}$
 is the normalization factor. With this band dispersion, we find that when  
\begin{eqnarray}\label{eq_SM:exactflatbandcondition}
\Delta E=\beta^2/\alpha,
\end{eqnarray}
the lower energy branch has 
\begin{eqnarray}
    \mathcal{E}_-=\frac{1}{2}(2E_c+ \Delta E +\alpha k^2-\sqrt{(\Delta E+\alpha k^2)^2})=E_c, 
\end{eqnarray}
which is an exact flat band. Therefore, we identify Eq.(\ref{eq_SM:exactflatbandcondition}) as the condition for the flat band in the THF model. 

The Berry curvature and quantum metric of the lower-energy branch $\ket{u_-(\kk)}$ can be calculated as 
\begin{eqnarray}\label{Eq:BC and QM}
\Omega_{xy} &=& -2 Im (\langle \partial_{k_x}u_{-}(\kk)| \partial_{k_y} u_{-}(\kk)\rangle ) =  i\frac{\bra{u_-(\kk)} \partial_{k_x} H_{HF}^{0}\ket{u_+(\kk)}\bra{u_+(\kk)} \partial_{k_y} H_{HF}^{0}\ket{u_-(\kk)}-h.c.}{(\mathcal{E}_--\mathcal{E}_+)^2} \nonumber \\
& = & \frac{2 \beta^2 (\Delta E + \alpha k^2)}{\left [ 4 \beta^2 k^2 + (\Delta E - \alpha k^2)^2\right ]^{3/2}},\\\
Tr(g) &=& \sum_{i=x,y} Re \left(\langle \partial_{k_i}u_{-}(\kk)| \partial_{k_i} u_{-}(\kk)\rangle - \langle\partial_{k_i}u_{-}(\kk)|u_{-}(\kk)\rangle \langle u_{-}(\kk)|\partial_{k_i} u_{-}(\kk)\rangle\right)
\nonumber\\
&=&
\sum_{i=x,y}\frac{\bra{u_-(\kk)} \partial_{k_i} H_{HF}^{0}\ket{u_+(\kk)}\bra{u_+(\kk)} \partial_{k_i} H_{HF}^{0}\ket{u_-(\kk)}}{(\mathcal{E}_--\mathcal{E}_+)^2} \nonumber \\
& = & \frac{2 \beta^2 \left((\Delta E)^2 + 2 \beta^2 k^2  + \alpha^2 k^4\right)}{\left [4 \beta^2 k^2 + (\Delta E   - \alpha k^2)^2\right ]^2}.
\end{eqnarray}

Using the above results, we find the ratio 
\begin{eqnarray}
\label{eq_SM:quantumgeometryratio}
\lambda=\frac{\Omega_{xy}(\kk)}{Tr(g(\kk))}=\frac{(\Delta E+\alpha k^2)\sqrt{4\beta^2k^2+(\Delta E-\alpha k^2)^2}}{\Delta E^2+2\beta^2k^2+\alpha^2 k^4}.
\end{eqnarray}
When the exact flat band condition in Eq.(\ref{eq_SM:exactflatbandcondition}) is satisfied, we also find $\lambda=1$ for any momentum $\kk$, corresponding to the "ideal quantum geometry" condition. 

Therefore, we conclude that the exact flat band with ideal quantum geometry can be realized in our THF model when the condition (\ref{eq_SM:exactflatbandcondition}) is satisfied.

To understand the above result, we choose $E_c=0$ and note that when $\Delta E=\beta^2/\alpha$, the matrix $H^0_{HF}$ can be written as
\begin{eqnarray} \label{Eq: matrix flat band}
    H^0_{HF}(\kk)=\begin{pmatrix}
        \alpha k^2&\beta k_+ \\ \beta k_-&\beta^2/\alpha
    \end{pmatrix} = \mathcal{D}^\dagger \mathcal{D}, 
\end{eqnarray}
where the matrix $\mathcal{D}$ is given by
\begin{eqnarray} \label{Eq: matrix flat bandd}
   \mathcal{D}^\dagger =\begin{pmatrix}
        \sqrt{\alpha} k_+ & 0 \\ \frac{\beta}{\sqrt{\alpha}} & 0
    \end{pmatrix} ; \quad \mathcal{D}=\begin{pmatrix}
        \sqrt{\alpha} k_- &\frac{\beta}{\sqrt{\alpha}}  \\ 0&0
    \end{pmatrix}. 
\end{eqnarray}  
Thus, we can consider the equation
\begin{eqnarray}\label{eqn_SM:zeromodeequation}
    \mathcal{D}\uu_0 = 0,
\end{eqnarray}
where $\uu_0 = (x, y)^T$ is a two-component vector. From Eq.(\ref{eqn_SM:zeromodeequation}), it is clear that $H^0_{HF}\uu_0 = \mathcal{D}^\dagger \mathcal{D} \uu_0 = 0$. Due to the form of $\mathcal{D}$, Eq.(\ref{eqn_SM:zeromodeequation}) gives 
\begin{eqnarray}
    \sqrt{\alpha} k_- x + \frac{\beta}{\sqrt{\alpha}} y = 0 \rightarrow 
    \frac{x}{y} = - \frac{\beta}{\alpha k_-} \rightarrow \uu_0 = \frac{1}{\sqrt{\beta^2 +\alpha^2 k^2}} \left(\begin{array}{c}
       -\beta \\
        \alpha k_- 
    \end{array}\right),
\end{eqnarray}
which gives the wavefunction of the zero mode. Furthermore, we notice that beside the normalization factor $\frac{1}{\sqrt{\beta^2 +\alpha^2 k^2}}$, $\left(\begin{array}{c} -\beta \\ \alpha k_- \end{array}\right)$ is a holomorphic function, implying its ideal quantum geometry property \cite{wang2021exact, ledwith2023vortexability}.

\subsection{Estimate of Kondo temperature in the Kondo regime}

 In the main text, we have demonstrated the existence of fractional Chern insulator phase in the projected regime of THF model with the Coulomb interaction weaker than the hybridization gap. Below we will consider the opposite limit, the Kondo regime with the Coulomb interaction stronger than the hybridization gap, in which the THF model can be transformed into the Kondo Hamiltonian through the Schrieffer–Wolff transformation \cite{hewson1997kondo,coleman2015introduction}. 

We consider the THF Hamiltonian in Eq.(\ref{Eq:HAM_HF}) with spin index and also include the onsite Hubbard interaction between f-electrons, and the full Hamiltonian is given by 
\begin{eqnarray}\label{Eq:HAM_HFperiodic}
&&\hat{H}_{HF}=\sum_{|\kk|<\Lambda_c, \sigma}\alpha k^2 c^{\dagger}_{\kk,\sigma}c_{\kk,\sigma}+\sum_{\RR}(\Delta E n_{f,\RR}+U_0 n_{\uparrow,\RR}n_{\downarrow,\RR})\nonumber\\
&&
+\sum_{|\kk|<\Lambda_c,\sigma,\RR}[e^{-i \kk \cdot \RR} V_\sigma (\kk)c^{\dagger}_{\kk,\sigma}f_{\RR,\sigma}+h.c.],
\end{eqnarray}
where we have set $E_c=0$, $\sigma=\uparrow,\downarrow$ represents the spin, $V_{\uparrow(\downarrow)} (\kk)=\beta(k_x+(-)ik_y )$, and $U_0$ denote the strength of onsite coulomb interaction between f electrons. The Eq.(\ref{Eq:HAM_HFperiodic}) is the periodic Anderson model. In the limit $U_0 \gg \beta \sqrt{\Delta E/\alpha}$, our system is projected in the Kondo regime. 

We below give a rough estimate of Kondo temperature of our system by considering a single-impurity limit (a single site for f-electron), in which the Hamiltonian takes the standard Anderson impurity model form,  
\begin{eqnarray}\label{Eq:HAM_HF spin}
\hat{H}^{one}_{HF}=\sum_{|\kk|<\Lambda_c, \sigma}\alpha k^2 c^{\dagger}_{\kk,\sigma}c_{\kk,\sigma}
\sum_{|\kk|<\Lambda_c,\sigma}[V_\sigma (\kk)c^{\dagger}_{\kk,\sigma}f_{\sigma}+h.c.]+\Delta E n_f+U_0 n_{\uparrow}n_{\downarrow}.
\end{eqnarray}
In the Kondo regime $U_0 \gg \beta \sqrt{\Delta E/\alpha}$ 
, we can treat $V_{\uparrow(\downarrow)}$ as a perturbation and perform the Schrieffer–Wolff transformation, which leads to the Kondo Hamiltonian 
\begin{eqnarray}
\hat{H}_{Kondo}=\sum_{\kk\sigma}\epsilon_\kk\,c_{\kk\sigma}^{\dagger}c_{\kk\sigma}
+\sum_{\kk,\kk'} J^{\alpha,\beta}_{\kk,\kk'}\,c_{\kk,\alpha}^{\dagger}\,\vec{\sigma}_{\alpha\beta}\,c_{\kk',\beta}\cdot\vec{S}_f,
\end{eqnarray}
where $\epsilon_\kk=\alpha \kk^2-\mu$ and $J^{\alpha,\beta}_{\kk,\kk'}=V_\beta({k'})^{*}V_\alpha(k)\left[\frac{1}{\Delta E-\mu+U_0}-\frac{1}{\Delta E-\mu }\right]
$ with $\mu$ as the chemical potential. 
The Kondo temperature can be estimated as $T_K=\frac{2U_0\Delta}{\pi} e^{-\frac{\pi U_0}{8 \Delta}}$ \cite{coleman2015introduction}. Here,  $\Delta=\pi \Omega_0 \rho(\Delta E  ) |V_\sigma(\KK_f )|^2$, where $\rho(\Delta E  )=\frac{1}{2\pi\alpha}$ is density of states of c electrons at $\Delta E$, $\KK_F=(\sqrt{\frac{\Delta E }{a}},0)$ and $\Omega_0=113.7$ nm$^2$ is the area of the moiré unit cell. We choose the interaction $U_0$ to be 30 meV, which is comparable to the value $U_0\approx29$ meV reported for TBG \cite{song2022magic,zhou2024kondo}. In Sec.\ref{sec:THFdelta}, we derived $\beta$ and $\alpha$ and $\Delta E$ analytically when a $\delta$-function-like potential is applied.  With the parameters in Sec.~\ref{sec:THFdelta}, namely $\alpha=1.238$   $\mathrm{eV\cdot nm^2}$, $\beta=0.120$ $\mathrm{eV\cdot nm^2}$, and $\Delta E=0.023$ $\mathrm{eV}$, we obtain a Kondo temperature of $T_K \approx 5.8$ $\mathrm{meV}$. This value is comparable to estimates of the Kondo temperature for twisted bilayer graphene, $T_K \approx 1.5$ $\mathrm{meV}$, in Ref.\cite{zhou2024kondo}. 

We note that our estimate here is for a single-impurity problem and a controlled, quantitative determination of the Kondo temperature for the periodic Anderson model typically requires a dynamical mean-field theory (DMFT) treatment (or related many-body approaches), which is beyond the scope of the present work. We noted that the estimate of Kondo temperature from a single impurity Anderson model in Ref.\cite{zhou2024kondo} is reasonably consistent with that calculated from DMFT in TBG. We also note that the heavy fermion physics has been demonstrated experimentally in TBG \cite{wong2020cascade, merino2025interplay, xiao2025interacting}, and thus we would expect the realization of Kondo physics is feasible in our system.

\section{Moir\'e superlattice with the periodic $\delta$-function-like potential}

In this section, we will discuss the moir\'e superlattice potential that takes the form of periodic $\delta$-function potential. 
In real space, the $ H _ \text M (\rr)$ 
takes the form
\begin{eqnarray}\label{Eq:HM_delta_rspace_periodic_}
    \hat{H}_{\text{m}} = \sum_{\RR,\rr} \Delta_0 \Omega_0 \delta (\rr - \RR) a_{\alpha \rr}^\dagger a_{\alpha \rr},
\end{eqnarray}
where  $\Omega_0=\frac{\sqrt{3}}{2}a^2_0$ is the area of the unit cell for the moir\'e superlattice potential.
In the momentum space, the moir\'e potential Hamiltonian takes a simple form 
\begin{eqnarray}\label{Eq:HM_delta_periodic}
&& \hat{H}_{\text{m}} = \Delta_0\sum_{\kk  \alpha } \sum_{\gs, \gs' \in \GG_M }  a_{\alpha \kk \gs}^\dagger a_{\alpha \kk \gs'}.
\end{eqnarray}
For the first shell approximation, the moir\'e potential Eq.(\ref{Eq:HM_kspace}) only couples two states with their momentum difference in a short range within the first shell $\GG^{\mathds{1}}_M$. In contrast, the $\delta$-function potential is a long range coupling in the momentum space, and the coupling between any two momenta related by a moir\'e reciprocal lattice vector has an equal strength. 



\subsection{Review of parabolic bands in a periodic $\delta$-function-like potential}\label{Sec:deltapotential_parabolicbands}
In this section, we will first provide a review of the free electron gas with the parabolic band in a periodic $\delta$-function potential. We are mostly interested in the strong bound state due to the confining $\delta$-function potential and its binding energy and localized wave function, which depend on the approximation that we made for the $\delta$-function potential  \cite{Laughlinnote}.

We start with the two-dimensional free electron gas in a periodic $\delta$-function potential with the Hamiltonian
\begin{eqnarray}   &&H_{2D}= H_0 + H_{\text{m}} = -\frac{\hbar^2}{2 m_e} \nabla^2 +\sum_{\RR} \Delta_0 \Omega_0 \delta (\rr - \RR), 
\end{eqnarray}
where $m_e$ is the electron effective mass, $\rr = (x, y)$ and $\RR$ are the superlattice vectors. We may regularize the length and the energy scale of this Hamiltonian by the moir\'e superlattice length $a_0$, and the energy unit is set as $E_0 = \frac{\hbar^2 }{2 m_e a_0^2}$. By defining $\tilde{\Delta}_0 = \frac{\Delta_0}{E_0}$, $(\tilde{x}, \tilde{y}) = (\frac{x}{a_0}, \frac{y}{a_0})$, $\tilde{\RR}=\frac{\RR}{a_0}$, and $\tilde{\Omega}_0 = \frac{\Omega_0}{a_0^2}$, we find 
\begin{eqnarray}  
&&\tilde{H}_{2D} = \frac{H_{2D}}{E_0} = - \tilde{\nabla}^2 +\sum_{\RR_M} \tilde{\Delta}_0 \tilde{\Omega}_0 \delta (\tilde{\rr} - \tilde{\RR}), 
\end{eqnarray}
so all the terms in $\tilde{H}_{2D}$ become dimensionless. Below we drop the notation $\sim$ for all the symbols. 

Next we transform this Hamiltonian into the momentum space by considering the basis of plane wave functions $|\kk + \gs\rangle$, which is defined as $\langle \rr|\kk + \gs\rangle=e^{i (\kk + \gs)\cdot \rr}$ and the Hamiltonian matrix element reads
\begin{eqnarray}\label{Eq:Ham2DEG_delta_potential}
&&\bra{\kk+\gs}H_{2D}\ket{\kk+\gs'}=|\kk+\gs|^2\delta_{\gs,\gs'}+\Delta_0,
\end{eqnarray}
for any $\gs, \gs'$. As we will discuss below, the binding energy of the bound state for the Hamiltonian (\ref{Eq:Ham2DEG_delta_potential}) diverges for a 2D periodic $\delta$-function potential, and thus we need to impose a momentum cut-off $G_\Lambda$, e.g. $|\gs| \le G_\Lambda$, for the moir\'e reciprocal lattice vectors. With the momentum cut-off, the Hamiltonian $H_{2D}$ can be re-written into a matrix form as
\begin{eqnarray}\label{Eq:H2DEG_momentum}
&&H_{2D}(\kk)=H_0(\kk)+N_g\Delta_0 P_0, \nonumber\\
&&H_0(\kk)=\begin{pmatrix}
E_1 (\kk)  & 0 & \dots& 0\\
0 & E_2 (\kk) & \dots& 0\\
\vdots & \vdots & \ddots&  \vdots\\
0 & 0 & \dots & E_{N_g} (\kk)
\end{pmatrix}, 
\end{eqnarray}
where $N_g$ is the number of $\gs$ within the momentum cut-off $G_\Lambda$,  $E_j(\kk)=|\kk+\gs_j|^2$ with $j=1, 2, ..., N_g$ labels all possible values of $\gs$, and the projection operator $P_0$ is
\begin{eqnarray}\label{Eq:projection0}
    P_0 = \ket{\phi_0}\bra{\phi_0}; \qquad
\ket{\phi_0}=\frac{1}{\sqrt{N_g}}\begin{pmatrix}
1\\
1\\
\vdots\\
1 
\end{pmatrix}. 
\end{eqnarray}
We call this form of moir\'e potential with a cut-off in the momentum space as the periodic $\delta$-function-like potential.

Below we will use the form of the full Green's function to solve the eigen-energies and eigen-wave functions of the Hamiltonian $H_{2D}$. The eigen-energies of $H_{2D}$ can be extracted from the poles of the full Green's function, $G=(E-H_{2D})^{-1}$ of $H_{2D}$, or equivalently, the trace of $G$, 
\begin{eqnarray}
    Tr(G)=Tr[(E-H_{2D}(\kk))^{-1}] 
    = \sum_j \bra{j} (E-H_{2D}(\kk))^{-1}\ket{j}, 
\end{eqnarray} 
which can be written on the eigenstate basis $\ket{j}$ of $H_0$ in Eq.(\ref{Eq:H2DEG_momentum}) with the eigen-energy $E_j$, e.g. $H_0(\kk)\ket{j}=E_j(\kk)\ket{j}$, where $j$ labels the wavefunction components for the reciprocal lattice vectors $\gs_j$. Since $H_0(\kk)$ is already diagonal in the momentum space, $\ket{j}$ is independent of the momentum $\kk$. Our aim is to find the analytical form of $Tr(G)$. To achieve that, we first can show the matrix element of the bare Green's function, defined as $\hat{g}(\kk)=(E-H_0(\kk))^{-1}$, connecting the state $\ket{\phi_0}$ is given by 
\begin{eqnarray}\label{Eq: bare GreenFunction_Relation_1}
    \mathbf{g}(\kk,E) = N_g\bra{\phi_0}\hat{g}(\kk)\ket{\phi_0}= N_g\bra{\phi_0}(E-H_{0}(\kk))^{-1}\ket{\phi_0} = \sum^{N_g}_{j=1}\frac{1}{E-E_j(\kk)}, 
\end{eqnarray}
where we have used the explicit expression of $|\phi_0\rangle$ in Eq.(\ref{Eq:projection0}) and included an additional factor $N_g$ in the definition of $\mathbf{g}(\kk,E)$.  
The full Green's function connecting the state $\ket{\phi_0}$ is given by 
\begin{eqnarray}\label{Eq:GreenFunction_Relation_0}
    N_g \bra{\phi_0}G\ket{\phi_0} &=& N_g\bra{\phi_0}(E-H_{0}(\kk)-N_g\Delta_0 P_0)^{-1}\ket{\phi_0}\nonumber\\&=&N_g\bra{\phi_0}(1-(E-H_{0}(\kk))^{-1}N_g\Delta_0 P_0)^{-1}(E-H_{0}(\kk))^{-1}\ket{\phi_0}\nonumber\\
    &=&N_g\bra{\phi_0}\sum^\infty_{n=0} (\hat{g}(\kk)N_g\Delta_0 P_0)^n\hat{g}(\kk)\ket{\phi_0}\nonumber\\&=&\sum^\infty_{n=0}\sum_i \bra{\phi_0}(\hat{g}(\kk)N_g\Delta_0 P_0)^n\ket{\phi_i}\bra{\phi_i}N_g\hat{g}(\kk)\ket{\phi_0}.
    \end{eqnarray}
In the last step, we insert a complete set of basis functions $\{\ket{\phi_i} \}$ that satisfies $\langle\phi_i | \phi_j\rangle=\delta_{i,j}$ and $\sum^{N_g-1}_{i=0} \ket{\phi_i}\bra{\phi_i}=1$, and $\ket{\phi_{i=0}}$ is just the wavefunction defined in Eq.(\ref{Eq:projection0}). 
The first factor in Eq.(\ref{Eq:GreenFunction_Relation_0}) can be calculated as \begin{eqnarray}\label{Eq:GreenFunction_Relation_02}
\bra{\phi_0}(\hat{g}(\kk)N_g\Delta_0P_0)^n\ket{\phi_i}&=&\bra{\phi_0}\hat{g}(\kk)N_g\Delta_0P_0 N_g\Delta_0P_0\dots\hat{g}(\kk)N_g\Delta_0 P_0\ket{\phi_i}\nonumber\\
&=&\bra{\phi_0}\hat{g}(\kk)N_g\Delta_0\ket{\phi_0}\bra{\phi_0}\dots\ket{\phi_0}\bra{\phi_0}\hat{g}(\kk)N_g\Delta_0 \ket{\phi_0}\langle\phi_0|\phi_i\rangle\nonumber\\
&=&(\Delta_0\mathbf{g}(\kk,E))^{n}\delta_{i,0},
\end{eqnarray}
where we have used the definition of the projection operator $P_0$ in Eq.(\ref{Eq:projection0}) in the second line of the above derivation.  Substituting Eq.(\ref{Eq:GreenFunction_Relation_02}) into Eq.(\ref{Eq:GreenFunction_Relation_0}), we find
\begin{eqnarray}\label{Eq:GreenFunction_Relation_1}
N_g \bra{\phi_0}G\ket{\phi_0} &=&\sum^\infty_{n=0}(\Delta_0\mathbf{g}(\kk,E))^{n}\mathbf{g}(\kk,E)\nonumber\\
&=&\frac{\mathbf{g}(\kk,E)}{1-\Delta_0 \mathbf{g}(\kk,E)}. 
\end{eqnarray}

We next consider the identity \begin{eqnarray}\label{Eq:1 free gas delta}
    [E-H_0-N_g\Delta_0 P_0](E-H_{2D})^{-1}= (E-H_{2D})^{-1}[E-H_0-N_g\Delta_0 P_0]=\mathcal{I},
\end{eqnarray}
where $\mathcal{I}$ is an identity matrix. Now we act on the wave functions $\ket{j}$ and $\ket{\phi_0}$ on the identity (\ref{Eq:1 free gas delta}) and obtain
\begin{eqnarray}
    \bra{\phi_0}(E-H_{2D})^{-1}[E-H_0-N_g\Delta_0 P_0]\ket{j}= \langle \phi_0 | j \rangle, \label{Eq:3 free gas delta}
\end{eqnarray}
where $\langle \phi_0 | j \rangle=\frac{1}{\sqrt{N_g}}$ according to the wavefunction form in Eq.(\ref{Eq:projection0}).
We can rearrange the terms in Eq.(\ref{Eq:3 free gas delta}) and obtain 
\begin{eqnarray}
&&(E-E_j)\bra{\phi_0}(E-H_{2D})^{-1}\ket{j}-\bra{\phi_0}(E-H_{2D})^{-1}N_g\Delta_0 \ket{\phi_0}\langle \phi_0 | j \rangle=\langle \phi_0 | j \rangle\nonumber\\
&&\bra{\phi_0}(E-H_{2D})^{-1}\ket{j}=\frac{\langle \phi_0 | j \rangle}{E-E_j}(1+N_g\Delta_0 \bra{\phi_0}(E-H_{2D})^{-1}\ket{\phi_0})
=\frac{1}{\sqrt{N_g}(E-E_j(\kk))}\frac{1}{1-\Delta_0 \mathbf{g}(\kk, E)}\label{Eq:4 free gas delta},
\end{eqnarray}
where we have used Eq.(\ref{Eq:GreenFunction_Relation_1}). 

We also act the wave function $\ket{j}$ on both sides of the equality (\ref{Eq:1 free gas delta}), leading to
\begin{eqnarray}
   && \bra{j} [E-H_0-N_g\Delta_0P_0](E-H_{2D})^{-1}\ket{j}=1 \rightarrow \nonumber \\
   && (E-E_j) \bra{j}(E-H_{2D})^{-1}\ket{j} = 1 + N_g\Delta_0 \langle j | \phi_0 \rangle \bra{\phi_0}(E-H_{2D})^{-1}\ket{j}
   \label{Eq:2 free gas delta}
\end{eqnarray}
Combined Eq.(\ref{Eq:2 free gas delta}) and Eq.(\ref{Eq:4 free gas delta}) give rise to
\begin{eqnarray} 
\bra{j} (E-H_{2D})^{-1}\ket{j}=\frac{1}{E-E_j(\kk)}+\frac{\Delta_0}{(E-E_j(\kk))^2}\frac{1}{1-\Delta_0 \mathbf{g}(\kk, E)}, \end{eqnarray}
which leads to the analytical form of $Tr(G)$, 
\begin{eqnarray} 
 Tr(G) = \sum_j \bra{j} (E-H_{2D})^{-1}\ket{j}=\mathbf{g}(\kk, E)+\sum_j\frac{\Delta_0}{(E-E_j)^2}\frac{1}{1-\Delta_0 \mathbf{g}(\kk, E)}. 
 \label{Eq:5 free e delat}
\end{eqnarray}

With Eq.(\ref{Eq:5 free e delat}), we can determine the energy spectrum from the poles of $Tr(G)$. At first sight, it seems that the poles of the free Green's function trace $\mathbf{g}(\kk, E)$, namely $E=E_j(\kk)$, should also be the poles of $Tr(G)$. However, this is {\it not} true. In the limit $E\rightarrow E_j$, Eq.(\ref{Eq:5 free e delat}) is approximated by 
\begin{eqnarray}
 lim_{E\rightarrow E_j}\left[Tr(G)\right]\approx lim_{E\rightarrow E_j}\left[\frac{1}{(E-E_j)}+\frac{\Delta_0}{(E-E_j)^2}\frac{1}{1-\Delta_0\frac{1}{(E-E_j)}}\right]=-\frac{1}{\Delta_0}, \label{Eq:6 free electron delta}
\end{eqnarray}
from which we find $Tr(G)$ is a constant but does not diverge so that $E=E_j$ is {\it not} a pole of $Tr(G)$ for any finite value of $\Delta_0$. 

\begin{figure}
\centering
\includegraphics[width=1\textwidth]{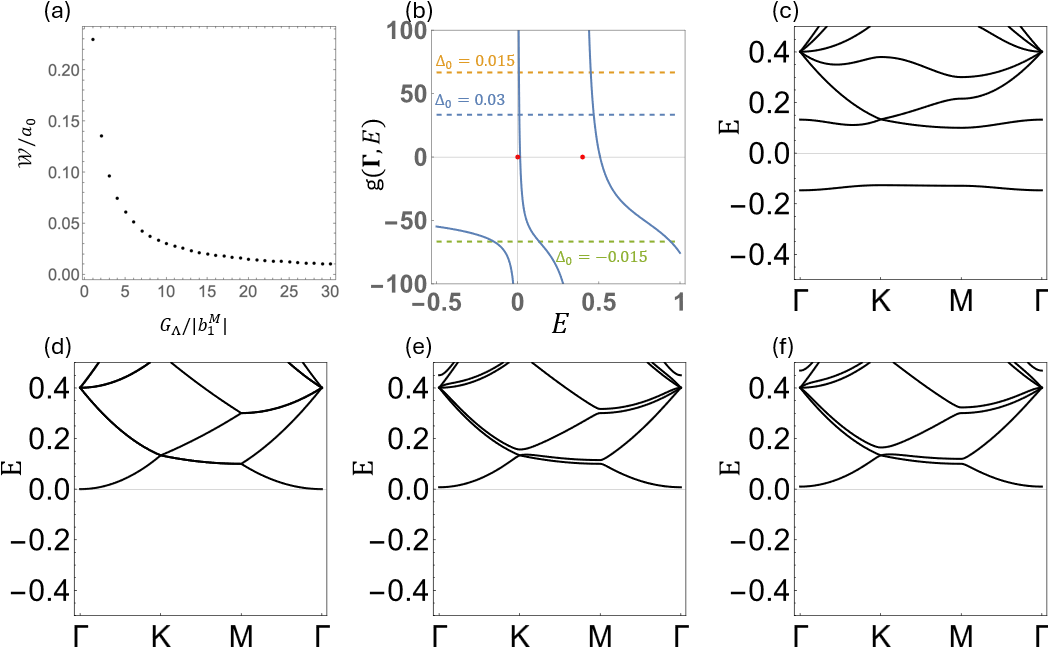}
\caption{\justifying \label{free e gas delta}  (a) The broadening $\mathcal{W}$ of $\delta$-function-like potential for different values of $G_\Lambda$. (b) $\mathbf{g}(\mathbf{\Gamma},E)$ as function of $E$. The yellow, blue and green dashed lines represent $1/\Delta_0$ for $\Delta_0$=0.015,  $\Delta_0$=0.03 and $\Delta_0$=-0.015, respectively. The red dots represent the eigenvalue of $H_{2D}(\kk)$ at the zero moir\'e potential $\Delta_0 = 0$. (c)-(f) show the band dispersion for $\Delta_0$=-0.015, $\Delta_0$=0, $\Delta_0$=0.015 and $\Delta_0$=0.03, respectively. Here the momentum cutoff is chosen as $G_{\Lambda}=23|b^M_1|$.}  
\end{figure}

From the second term in Eq.(\ref{Eq:5 free e delat}), we can see $Tr(G)$ diverges when the condition 
\begin{eqnarray} \label{Eq:equation_poles}
    1=\Delta_0\mathbf{g}(\kk, E)
\end{eqnarray}
is satisfied, which determines all poles of $Tr(G)$ in Eq.(\ref{Eq:5 free e delat}) and gives the energy dispersion of the full Hamiltonian $H_{2D}$. Fig.\ref{free e gas delta}(b) shows $\mathbf{g}(\kk=\mathbf{\Gamma}, E)$ as a function of $E$. In this plot, the intersections between the function $\mathbf{g}(\mathbf{\Gamma}, E)$ and the horizontal line $1/\Delta_0$ are the solutions of Eq.(\ref{Eq:equation_poles}), and thus determine the eigen-energy spectrum of $H_{2D}(\kk=\mathbf{\Gamma})$, e.g. $\mathcal{E}_n(\kk=\mathbf{\Gamma})$. One can see the energy spectrum is very different for $\Delta_0 > 0$ and $\Delta_0 < 0$. For $\Delta_0 \rightarrow 0^{\pm}$ (or $1/\Delta_0 \rightarrow \pm \infty$), all the curves of the function $\mathbf{g}(\mathbf{\Gamma}, E)$ merge to the energy values of a parabolic band after band folding at ${\bf \Gamma}$ in the mBZ. Here the momentum cutoff is chosen to be $G_\Lambda=23|b_1^M|$. For $\Delta_0>0$, we notice in Fig.\ref{free e gas delta}(b) all the eigen-energies remain positive, so there is no bound state with negative energy, and the lowest eigen-energy has the value of $\sim 0.008$ for $\Delta_0=0.015$ and $\sim 0.01$ for $\Delta_0=0.03$, thus remaining close to zero. Fig.\ref{free e gas delta}(d), (e) and (f) show the energy dispersion of $H_{2D}$ for $\Delta_0$=0, $\Delta_0$=0.015 and  $\Delta_0$=0.03, respectively. One can see the energy dispersion of the lowest conduction band remains almost unchanged for these different $\Delta_0$ values. For $\Delta_0<0$, a bound state with negative eigen-energy appears, as shown by the intersection between the dashed green horizontal line and the function $\mathbf{g}(\kk={\bf \Gamma},E)$ in Fig.\ref{free e gas delta}(b) for $\Delta_0$=-0.015. The corresponding energy dispersion for $\Delta_0$=-0.015 is shown in Fig.\ref{free e gas delta}(c), where all the low-energy bands are substantially influenced by moir\'e potential, in sharp contrast to the almost unchanged low-energy band dispersion for $\Delta_0\geq 0$ in Fig.\ref{free e gas delta}(d)-(f).


We emphasize that the binding energy of the bound state, denoted as $\mathcal{E}_0$, depends on the momentum cut-off $G_{\Lambda}$ (or equivalently $N_g$), 
and $\mathcal{E}_0$ diverges as $G_{\Lambda}\rightarrow \infty$. To see that, we can consider $\mathbf{g}(\kk,E)$ at $\kk=\mathbf{\Gamma}$ and change the summation of the index $j$ to the integral over the momentum as 
\begin{eqnarray} 
    &&\mathbf{g}(\mathbf{\Gamma},E)=\sum_j\frac{1}{E-g^2_j}=\frac{1}{\Omega_B }\int^{G_\Lambda}_0 \frac{2\pi g}{E-g^2} dg=-\frac{\pi}{\Omega_B} ln\left(\frac{E-G^2_\Lambda}{E}\right),
\end{eqnarray}
where $\Omega_B$ is the area of mBZ. Combining Eq.(\ref{eq_SM:boundstate_energy1}) with Eq.(\ref{Eq:equation_poles}) yields 
\begin{eqnarray} \label{eq_SM:boundstate_energy1}
    \mathbf{g}(\mathbf{\Gamma},E = \mathcal{E}_0)=\frac{1}{\Delta_0} \rightarrow \mathcal{E}_0=-\frac{G_\Lambda^2}{-1+e^{-\frac{\Omega_B}{\Delta_0 \pi}}}. 
\end{eqnarray}

This divergence originates from the mathematical simplification of the $\delta$-function. In the numerical calculations, we always choose a cutoff, so that the moir\'e potential  is written as 
\begin{eqnarray} \label{Eq: potential broadening cutoff}
    &&H_{\text{m}}(\rr) =\sum_{\gs\leq|G_\Lambda|} \Delta_{0} e^{i \gs \cdot \rr}\nonumber
\end{eqnarray}
in the real space. 
When $|G_\Lambda|\rightarrow\infty$, the potential recovers the periodic $\delta$ potential. For a finite $|G_\Lambda|$, the potential $H_{\text{m}}(\rr)$ decays as a function of $r$ and we define the broadening parameter $\mathcal{W}$ as the distance at which $H_{\text{m}}((\mathcal{W},0))$ is half of $H_{\text{m}}(\rr=0)$, $\frac{H_{\text{m}}((\mathcal{W},0))}{H_{\text{m}}(\rr=0)}=0.5$. Fig.\ref{free e gas delta}(a) shows $\mathcal{W}$ as a function of $G_\Lambda$, from which one can see that the broadening parameter $\mathcal{W}$ decays rapidly when increasing the cut-off $G_\Lambda$.  For a realistic periodic $\delta$-potential-like  $H_{\text{m}}(\rr)$, the cut-off $G_\Lambda$ can be determined by the spatial broadening of $H_{\text{m}}(\rr)$.

Furthermore, we can also extract the eigen wavefunction form of the bound state from the above formalism. We denote the eigen wave function of $H_{2D}$ as $\ket{n}$, namely $H_{2D}\ket{n,\kk} = \mathcal{E}_n(\kk) \ket{n,\kk}$ with $\mathcal{E}_{n}(\kk)$ determined from Eq.(\ref{Eq:equation_poles}). One can show  
\begin{eqnarray}
\text{lim}_{E\rightarrow \mathcal{E}_n(\kk)}\bra{j} (E-H_{2D})^{-1}\ket{\phi_0}&=&\sum_{n'}\text{lim}_{E\rightarrow \mathcal{E}_n(\kk)}\bra{j} (E-H_{2D})^{-1}\ket{n',\kk}\langle n',\kk|\phi_0\rangle\nonumber\\
&\approx&\text{lim}_{E\rightarrow \mathcal{E}_n(\kk)} (E-\mathcal{E}_n(\kk))^{-1} \langle j|n,\kk\rangle \langle n,\kk|\phi_0\rangle
.\label{Eq:wavefunction 1d 0}
\end{eqnarray}
From Eq.(\ref{Eq:wavefunction 1d 0}), we have
\begin{eqnarray}
 \langle j | n,\kk \rangle \langle n,\kk | \phi_0 \rangle&=&\text{lim}_{E\rightarrow \mathcal{E}_n(\kk)}(E- \mathcal{E}_n(\kk))\bra{j} (E-H_{2D})^{-1}\ket{\phi_0}\nonumber\\
 &=&\text{lim}_{E\rightarrow \mathcal{E}_n(\kk)}\frac{1}{ \sqrt{N_g}}\frac{E- \mathcal{E}_n(\kk)}{1-\Delta_0 \mathbf{g}(\kk, E)}\frac{1}{E-E_j(\kk)},
  \label{Eq:wavefunction 1d}
\end{eqnarray}
where we have used Eq.(\ref{Eq:4 free gas delta}) in the last step. When $E\rightarrow\mathcal{E}_n$, Eq.(\ref{Eq:equation_poles}) should be satisfied so both the numerator $E- \mathcal{E}_n(\kk)$ and the denominator $1-\Delta_0 \mathbf{g}(\kk, E)$ in Eq.(\ref{Eq:wavefunction 1d}) are approaching zero. This limit can be evaluated via the L'Hospital's rule as
\begin{eqnarray} 
\text{lim}_{E\rightarrow \mathcal{E}_n(\kk)}\frac{E- \mathcal{E}_n(\kk)}{1-\Delta_0 \mathbf{g}(\kk, E)}&=&-\left.\frac{1}{\Delta_0\partial\mathbf{g}(\kk,E)/\partial E}\right|_{E=\mathcal{E}_n(\kk)}
= \frac{1}{\Delta_0\sum_{j'} (\mathcal{E}_n(\kk)-E_{j'}(\kk))^{-2}}.
  \label{Eq:limit wave}
\end{eqnarray}
 Although some $\mathcal{E}_n(\kk)$ are very close to $E_j(\kk)$, they are not equal, i.e., $\mathcal{E}_n(\kk) \neq E_j(\kk)$, as proven in Eq.~(\ref{Eq:6 free electron delta}). Therefore, the limit in Eq.(\ref{Eq:limit wave}) converges and 
\begin{eqnarray}\label{Eq:wavefunction 1d1}
\langle j | n,\kk \rangle \langle n,\kk | \phi_0 \rangle = \frac{1}{ \sqrt{N_g} } \frac{1}{\Delta_0\sum_{j'} (\mathcal{E}_n(\kk)-E_{j'}(\kk))^{-2}} \frac{1}{\mathcal{E}_n(\kk)-E_j(\kk)}. \\
 \langle j | n,\kk \rangle = \frac{1}{ \sqrt{N_g} \langle n,\kk | \phi_0 \rangle} \frac{1}{\Delta_0\sum_{j'} (\mathcal{E}_n(\kk)-E_{j'}(\kk))^{-2}} \frac{1}{\mathcal{E}_n(\kk)-E_j(\kk)}. \label{Eq:wavefunction 1d2}
\end{eqnarray}
In the last step, we divide $ \langle n,\kk | \phi_0 \rangle$ in both sides, which requires $\langle n,\kk | \phi_0 \rangle \neq 0$.
Supposing $\langle n,\kk | \phi_0 \rangle = 0$, we will have $H_{2D}\ket{n,\kk}=(H_0(\kk)+P_0)\ket{n,\kk}=H_0(\kk)\ket{n,\kk}$ using the form of $P_0$ in Eq.(\ref{Eq:projection0}). Since $\ket{n,\kk}$ is an eigenstate of $H_{2D}(\kk)$, e.g. $H_{2D}(\kk)\ket{n,\kk}=\mathcal{E}(\kk)\ket{n,\kk}$, we find $\mathcal{E}(\kk)$ has also to be an eigenstate of $H_0(\kk)$. However, we have previously proven any eigen-energy $E_j(\kk)$ of $H_0(\kk)$ is {\it not} a pole of $Tr(G)$ and thus cannot be an eigenstate of $H_{2D}$, and thus we find a contradiction, which means $\langle n,\kk | \phi_0 \rangle$ cannot be zero, $\langle n,\kk | \phi_0 \rangle\neq0$.  
We note that only $\frac{1}{\mathcal{E}_n(\kk)-E_j}$ depends on the index $j$, while the whole factor $\frac{1}{ \sqrt{N_g} \langle n,\kk | \phi_0 \rangle} \frac{1}{\Delta_0\sum_{j'} (\mathcal{E}_n(\kk)-E_{j'}(\kk))^{-2}}$ in front of $\frac{1}{\mathcal{E}_n(\kk)-E_j}$ in Eq.(\ref{Eq:wavefunction 1d2}) is independent of $j$ and thus just the normalization factor. Therefore, we can write the wavefunction as 
\begin{equation} \label{Eq:wavefunction 1d3}
    \langle j | n, \kk\rangle=N_{2D} \frac{1}{\mathcal{E}_n(\kk)-E_j(\kk)}
\end{equation} 
where $N_{2D}=\frac{1}{ \sqrt{N_g} \langle n,\kk | \phi_0 \rangle} \frac{1}{\Delta_0\sum_{j'} (\mathcal{E}_n(\kk)-E_{j'}(\kk))^{-2}}$ is regarded as the normalization factor. 

We denote the localized eigen-wavefunction of the Hamiltonian in Eq.(\ref{Eq:H2DEG_momentum}) of bounded state as $\ket{0,\kk}$. For the wavefunction at $\mathbf{\Gamma}$, namely $\ket{0,\mathbf{\Gamma}}$, its real space form can be approximated as
\begin{eqnarray}
    \langle \rr|0,\mathbf{\Gamma}\rangle = \sum_j \langle j|0,\mathbf{\Gamma}\rangle e^{i \gs_j\cdot \rr}  \approx \frac{1}{\Omega_B} \int d\gs e^{i \gs \cdot \rr} \langle \gs|0,\mathbf{\Gamma}\rangle,
\end{eqnarray}
where we change the summation over the discrete $\gs_j$ to the integral over the continuous variable $\gs$. Using Eq.(\ref{Eq:wavefunction 1d2}), we find 
\begin{eqnarray}
   \langle \rr|0,\mathbf{\Gamma}\rangle = \frac{N_{2D}}{(2\pi)^2}\int\frac{1}{{\mathcal{E}_0(\mathbf{\Gamma})-\gs^2}}e^{i \gs\cdot \rr} d \gs = -N_{2D}K_0(\sqrt{-\mathcal{E}_0(\mathbf{\Gamma})} |\rr|),
\end{eqnarray}
where $K_0(\rr)$ is the modified Bessel function of the second kind. In the limit $|\rr| \gg 1$, the asymptotic behavior of $K_{0}(\rr)$ is given by 
$K_{0}(\rr) \approx \sqrt{\frac{\pi}{2|\rr|}}\;e^{-\sqrt{-\mathcal{E}_0(\mathbf{\Gamma})}|\rr|}$, thus corresponding to an exponentially localized wavefunction in periodic $\delta$-like-function.

\begin{figure}
\centering
\includegraphics[width=1\textwidth]{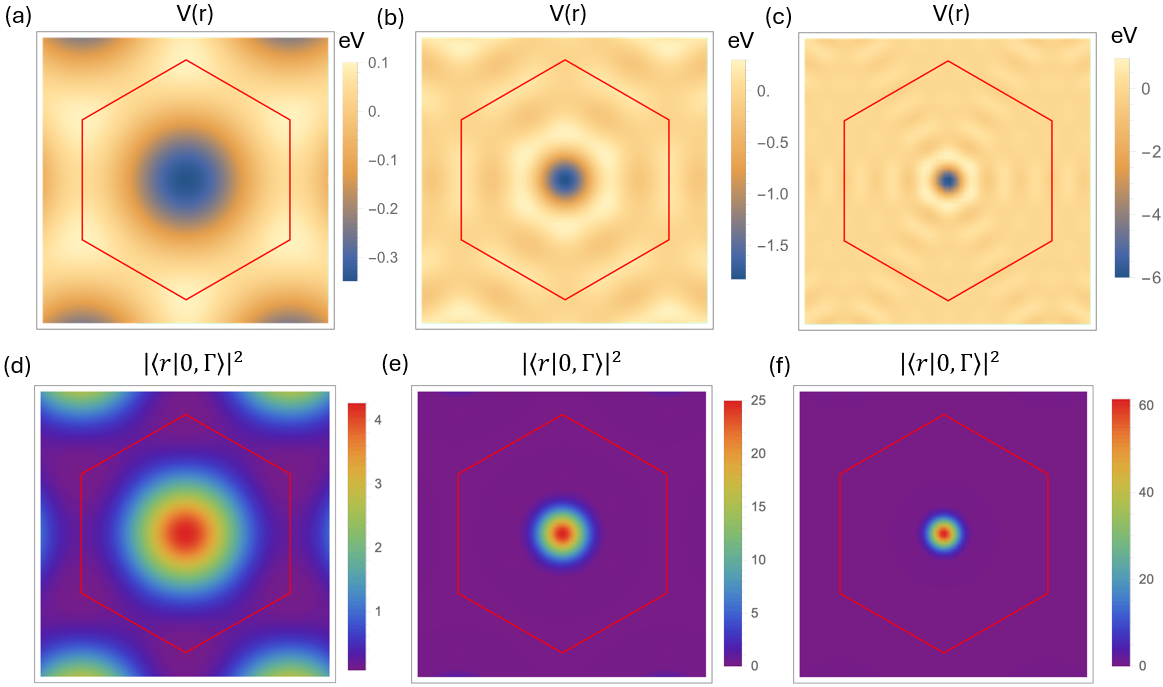}
\caption{\justifying \label{Fig: Wave_function_ delta} 
(a)-(c) V$(\rr)$ with $\Delta_0$=-0.05 at $G_\Lambda$= $|b^M_1|$, 3$|b^M_1|$ and 6$|b^M_1|$. (d)-(f) $|\langle r|0,\mathbf{\Gamma}\rangle|^2$ with $\Delta_0$=-0.05 at $G_\Lambda$= $|b^M_1|$, 3$|b^M_1|$ and 6$|b^M_1|$. }
\end{figure}

From the above form (\ref{Eq:wavefunction 1d2}) of the eigen wavefunction, we also numerically examine the localized wavefunction form of the bound state, denoted as $\ket{0,\kk}$, for different values of the cut-off $G_\Lambda$ in  Fig.\ref{Fig: Wave_function_ delta}. Fig.\ref{Fig: Wave_function_ delta}(a)-(c) shows the moir\'e superlattice potential form for different cut-off values of $G_\Lambda$, from which one can see the confining potential becomes more and more similar to a $\delta$-function potential as  $G_\Lambda$ is increased. At the same time, the real space form of the bound state eigen wavefunction, denoted by $|\langle \rr|0,\mathbf{\Gamma}\rangle|^2$, also becomes more and more localized around 1a Wyckoff position of moir\'e unit cell with increasing $G_\Lambda$ in Fig.\ref{Fig: Wave_function_ delta}(d)-(f). Here we keep $\Delta_0=-0.05$. 

\subsection{Moir\'e Chern-band model with a periodic $\delta$-function potential} \label{section:delta chern band}
Now we apply the periodic $\delta$-function potential to the Chern-band model, and the full Hamiltonian is $H_{mC}$ in Eq.(\ref{Eq:HammC}) with the Chern-band Hamiltonian $H_{C}$ in Eq.(\ref{Eq:Ham_Chernband}). In the Sec.\ref{Sec:deltapotential_parabolicbands}, we have shown that for a small positive $\Delta_0$, the band dispersion around the conduction band bottom remains almost unchanged. Therefore, for the convenience of deriving a compact formalism, we apply the periodic $\delta$-function-like potential to both the conduction and valence bands with $\Delta_0>0$. $H_{m}$ is given by the periodic $\delta$-function-like  potential in Eq.(\ref{Eq:HM_delta_periodic}). 
In the momentum space, the full Hamiltonian reads
\begin{eqnarray}\label{Eq:HammC_delta_momentum}
&&H_{mC}(\kk) =H_0(\kk) + H_{\text{m}} = H_0(\kk) +N_g\Delta_0 P_0 \\
&&H_0(\kk)=\begin{pmatrix}
H_{C} (\kk + \gs_1)  & 0 & \dots& 0\\
0 & H_{C} (\kk + \gs_2) & \dots& 0\\
\vdots & \vdots & \ddots&  \vdots\\
0 & 0 & \dots & H_{C} (\kk + \gs_{N_g})
\end{pmatrix},
\end{eqnarray}
where $H_{C} (\kk)$ is a $2\times 2$ matrix defined in Eq.(\ref{Eq:Ham_Chernband}), $\gs_j$ is the moir\'e reciprocal lattice vectors within the momentum cut-off $G_\Lambda$ ($j = 1, 2,..., N_g$), and 
\begin{eqnarray} \label{eq_SM:P0_projector}
    P_0 = \ket{\phi_0}\bra{\phi_0} \otimes 1_{2\times2}
\end{eqnarray} is the projection operator with $\ket{\phi_0}$ defined in Eq.(\ref{Eq:projection0}) and  $1_{2\times 2}$ is the 2-by-2 identity matrix. 

Below we define $\ket{\phi_0,\alpha}=\ket{\phi_0}\otimes\ket{\alpha}$, with $\alpha = 1, 2$ indexing two orbital basis of $H_{C}$. The Moir\'e term can be written as $H_{\text{m}}=N_g\Delta_0  (\ket{\phi_0,1}\bra{\phi_0,1}+\ket{\phi_0,2}\bra{\phi_0,2})$. For the derivation below, we define $\ket{j,\alpha}$ as $\ket{j,\alpha}=\ket{j}\otimes\ket{\alpha}$ for $j$ indexing the moir\'e reciprocal lattice vector, so that we have  $\bra{j,\alpha}H_0(\kk)\ket{j',\beta}= H_{C}^{\alpha\beta}(\kk+\gs_j)\delta_{j'j}$. $\ket{j,\alpha}$ is independent of the momentum $\kk$.

The bare Green's function is defined by the matrix inverse
\begin{eqnarray}
    \hat{g}(\kk)=(E 1_{2N_g\times 2N_g}-H_{0}(\kk))^{-1}
\end{eqnarray}
and its component connecting the state $\ket{\phi_0,\alpha}$ and $\ket{\phi_0,\beta}$ is given by 
\begin{eqnarray}\label{Eq: bare GreenFunction_Relation_BHZ_0}
    \mathbf{g}^{0}_{\alpha,\beta}(\kk,E) = N_g\bra{\phi_0,\alpha}\hat{g}\ket{\phi_0,\beta} = N_g\bra{\phi_0,\alpha}(E 1_{2N_g\times 2N_g}-H_{0}(\kk))^{-1}\ket{\phi_0,\beta}. 
\end{eqnarray}

By using the form of $|\phi_0\rangle$ in Eq.(\ref{Eq:projection0}), we can further define the matrix form $\mathbf{g}(k,E)$ as 
\begin{eqnarray}
    \mathbf{g}(\kk, E)=\sum_j \mathbf{g}^0(\kk+\gs_j, E), 
\end{eqnarray}
where
\begin{eqnarray}
   \mathbf{g}^0(\kk+\gs_j, E)=\frac{1}{E 1_{2\times 2}-H_{C}(\kk+\gs_j)}
\end{eqnarray}
is a $2\times 2$ matrix. The full Green's function is defined as 
\begin{eqnarray}
    G = (E 1_{2N_g\times 2N_g}-H_{mC}(\kk))^{-1}
\end{eqnarray}
and its component connecting the state $\ket{\phi_0,\alpha}$ and $\ket{\phi_0,\beta}$ is given by 
\begin{eqnarray}\label{Eq:GreenFunction_Relation_0_BHZ}
    N_g \bra{\phi_0,\alpha}G\ket{\phi_0,\beta} &=& N_g\bra{\phi_0,\alpha}(E 1_{2N_g\times 2N_g}-H_{0}(\kk)-H_{\text{m}})^{-1}\ket{\phi_0,\beta}\nonumber\\&=&N_g\bra{\phi_0,\alpha}(1_{2N_g\times 2N_g}-(E 1_{2N_g\times 2N_g}-H_{0}(\kk))^{-1}H_{\text{m}})^{-1}(E 1_{2N_g\times 2N_g}-H_{0}(\kk))^{-1}\ket{\phi_0,\alpha}\nonumber\\
    &=&N_g\bra{\phi_0,\alpha}\sum^\infty_{n=0} (\hat{g}(\kk)H_{\text{m}})^n\hat{g}(\kk)\ket{\phi_0,\beta}\nonumber\\
    &=&N_g\sum^\infty_{n=0}\sum_{i,\gamma} \bra{\phi_0,\alpha}(\hat{g}(\kk)H_{\text{m}})^n\ket{\phi_i,\gamma}\bra{\phi_i,\gamma}\hat{g}(\kk)\ket{\phi_0,\beta},
    \end{eqnarray}
where $\alpha, \beta, \gamma = 1, 2$ and $\phi_i$ indexing a complete set of basis wavefunction with $\phi_{i=0}$ given by Eq.(\ref{Eq:projection0}).
In Eq.(\ref{Eq:GreenFunction_Relation_0_BHZ}), we have
\begin{eqnarray}\label{Eq:GreenFunction_Relation_02_BHZ}
\bra{\phi_0,\alpha}(\hat{g}(\kk)H_{\text{m}})^n\ket{\phi_i,\gamma}&=&\sum_j\sum_{\gamma'}\dots\sum_k\sum_{\gamma''}\bra{\phi_0,\alpha}\hat{g}(\kk)H_{\text{m}}\ket{\phi_j,\gamma'}\bra{\phi_j,\gamma'}\dots\ket{\phi_k,\gamma''}\bra{\phi_k,\gamma''}\hat{g}(\kk)H_{\text{m}}\ket{\phi_i,\gamma}\nonumber\\
&=&\Delta_0^n[\mathbf{g}^{n}(\kk,E)]_{\alpha,\gamma}\delta_{i,0},
\end{eqnarray}
where we have used the form $H_{\text{m}} = N_g \Delta_0 P_0$ with $P_0$ given by Eq.(\ref{eq_SM:P0_projector}) in deriving the last step. Substituting Eq.(\ref{Eq:GreenFunction_Relation_02_BHZ}) into Eq.(\ref{Eq:GreenFunction_Relation_0_BHZ}), we get
\begin{eqnarray}\label{Eq:GreenFunction_Relation_1_BHZ}
N_g \bra{\phi_0,\alpha}G\ket{\phi_0,\beta} &=&\sum^\infty_{n=0}\sum_\gamma \Delta^n_1[\mathbf{g}^n(\kk,E)]_{\alpha,\gamma}\mathbf{g}(\kk,E)_{\gamma,\beta}\nonumber\\
&=& \sum_\gamma (1_{2\times 2} -\Delta_0 \mathbf{g}(\kk,E))^{-1}_{\alpha,\gamma} [\mathbf{g}(\kk,E)]_{\gamma,\beta}
\end{eqnarray}
The corresponding $2\times 2$ matrix form can be written as
\begin{eqnarray}\label{Eq:GreenFunction_Relation_1_BHZ_matrix}
N_g \bra{\phi_0}G\ket{\phi_0} &=&\sum^\infty_{n=0}\Delta_0^n\mathbf{g}^n(\kk,E)\mathbf{g}(\kk,E)\nonumber\\
&=&(1_{2\times 2} -\Delta_0 \mathbf{g}(\kk,E))^{-1}\mathbf{g}(\kk,E).
\end{eqnarray}
Eq.(\ref{Eq:GreenFunction_Relation_1_BHZ_matrix}) is the generalization of Eq.(\ref{Eq:GreenFunction_Relation_1}) to the moir\'e Chern-band model.

We further introduce the projection operator 
\begin{eqnarray}
\PP_{\kk+\gs_j, \eta}=\ket{\kk+\gs_j,\eta}\bra{\kk+\gs_j,\eta}, 
\end{eqnarray}
where $\ket{\kk+\gs_j,\eta}$ is the eigenstate of $H_{C}(\kk+\gs_j)$, 
\begin{eqnarray}\label{Eq:HamC_eigen_basis}
H_{C}(\kk+\gs_j)\ket{\kk+\gs_j,\eta}=E_{\eta}(\kk+\gs_j)\ket{\kk+\gs_j,\eta}
\end{eqnarray}
with $\eta = 1, 2$ labelling two eigenstates. The project operator can be written in the Pauli matrix form, 
\begin{eqnarray}\label{Eq: Projection Pauli} 
&&\PP_{\kk,\eta}=\sum^3_{i=0}d^{\eta}_{i,\kk}\sigma_i\nonumber\\
&&d^{\eta}_{0,\kk}=\frac{1}{2}; d^{\eta}_{1,\kk}=\frac{\eta A k_x}{2 D_{0\kk}};d^{\eta}_{2,\kk}=\frac{-\eta  Ak_y}{2 D_{0\kk}};d^{\eta}_{3,\kk}=\frac{\eta\mathcal{M}(\kk)}{2 D_{0\kk}},\nonumber\\
&& D_{0\kk}=\sqrt{\mathcal{M}^2(\kk)+A^2 k^2}
\end{eqnarray}
for the Chern-band model Hamiltonian $H_C$ in Eq.(\ref{Eq:Ham_Chernband}). With the projection operator, we can rewrite 
\begin{eqnarray}\label{Eq:freeGreenFunction_BHZ}
    \mathbf{g}^0(\kk+\gs_j, E) = \sum_{\eta} \frac{\PP_{\kk+\gs_j,\eta}}{E-E_{\eta}(\kk+\gs_j)};  \quad \mathbf{g}(\kk, E)=\sum_j \mathbf{g}^0(\kk+\gs_j, E). 
\end{eqnarray}

We next need to find the trace of the full Green's function. We start with  \begin{eqnarray}\label{Eq:identity BHZ}
    &&[E 1_{2N_g\times 2N_g} -H_0-N_g\Delta_0  P_0](E 1_{2N_g\times 2N_g}-H_{mC})^{-1} =  (E 1_{2N_g\times 2N_g} - H_{mC})^{-1}[E 1_{2N_g\times 2N_g}-H_0-N_g\Delta_0  P_0]\nonumber \\
    &&=1_{2N_g\times 2N_g}
\end{eqnarray}
Using Eq.(\ref{Eq:identity BHZ}), we have
\begin{eqnarray}\label{Eq:alphaj BHZ 1}
   &&\bra{\phi_0,\alpha}(E 1_{2N_g\times 2N_g} - H_{mC})^{-1}(E 1_{2N_g\times 2N_g} -H_0)\ket{j, \beta}-\sum^2_{\gamma=1}N_g\Delta_0\bra{\phi_0,\alpha}(E 1_{2N_g\times 2N_g}-H_{mC})^{-1}\ket{\phi_0,\gamma}\langle \phi_0,\gamma | j,\beta \rangle\nonumber \\
   && =\langle \phi_0,\alpha | j,\beta \rangle.
\end{eqnarray}  
We insert the identity $1 = \sum_{j', \gamma}\ket{j',\gamma}\bra{j',\gamma}$ into the first term of Eq.(\ref{Eq:alphaj BHZ 1}) and obtain
\begin{eqnarray}
 &&\sum_{j',\gamma}\bra{\phi_0,\alpha}(E 1_{2N_g\times 2N_g}-H_{mC})^{-1}\ket{j',\gamma}\bra{j',\gamma}(E 1_{2N_g\times 2N_g}-H_0)\ket{j,\beta}\nonumber \\
 && =\sum^2_{\gamma=1}\bra{\phi_0,\alpha}(E 1_{2N_g\times 2N_g}-H_{mC})^{-1} \ket{j,\gamma}(E \delta_{\beta \gamma} - H^{\gamma\beta}_{C}(\kk+\gs_j)).
\end{eqnarray} 
Therefore, Eq.(\ref{Eq:alphaj BHZ 1}) changes to 
\begin{eqnarray}
    && \sum^2_{\gamma=1}\bra{\phi_0,\alpha}(E 1_{2N_g\times 2N_g}-H_{mC})^{-1} \ket{j,\gamma}(E \delta_{\beta \gamma} - H^{\gamma\beta}_{C}(\kk+\gs_j)) = \nonumber \\
    && \langle \phi_0,\alpha | j,\beta \rangle + \sum^2_{\gamma=1}N_g\Delta_0\bra{\phi_0,\alpha}(E 1_{2N_g\times 2N_g}-H_{mC})^{-1}\ket{\phi_0,\gamma}\langle \phi_0,\gamma | j,\beta \rangle . \label{alphaj BHZ 1 component}
\end{eqnarray}

We can further simplify Eq.(\ref{alphaj BHZ 1 component}) by the matrix inverse,
\begin{eqnarray}
    && \sum^2_{\beta,\gamma=1}\bra{\phi_0,\alpha}(E 1_{2N_g\times 2N_g}-H_{mC})^{-1} \ket{j,\gamma}(E 1_{2\times2} - H_{C}(\kk+\gs_j))_{\gamma,\beta}(E 1_{2\times2}- H_{C}(\kk+\gs_j))^{-1}_{\beta,\eta}  \nonumber \\
    && =\sum^2_{\beta=1}\left(\frac{\delta_{\alpha,\beta}}{\sqrt{N_g}}+\sum^2_{\gamma=1}N_g\Delta_0\bra{\phi_0,\alpha}(E 1_{2N_g\times 2N_g}-H_{mC})^{-1}\ket{\phi_0,\gamma}\frac{\delta_{\gamma,\beta}}{\sqrt{N_g}}\right)(E 1_{2\times2}- H_{C}(\kk+\gs_j))^{-1}_{\beta,\eta},  \nonumber
    \end{eqnarray}
    
 so that
    \begin{eqnarray}
     && \bra{\phi_0,\alpha}(E 1_{2N_g\times 2N_g}-H_{mC})^{-1} \ket{j,\eta} =\frac{1}{\sqrt{N_g}} \sum^2_{\beta=1} ( \delta_{\alpha,\beta} +N_g\Delta_0\bra{\phi_0,\alpha}G\ket{\phi_0,\beta} ) (E 1_{2\times2}- H_{C}(\kk+\gs_j))^{-1}_{\beta,\eta} \nonumber \\
     && = \frac{1}{\sqrt{N_g}} \sum^2_{\beta=1} ( \delta_{\alpha,\beta} +\sum_\gamma (1_{2\times 2} - \Delta_0 \mathbf{g}(\kk,E))^{-1}_{\alpha,\gamma} [\Delta_0 \mathbf{g}(\kk,E)]_{\gamma,\beta}) (E 1_{2\times2}- H_{C}(\kk+\gs_j))^{-1}_{\beta,\eta} \nonumber \\
     && = \frac{1}{\sqrt{N_g}} \sum^2_{\beta=1} (1_{2\times 2} - \Delta_0 \mathbf{g}(\kk,E))^{-1}_{\alpha,\beta}  (E 1_{2\times2}- H_{C}(\kk+\gs_j))^{-1}_{\beta,\eta}, 
     \label{alphaj BHZ 2 component}
\end{eqnarray}
where we have used $\langle\phi_0, \alpha|j, \beta\rangle=\frac{1}{\sqrt{N_g}} \delta_{\alpha\beta}$ and Eq.(\ref{Eq:GreenFunction_Relation_1_BHZ_matrix}). The above expression can be rewritten into a $2\times 2$ matrix form
\begin{eqnarray} \label{Eq:alphaj BHZ 1 Matrix} 
&& \bra{\phi_0}(E 1_{2N_g\times 2N_g} -H_{mC})^{-1}\ket{j}= \frac{1}{\sqrt{N_g}} (1_{2\times 2} - \Delta_0 \mathbf{g}(\kk,E))^{-1}  (E 1_{2\times2}- H_{C}(\kk+\gs_j))^{-1}. \label{Eq:alphaj BHZ 2 Matrix} 
\end{eqnarray}
Eq.(\ref{alphaj BHZ 2 component}) can be regarded as a generalization of Eq.(\ref{Eq:4 free gas delta}) to the moir\'e Chern-band model. 

We next generalize Eq.(\ref{Eq:2 free gas delta}) as
\begin{eqnarray} \label{Eq: JJ BHZ 1}
&& \bra{j,\alpha} \left[E 1_{2N_g\times 2N_g} -H_0- \sum^2_{\gamma=1}N_g\Delta_0\ket{\phi_0,\gamma}\bra{\phi_0,\gamma}\right](E 1_{2N_g\times 2N_g}-H_{mC})^{-1}\ket{j,\beta}=\delta_{\alpha,\beta}\nonumber\\
\end{eqnarray}
which can be simplified as
\begin{eqnarray} \label{Eq:2d generalization SC16}
&&\sum^{2}_{\gamma=1}\sum_{j'}\bra{j,\alpha} (E 1_{2N_g\times 2N_g} -H_0)\ket{j',\gamma}\bra{j',\gamma}(E 1_{2N_g\times 2N_g} -H_{mC})^{-1}\ket{j,\beta} = \nonumber \\
&& \delta_{\alpha,\beta}+ \sum^2_{\gamma=1}N_g\Delta_0\langle j,\alpha|\phi_0,\gamma\rangle\bra{\phi_0,\gamma}(E 1_{2N_g\times 2N_g}-H_{mC})^{-1}\ket{j,\beta},
\nonumber\\
&&\sum_\gamma(E 1_{2\times 2} -H_{C}(\kk+\gs_j))_{\alpha\gamma} \bra{j,\gamma} (E 1_{2N_g\times 2N_g} -H_{mC})^{-1}\ket{j,\beta} = \nonumber \\
&& \delta_{\alpha,\beta}+\sum_{\eta}\Delta_0 (1_{2\times2}-\Delta_0 \mathbf{g}(\kk,E))^{-1}_{\alpha\eta}(E 1_{2\times 2} -H_{C}(\kk+\gs_j))^{-1}_{\eta\beta},
\end{eqnarray}
where we have used Eq.(\ref{alphaj BHZ 2 component}) to simplify the right side of the first equation in Eq.(\ref{Eq:2d generalization SC16}). With matrix inversion, we find
\begin{eqnarray}
    &&\bra{j,\gamma} (E 1_{2N_g\times 2N_g} -H_{mC})^{-1}\ket{j,\beta}=(E 1_{2\times 2} -H_{C}(\kk+\gs_j))^{-1}_{\gamma\beta} +\nonumber\\
&&\sum_{\eta,\alpha} (E 1_{2\times 2} - H_{C}(\kk+\gs_j))^{-1}_{\gamma\alpha}\Delta_0 (1_{2\times2}-\Delta_0 \mathbf{g}(\kk,E))^{-1}_{\alpha\eta}(E-H_{C}(\kk+\gs_j))^{-1}_{\eta\beta}.
\label{Eq:trace BHZ delta comp}
\end{eqnarray}
The matrix form of Eq.(\ref{Eq:trace BHZ delta comp}) is 
\begin{eqnarray}
&& \bra{j} (E 1_{2N_g\times 2N_g} -H_{mC}(\kk))^{-1}\ket{j}=\frac{1}{E 1_{2\times 2}-H_{C}(\kk+\gs_j)}+\frac{\Delta_0}{E 1_{2\times 2}-H_{C}(\kk+\gs_j)}\frac{1}{1_{2\times 2}-\Delta_0\mathbf{g}(\kk,E)}\frac{1}{E 1_{2\times 2}-H_{C}(\kk+\gs_j)}.\nonumber \\\label{Eq:trace BHZ delta0}
\end{eqnarray}
By summing over $j$ on both sides, we get the trace of the full Green's function
\begin{eqnarray}
&& Tr(G) = Tr_{\sigma}\left(\sum_j \bra{j} (E 1_{2N_g\times 2N_g} -H_{mC}(\kk))^{-1}\ket{j}\right)  \nonumber \\
&& = Tr_{\sigma}\left(\mathbf{g}(\kk,E)+\sum_j\frac{\Delta_0}{E 1_{2\times 2}-H_{C}(\kk+\gs_j)}\frac{1}{1_{2\times 2}-\Delta_0\mathbf{g}(\kk,E)}\frac{1}{E 1_{2\times 2}-H_{C}(\kk+\gs_j)}\right) \nonumber \\
&& = Tr_{\sigma}\left(\mathbf{g}(\kk,E)+\Delta_0 \sum_j \mathbf{g}^0(\kk+\gs_j, E) \frac{1}{1_{2\times 2}-\Delta_0\mathbf{g}(\kk,E)} \mathbf{g}^0(\kk+\gs_j, E)\right),\label{Eq:trace BHZ delta}
\end{eqnarray}
where $Tr_{\sigma}$ is taken for the $2\times 2$ matrices.
Eq.(\ref{Eq:trace BHZ delta}) can be viewed as the generalization of Eq.(\ref{Eq:5 free e delat}).

We next would like to identify the poles of $Tr(G)$. We will first show the poles of $\mathbf{g}(\kk,E)$ are {\it not} the poles of $Tr(G)$. From Eq.(\ref{Eq:freeGreenFunction_BHZ}), 
the pole of $\mathbf{g}(\kk,E)$ is given by $E=E_{\eta}(\kk+\gs_j)$. When $E = E_{\eta}(\kk+\gs_j) + \delta E$ with $\delta E \rightarrow 0$, we only need to keep the divergent terms so that 
\begin{eqnarray}
    \mathbf{g}(\kk, E) \approx \mathbf{g}^0(\kk+\gs_{j}, E) \approx \frac{\PP_{\kk+\gs_j,\eta}}{\delta E}
\end{eqnarray}
and  $Tr(G)$ is approximated by 
\begin{eqnarray}
\lim_{E\rightarrow E_{\eta}(\kk+\gs_j)} Tr_{\sigma}(G) &\approx&
\lim_{\delta E\rightarrow 0} Tr_{\sigma}\left[ \frac{\PP_{\kk+\gs_j,\eta}}{\delta E} + \Delta_0 \frac{\PP_{\kk+\gs_j,\eta}}{\delta E} (1_{2\times 2} - \Delta_0 \frac{\PP_{\kk+\gs_j,\eta}}{\delta E})^{-1} \frac{\PP_{\kk+\gs_j,\eta}}{\delta E} \right]\nonumber\\
&= &\lim_{\delta E\rightarrow0} Tr_{\sigma}\left[\frac{\PP_{\kk+\gs_j, \eta}}{\delta E}\cdot \left(1_{2\times 2} + (\delta E 1_{2\times 2} -\Delta_0 \PP_{\kk+\gs_j, \eta})^{-1}\Delta_0 \PP_{\kk+\gs_j, \eta}\right)\right]\nonumber\\
& =&\lim_{\delta E\rightarrow 0}Tr_{\sigma}\left[\PP_{\kk+\gs_j, \eta}\cdot(\delta E 1_{2\times 2}-\Delta_0 \PP_{\kk+\gs_j, \eta}
)^{-1}\right]\nonumber\\
& =& \lim_{\delta E\rightarrow0}Tr_{\sigma} \left[\frac{(\sum^3_{i=0} d^{\eta}_{i,\kk+\gs_j}\sigma_i)[(\delta  E-\Delta_0d^{\eta}_{0,\kk+\gs_j})\sigma_0+\sum^3_{i=1} \Delta_0d^{\eta}_{i,\kk+\gs_j}\sigma_i]}{(\delta E-\Delta_0 d^{\eta}_{0,\kk+\gs_j})^2-\sum^3_{i=1} (\Delta_0 d^{\eta}_{i,\kk+\gs_j})^2} \right]\nonumber\\& =&\lim_{\delta E\rightarrow0}2\frac{d^{\eta}_{0,\kk+\gs_j}(\delta  E-\Delta_0d^{\eta}_{0,\kk+\gs_j})+\sum^3_{i=1} \Delta_0(d^{\eta}_{i,\kk+\gs_j})^2}{(\delta E-\Delta_0 d^{\eta}_{0,\kk+\gs_j})^2-\sum^3_{i=1} (\Delta_0 d^{\eta}_{i,\kk+\gs_j})^2}.
\label{Eq:8}
    \end{eqnarray}
From Eq.(\ref{Eq: Projection Pauli}), we know $(d^{\eta}_{0,\kk+\gs_j})^2-\sum^3_{i=1}(d^{\eta}_{i,\kk+\gs_j})^2=0$, so that 
\begin{eqnarray}
\lim_{E\rightarrow E_{\eta}(\kk+\gs_j)} Tr(G) &\approx &\lim_{\delta E\rightarrow 0}\frac{2 d^{\eta}_{0,\kk+\gs_j}\delta E}{-2d^{\eta}_{0,\kk+\gs_j}\Delta_0\delta E+(\delta E)^2}\approx -\frac{1}{\Delta_0}.
\label{Eq:BHZ trace0}
\end{eqnarray}
Thus, we have proven that the pole of $\mathbf{g}(\kk, E)$ is {\it not} the pole of $Tr(G)$ for any finite $\Delta_0$. Eq.(\ref{Eq:BHZ trace0}) is the generalization of Eq.(\ref{Eq:6 free electron delta}), and we get the same conclusion as that in 2D free electron gas with periodic $\delta$-function potential.  
Since we can always choose any complete set of basis wavefunction for the computation of the trace, we have $Tr(\PP_{\kk+\gs_j, \eta}) = 1$ and 
\begin{eqnarray} \label{Eq: BHZ trace 1}
Tr(G) &=& Tr\left(\mathbf{g}(\kk,E)\right)+\Delta_0 \sum_j Tr\left( \mathbf{g}^0(\kk+\gs_j, E) \frac{1}{1_{2\times 2}-\Delta_0\mathbf{g}(\kk,E)} \mathbf{g}^0(\kk+\gs_j, E)\right)\nonumber \\
&=&  \sum_{j, \eta} \frac{1}{E-E_{\eta}(\kk+\gs_j)} + \Delta_0 \sum_{j,\eta}  \frac{1}{(E -E_\eta(\kk+\gs_j))^2}\bra{\kk+\gs_j,\eta}\frac{1}{1_{2\times 2}-\Delta_0\mathbf{g}(\kk,E)}\ket{\kk+\gs_j,\eta},
\end{eqnarray}
in which we use the basis function $\ket{\kk+\gs_j,\eta}$ defined in Eq.(\ref{Eq:HamC_eigen_basis}).



As $E_\eta(\kk+\gs_j)$ is not a pole of $Tr(G)$, the poles of $Tr(G)$ can only come 
from $\bra{\kk+\gs_j,\eta}\frac{1}{1_{2\times 2}-\Delta_0\mathbf{g}(\kk,E)}\ket{\kk+\gs_j,\eta}$. Without losing generality, ${1_{2\times 2}-\Delta_0\mathbf{g}(\kk,E)}$ and $\frac{1}{1_{2\times 2}-\Delta_0\mathbf{g}(\kk,E)}$ can be expressed as 
\begin{eqnarray}
    {1_{2\times 2}-\Delta_0\mathbf{g}(\kk,E)}=\sum^3_{i=0}\mathbf{\hat{g}}^i(\kk,E)\sigma_i;\quad \frac{1}{1_{2\times 2}-\Delta_0\mathbf{g}(\kk,E)}=\frac{\mathbf{\hat{g}}^0(\kk,E)\sigma_0-\sum^3_{i=1}\mathbf{\hat{g}}^i(\kk,E)\sigma_i}{Det (1_{2\times 2}-\Delta_0\mathbf{g}(\kk,E))}. \label{Eq:inverse_poleequation}
\end{eqnarray}
Here we expand the 2-by-2 matrix of the expression ${1_{2\times 2}-\Delta_0\mathbf{g}(\kk,E)}$ into the basis of identity matrix and Pauli matrices with the expansion coefficients defined as $\mathbf{\hat{g}}^i(\kk,E)$. Therefore, the poles of $Tr(G)$ is given by 
\begin{eqnarray} \label{Eq: BHZ E 0}
Det(1_{2\times 2}-\Delta_0 \mathbf{g}(\kk,\mathcal{E}_n))=0,
\end{eqnarray}
from which we can solve the full energy spectrum $\mathcal{E}_n(\kk)$ of our model. 

Finally, we will generalize the derivation of Eq.(\ref{Eq:wavefunction 1d}) to solve the eigen-wavefunctions. For the eigen-energy $\mathcal{E}_n(\kk)$, we consider
\begin{eqnarray}
\lim_{E\rightarrow \mathcal{E}_n(\kk)}\bra{j,\beta} (E 1_{{2N_g\times 2N_g}}-H_{mC})^{-1}\ket{\phi_0,\alpha}=\lim_{E\rightarrow \mathcal{E}_n(\kk)}(E - \mathcal{E}_n(\kk) )^{-1}\langle j,\beta | n, \kk \rangle \langle n, \kk | \phi_0,\alpha \rangle,\label{Eq:wavefunction BHZ1}
\end{eqnarray}
where $\ket{n, \kk}$ is the eigenstate of $H_{mC}$ defined as $H_{mC}(\kk)\ket{n,\kk}=\mathcal{E}_n(\kk)\ket{n, \kk}$.

On the other hand, the Hermitian conjugation of Eq.(\ref{Eq:alphaj BHZ 2 Matrix}) gives
\begin{eqnarray}
  &&\bra{j,\beta}(E 1_{2N_g\times 2N_g}-H_{mC})^{-1} \ket{\phi_0,\alpha} =\frac{1}{\sqrt{N_g}} \sum^2_{\eta=1} (E 1_{2\times2}- H_{C}(\kk+\gs_j))^{-1}_{\beta,\eta} (1_{2\times 2} - \Delta_0 \mathbf{g}(\kk,E))^{-1}_{\eta, \alpha}. 
  \label{wavebhz0}
\end{eqnarray}

Comparing Eq.(\ref{Eq:wavefunction BHZ1}) with Eq.(\ref{wavebhz0}), we find
\begin{eqnarray}
&& \langle j,\beta | n, \kk \rangle \langle n, \kk | \phi_0,\alpha\rangle=lim_{E\rightarrow \mathcal{E}_n(\kk)}(E- \mathcal{E}_n(\kk))\bra{j,\beta} (E 1_{{2N_g\times 2N_g}}-H_{mC})^{-1}\ket{\phi_0,\alpha}\nonumber\\
&=&lim_{E\rightarrow \mathcal{E}_n(\kk)}(E- \mathcal{E}_n(\kk))
\frac{1}{\sqrt{N_g}} \sum^2_{\eta=1} (E 1_{2\times2}- H_{C}(\kk+\gs_j))^{-1}_{\beta,\eta} (1_{2\times 2} - \Delta_0 \mathbf{g}(\kk,E))^{-1}_{\eta, \alpha}
\label{Eq:wavefunction BHZ2}
\end{eqnarray}

In the limit $E\rightarrow \mathcal{E}_n(\kk)$, both $E- \mathcal{E}_n(\kk)$ and $Det(1_{2\times 2} - \Delta_0 \mathbf{g}(\kk,E))$ equal to zero, so we need to use the L'Hospital's rule to take the limit as
\begin{eqnarray}
&& \langle j,\beta | n, \kk \rangle \langle n, \kk | \phi_0,\alpha\rangle
= \lim_{E\rightarrow \mathcal{E}_n(\kk)}(E- \mathcal{E}_n(\kk))
\frac{1}{\sqrt{N_g} } \sum^2_{\eta=1} (E 1_{2\times2}- H_{C}(\kk+\gs_j))^{-1}_{\beta,\eta} \left[\frac{\mathbf{\hat{g}}^0(\kk,E)\sigma_0-\sum^3_{i=1}\mathbf{\hat{g}}^i(\kk,E)\sigma_i}{Det (1_{2\times 2}-\Delta_0\mathbf{g}(\kk,E))}\right]_{\eta, \alpha} \nonumber \\
&& = \frac{1}{\sqrt{N_g}} \left[\frac{1}{\partial_E Det (1_{2\times 2}-\Delta_0\mathbf{g}(\kk,E))}\right]_{E\rightarrow \mathcal{E}_n(\kk)} \left[(\mathcal{E}_n(\kk) 1_{2\times2}- H_{C}(\kk+\gs_j))^{-1}(\mathbf{\hat{g}}^0(\kk,\mathcal{E}_n(\kk))\sigma_0-\sum^3_{i=1}\mathbf{\hat{g}}^i(\kk,\mathcal{E}_n(\kk))\sigma_i)\right]_{\beta, \alpha} \nonumber \\
&& = \frac{1}{\sqrt{N_g} } \left[\frac{1}{\partial_E Det (1_{2\times 2}-\Delta_0\mathbf{g}(\kk,E))}\right]_{E\rightarrow \mathcal{E}_n(\kk)} \sum_{\eta} \frac{\left[\PP_{\kk+\gs_j,\eta}(\mathbf{\hat{g}}^0(\kk,\mathcal{E}_n(\kk))\sigma_0-\sum^3_{i=1}\mathbf{\hat{g}}^i(\kk,\mathcal{E}_n(\kk))\sigma_i)\right]_{\beta, \alpha} }{\mathcal{E}_n(\kk) -E_{\eta}(\kk+\gs_j)} \nonumber \\
&& = \frac{1}{\sqrt{N_g} } \left[\frac{1}{\partial_E Det (1_{2\times 2}-\Delta_0\mathbf{g}(\kk,E))}\right]_{E\rightarrow \mathcal{E}_n(\kk)} \mathcal{F}^{\beta,\alpha}(\kk,\gs_j), \label{Eq:wave function delta BHZ}
\end{eqnarray}
where $\mathcal{F}^{\beta,\alpha}(\kk,\gs_j) =\sum_{\eta} \frac{\left[\PP_{\kk+\gs_j,\eta}(\mathbf{\hat{g}}^0(\kk,\mathcal{E}_n(\kk))\sigma_0-\sum^3_{i=1}\mathbf{\hat{g}}^i(\kk,\mathcal{E}_n(\kk))\sigma_i)\right]_{\beta, \alpha} }{\mathcal{E}_n(\kk) -E_{\eta}(\kk+\gs_j)} $. 

Since $\langle n,\kk | \phi_0,\alpha\rangle$ at the left side of Eq.(\ref{Eq:wave function delta BHZ}) is independent of $j, \beta$ and the states $\ket{j,\beta}$ form a complete set of basis wavefunctions, we can take the norm for both sides and the sum over both $j$ and $\beta$ and obtain \begin{eqnarray}
&& \sum_{j,\beta}|\langle j,\beta | n, \kk \rangle |^2|\langle n, \kk | \phi_0,\alpha\rangle |^2= \frac{1}{N_g } \left[\frac{1}{\partial_E Det (1_{2\times 2}-\Delta_0\mathbf{g}(\kk,E))}\right]^2_{E\rightarrow \mathcal{E}_n(\kk)} \sum_{j,\beta} |\mathcal{F}^{\beta,\alpha}(\kk,\gs_j) |^2\\
&& |\langle n, \kk | \phi_0,\alpha\rangle| = \frac{1}{\sqrt{N_g} } \left|\frac{1}{\partial_E Det (1_{2\times 2}-\Delta_0\mathbf{g}(\kk,E))}\right|_{E\rightarrow \mathcal{E}_n(\kk)} \sqrt{\sum_{j,\beta} |\mathcal{F}^{\beta,\alpha}(\kk,\gs_j) |^2},\label{Eq:wave function delta BHZ2}
\end{eqnarray}
where we use $\sum_{j,\beta}|\langle j,\beta | n, \kk \rangle |^2=1$ as $\ket{j,\beta}$ is a complete set of basis. 

\begin{figure}
\centering
\includegraphics[width=0.8\textwidth]{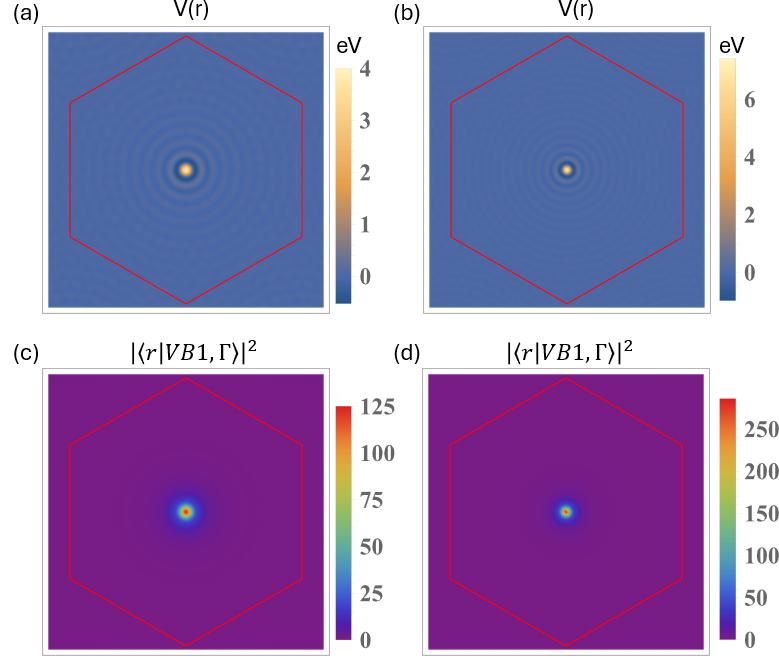}
\caption{\justifying \label{Fig: Wave_function_delta_BHZ} 
(a)-(b) V$(\rr)$ with $\Delta_0$=3.85 meV at $G_\Lambda$= 17$|b^M_1|$ and 23$|b^M_1|$. (c)-(d) $|\langle r|VB
1,\mathbf{\Gamma}\rangle|^2$ with $\Delta_0$=3.85 meV at $G_\Lambda$= 17$|b^M_1|$ and 23$|b^M_1|$. }
\end{figure}

With Eqs.(\ref{Eq:wave function delta BHZ}) and (\ref{Eq:wave function delta BHZ2}), we can derive the form of $\langle j,\beta | n, \kk \rangle$ up to a phase factor for any $j, \beta$ once $\langle n,\kk | \phi_0,\alpha\rangle\neq 0$. Below we will show that there must exist at least one $\alpha$ to satisfy $\langle n,\kk | \phi_0,\alpha\rangle\neq 0$. To see that we can assume $\langle n,\kk | \phi_0, \alpha \rangle = 0$ for both $\alpha = 1, 2$, and from Eq.(\ref{eq_SM:P0_projector}), we have 
\begin{eqnarray} \label{Eq: nozero delta}
    H_{mC}(\kk) \ket{n,\kk}=H_{0}(\kk) \ket{n,\kk}=\mathcal{E}_n(\kk) \ket{n,\kk}.
\end{eqnarray}  
Eq.(\ref{Eq: nozero delta}) indicates that  $\mathcal{E}_n(\kk)$ is the eigenvalue of $H_{0}(\kk)$. However, we have proven that the pole of $\mathbf{g}(\kk, E)$ is {\it not} the pole of $Tr(G)$ for any finite $\Delta_0$ in the above and thus we find a contradiction, which means there must be one $\alpha$ with non-zero $\langle n,\kk | \phi_0, \alpha \rangle$. By substituting Eq.(\ref{Eq:wave function delta BHZ2}) into Eq.(\ref{Eq:wave function delta BHZ}), we find 
\begin{eqnarray}
\langle j, \beta | n,\kk\rangle=\frac{\mathcal{F}^{\beta,\alpha}(\kk,\gs_j)}{\sqrt{\sum_{j,\beta} |\mathcal{F}^{\beta,\alpha}(\kk,\gs_j) |^2}}    
\end{eqnarray}
for any $j, \beta$, thus solving the form of the eigen-wavefunction $|n,\kk\rangle$. Here, we have chosen the phase factor of $\langle n,\kk | \phi_0, \alpha \rangle$ to be zero.
For $H_{mC}(\kk)$ in Eq.(\ref{Eq:HammC_delta_momentum}), we denote the wave function of the VB1 band at the $\mathbf{\Gamma}$ point as $\ket{\text{VB1}, \mathbf{\Gamma}}$ and the spatial distribution, namely $|\langle\mathbf{r}\mid VB1, \mathbf{\Gamma} \rangle|^2$. Under the $\delta$-function-like potential shown in Fig.\ref{Fig: Wave_function_delta_BHZ}(a) and (b) for different values of $G_\Lambda$ can be computed and shown in Fig.\ref{Fig: Wave_function_delta_BHZ}(c) and (d). We find that, similar to the case of a free electron gas in a $\delta$-function-like potential, the wave function $\ket{VB1, \mathbf{\Gamma}}$ is highly localized near the Wyckoff position 1a for a large $G_{\Lambda}$. 


Finally, the spectrum function $\mathcal{A}(\kk,E)$ of the moir\'e Chern-band model with a periodic $\delta$-function-like potential is defined as
\begin{eqnarray}
    \mathcal{A}(\kk,E)=-\Im(Tr(G(\kk,E+i \xi^+ )))/\pi
\end{eqnarray}
with the broadening parameter $\xi^+$ chosen to be 5 meV for our calculations. The peak of the spectrum function $\mathcal{A}(\kk,E)$ describes the eigen-energy spectrum of $H_{mC}$, which is shown in Fig.3(a) and (b) of the main text with the cutoff 
$G_\Lambda=23|b_1^M|$. 



\subsection{Mapping to Topological heavy fermion model}\label{sec:THFdelta}
\begin{figure}
\centering
\includegraphics[width=1\textwidth]{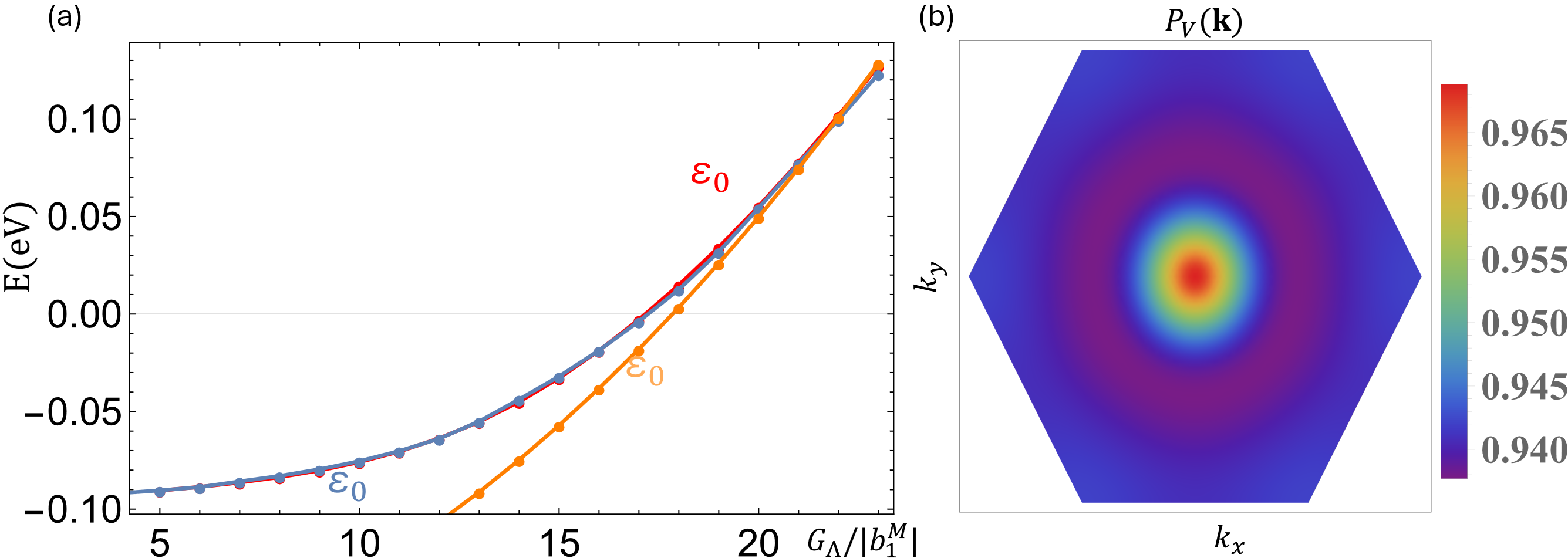}
\caption{\justifying \label{Fig：BHZ delta TF} 
(a) The solution $\mathcal{E}_0$ with $\Delta_0=4.05$ meV of Eq.(\ref{Eq: BHZ energy approximation 2}), Eq.(\ref{Eq: BHZ E 0}) and Eq.(\ref{Eq:BHZ delta e ana}) for different $G_\Lambda$ are labeled by red, blue and orange dots respectively. These three agree with each other well in the large $G_\Lambda$ . (b) The projection onto the valence band of the Chern-band model, $ P_V(\kk) $, for VB1 with $\Delta_0 = 3.85$ meV and $G_\Lambda = 23 |\bb^M_1|$.
}
\end{figure}

The numerical calculations in Sec.\ref{section:delta chern band} have demonstrated that the topological flat bands can also emerge in the Chen-band model with the periodic $\delta$-function-like potential. Intuitively, the strong periodic $\delta$-function-like potential induces a periodic lattice formed by the localized bound state in each moir\'e unit cell for the Chern-band model. Below we will numerically show this localized bound state originating from the valence band of the Chern-band model. To see that, we examine the wavefunctions of the trivial flat VB1 in Fig. 3(a) of the main text, which can be expanded as $\ket{VB1,\kk}=\sum_{\gs,\alpha} u_{\alpha,\kk,\gs}\ket{\alpha,\kk,\gs}$, where $u_{\alpha,\kk,\gs}$ is a coefficient. Recall that for $\ket{\alpha, \kk, \gs}$, $\alpha = 1, 2$ represents the basis functions of the Hamiltonian given in Eq.(\ref{Eq:HammC_delta_momentum}). We project the VB1 band into the $\alpha=2$ component, which corresponds to the valence band in the Chern-band model. The projection is defined as $ P_V(\kk) = \sum_{\gs} |u_{2,\kk,\gs}|^2 $. The distribution of $ P_V(\kk) $ over the entire  mBZ is shown in Fig.\ref{Fig：BHZ delta TF}(b), where it exceeds 0.9 throughout. This confirms the origin of the VB1 from the valence band of the Chern-band model. These bound states will form a trivial flat band. With increasing periodic $\delta$-function-like potential strength, the energy of this flat band is increased and can be higher than the conduction band bottom at $\mathbf{\Gamma}$, leading to band inversion due to the different  irreps of the states between the valence band top and conduction band bottom of the Chern-band model. In this section, we will first provide an approximate analytical solution for the bound states of the valence band due to the periodic $\delta$-function-like potential and then construct the THF model based on this analytical solution and the conduction band of the moir\'e Chern-band model. 

We start from the moir\'e Chern-band Hamiltonian $H_{mC}$ with periodic $\delta$-function-like potential and the Hamiltonian form in the momentum space is given in Eq.(\ref{Eq:HammC_delta_momentum}). For convenience of implementing the perturbation theory, we first rewrite $H_{mC}$ into the form
\begin{eqnarray}
&&H_{mC}(\kk)=\begin{pmatrix}
H_{+}(\kk)  & H_{+-}(\kk)  \\
H^\dagger_{+-}(\kk)  & H_{-}(\kk)
\end{pmatrix},
\end{eqnarray}
where 
\begin{eqnarray}\label{eq_SM:Hampm_1}
&&H_{\pm}(\kk)=\begin{pmatrix}
F_{\pm}(\kk +\gs_1)  & 0 & \dots& 0\\
0 & F_{\pm} (\kk + \gs_2) & \dots& 0\\
\vdots & \vdots & \ddots&  \vdots\\
0 & 0 & \dots & F_{\pm} (\kk + \gs_{N_g})
\end{pmatrix}  + N_g \Delta_0 P_0
\end{eqnarray}
with $F_{\pm}(\kk)=\epsilon(\kk) \pm \mathcal{M}(\kk)$, $\mathcal{M}(\kk)=M+B k^2$, $\epsilon(\kk)=D k^2$ and the projection operator $P_0$ is given in Eq.(\ref{Eq:projection0}). The first term in $H_{\pm}(\kk)$ is diagonal. The Hamiltonian matrix $H_{+-}(\kk)$ is given by  
\begin{eqnarray}
&&H_{+-}(\kk)=\begin{pmatrix}
A (k_+ + (\gs_1)_+)  & 0 & \dots& 0\\
0 & A (k_+ + (\gs_2)_+ ) & \dots& 0\\
\vdots & \vdots & \ddots&  \vdots\\
0 & 0 & \dots & A (k_+ + (\gs_{N_g})_+ )
\end{pmatrix}  
\end{eqnarray}
with $k_\pm=k_x\pm i k_y $ and $(\gs_j)_{\pm} = g_{j,x} \pm i g_{j,y}$ for different $j$ values. Below, we will set the moir\'e reciprocal lattice vector $\gs_1 = \bs{0}$. 

From the above numerical calculations of the wave function from Eq.(\ref{Eq:wave function delta BHZ2}), we know that the c-electrons should mainly come from the first diagonal matrix element $F_+(\kk+\gs_1)=F_+(\kk)$ in the block $H_+(\kk)$, while the f-electron is mainly from the block $H_-(\kk)$. Since c-electrons should weakly depend on moir\'e potential for $\Delta_0>0$, we make the approximation of neglecting 
the $N_g\Delta_0 P_0$ terms in $H_+(\kk)$ of Eq.(\ref{eq_SM:Hampm_1}) below, which corresponds to the moir\'e potential appearing only in layer B, but not in layer A. The Hamiltonian $H_{mC}(\kk)$ is rewritten into the form
\begin{eqnarray}\label{eq_SM:HmC_HFT_form1}
&&H_{mC}(\kk)=\begin{pmatrix}
H_{c}(\kk)  & \tilde{H}_{cf}(\kk)  \\
\tilde{H}^\dagger_{cf}(\kk)  & \tilde{H}_f(\kk)
\end{pmatrix}, \\
&& H_c(\kk)=F_+(\kk+\gs_1=\kk) + \Delta_0 \\
&& \tilde{H}_{cf}(\kk) = \begin{pmatrix} 0 & \dots & 0 & A k_+ & 0 & \dots & 0 \end{pmatrix},
\end{eqnarray}
where $H_{cf}$ is a $1 \times (2N_g-1)$ vector with the only non-zero component $A k_+$ appearing at the position $(1, N_g)$, and
\begin{eqnarray}
    && \tilde{H}_f(\kk)=\begin{pmatrix}
\tilde{H}_{+}(\kk)  & \tilde{H}_{+-}(\kk)  \\
\tilde{H}^\dagger_{+-}(\kk)  & H_{-}(\kk)
\end{pmatrix}
\end{eqnarray}
where 
\begin{eqnarray}
    &&\tilde{H}_{+}(\kk)=\begin{pmatrix}
F_+ (\kk + \gs_2)  & \dots& 0\\
\vdots & \ddots&  \vdots\\
0 & \dots & F_+ (\kk + \gs_{N_g})
\end{pmatrix}_{(N_g-1)\times(N_g-1)}, \\
&&\tilde{H}_{+-}(\kk)=\begin{pmatrix}
0 & A (k_+ + (\gs_2)_+ ) & \dots& 0\\
\vdots & \vdots & \ddots&  \vdots\\
0 & 0 & \dots & A (k_+ + (\gs_{N_g})_+ )
\end{pmatrix}_{(N_g-1)\times N_g}. \label{Eq:perturbation matrix1}
\end{eqnarray}






Now we want to apply the quasi-degenerate perturbation theory (also known as L\"owdin perturbation theory) \cite{winkler2003spin} to the Hamiltonian $\tilde{H}_f(\kk)$ around $\mathbf{\Gamma}$ above. We divide the whole basis wavefunctions into two subsets $\mathcal{A}$ and $\mathcal{B}$, which include the basis wavefunctions for $H_-(\kk)$ and $\tilde{H}_+(\kk)$, respectively. We hope to project out $\tilde{H}_+(\kk)$ in $\tilde{H}_f(\kk)$ and thus divide the whole Hamiltonian $\tilde{H}_f(\kk)$ into three parts,
\begin{eqnarray}
    && \tilde{H}_f=H^0+H^1+H^2
\end{eqnarray}
where
\begin{eqnarray}    
    && H^0=\begin{pmatrix}
 F_+ (\kk + \gs_2) & \cdots  &  0 & 0 & 0 & \cdots & 0   \\
 \vdots & \ddots & \vdots & \vdots & \vdots & \ddots & \vdots \\
 0 & \cdots & F_+ (\kk + \gs_{N_g}) & 0 & \cdots & \cdots & \cdots \\
0 & \cdots & 0  & F_- (\kk ) + \Delta_0 & 0  & \cdots & 0 \\
0 & \cdots & 0 & 0 & F_- (\kk + \gs_{2} ) + \Delta_0 & \cdots & 0 \\
\vdots & \ddots & \vdots & \vdots & \vdots & \ddots & \vdots \\
0 & \cdots & 0 & 0 & 0 & \cdots &  F_- (\kk + \gs_{N_g}) + \Delta_0  
\end{pmatrix}\nonumber \\
    && H^1 = \begin{pmatrix}
 \bs{0}_{(N_g-1)\times (N_g-1)}  &  \bs{0}_{(N_g-1)\times 1} & \bs{0}_{(N_g-1)\times 1} & \cdots & \bs{0}_{(N_g-1)\times 1}   \\
\bs{0}_{1\times (N_g-1)}  & 0 & \Delta_0  & \cdots & \Delta_0 \\
\bs{0}_{1\times (N_g-1)}  & \Delta_0 & 0 & \cdots & \Delta_0 \\
\vdots & \vdots & \vdots & \ddots & \vdots \\
\bs{0}_{1\times (N_g-1)} & \Delta_0 & \Delta_0 & \cdots & 0 
\end{pmatrix}\nonumber \\
&& H^2 = \begin{pmatrix}
\bs{0}_{(N_g-1)\times (N_g-1)}  & \tilde{H}_{+-}(\kk)  \\
\tilde{H}^\dagger_{+-}(\kk)  & \bs{0}_{N_g\times N_g}
\end{pmatrix}. 
\end{eqnarray}
Here $H^0$ is a diagonal matrix, $H^1$ includes all the off-diagonal components within  subsets $\mathcal{A}$ and $\mathcal{B}$ and $H^2$ includes all the off-diagonal components between  subsets $\mathcal{A}$ and $\mathcal{B}$. Our aim is to remove the off-diagonal components in $H^2$ perturbatively  by applying the L\"owdin perturbation \cite{winkler2003spin}. 
The effective Hamiltonian for f-electron up to the second-order perturbation terms reads
\begin{eqnarray}
&&H^{eff}_f(\kk)=\begin{pmatrix}
 F_- (\kk ) + \Delta_0 & \Delta_0 & \cdots & \Delta_0 \\
 \Delta_0 & F_- (\kk + \gs_{2} ) + \Delta_0 & \cdots & \Delta_0 \\
 \vdots & \vdots & \ddots & \vdots \\
\Delta_0 & 0 & \cdots &  F_- (\kk + \gs_{N_g}) + \Delta_0  
\end{pmatrix}+H^{eff(2)}_f(\kk)\\
&&[H^{eff(2)}_f(\kk)]_{ll’}=\frac{1}{2}\sum_m [\tilde{H}^\dagger_{+-}(\kk)]_{lm} [\tilde{H}_{+-}(\kk)]_{ml'}\left(\frac{1}{F_- (\kk + \gs_{l} )-F_+ (\kk + \gs_{m} )}+\frac{1}{F_- (\kk + \gs_{l'} )-F_+ (\kk + \gs_{m} )}\right),\nonumber \label{Eq: Heff(2)f}
\\ 
\end{eqnarray}
where $\tilde{H}_{+-}(\kk)_{ml}=\delta_{m+1,l}A (k_+ + (\gs_{l})_+ )$ from Eq.(\ref{Eq:perturbation matrix1}), so that $H^{eff(2)}_f(\kk)$ only has diagonal terms.


Based on the above discussion, we  find the effective Hamiltonian 
\begin{eqnarray}\label{eq_SM:Heff_fk1}
&&H^{eff}_f(\kk)=\tilde{H}^0_f(\kk)+ \Delta_0\begin{pmatrix}
1  & 1& \dots& 1\\
1 & 1 & \dots& 1\\
\vdots & \vdots & \ddots&  \vdots\\
1 & 1 & \dots &1
\end{pmatrix},\\
&&\tilde{H}^0_{f,ij}(\kk)=\delta_{i,j}E_v(\kk+\gs_j),
\end{eqnarray}
where $i,j = 1, \cdots, N_g$ and  $E_v(\kk+\gs_j)$ are obtained from the second order perturbation as
\begin{eqnarray}\label{Eq_SM:EvApprox0}
&&E_v(\kk+\gs_j)=\begin{cases}
    (D-B)|\kk+\gs_j|^2-M-\frac{A^2|\kk+\gs_j|^2}{2B|\kk+\gs_j|^2+2M-\Delta_0}. 
    & j\neq 1\\
   (D-B)\kk^2-M.& j=1
\end{cases}
\label{Eq:HBHZ valance band}
\end{eqnarray}
The term $\frac{A^2|\kk+\gs_j|^2}{2B|\kk+\gs_j|^2+2M-\Delta_0}$ is from the second order perturbation in Eq.(\ref{Eq: Heff(2)f}). Based on the above discussion, similar to Eq.(\ref{eq_SM:boundstate_energy1}), the eigen-energy $\mathcal{E}_0$ of the bound states at $\kk=\bs{0}$ can be solved from the equation
\begin{eqnarray}
\label{Eq: BHZ energy approximation}
    \sum_{|\gs_j|\leq G_\Lambda} \frac{1}{\mathcal{E}_0-E_v(\gs_j)}=\frac{1}{\Delta_0}.
\end{eqnarray}
where $G_\Lambda$ is the momentum cut-off. 

We choose the parameters to make the term $B|\kk + \gs_j|^2$ increasing rapidly due to the small effective mass, so that we can introduce another critical value $G_0$, for which $B |\gs_j|^2\gg |M|$ and $|\Delta_0|$ when $|\gs_j|>G_0$. Based on this approximation, the $E_v$ in Eq.(\ref{Eq_SM:EvApprox0}) can be simplified as
\begin{eqnarray}
&&E_v(\kk+\gs_j)\approx (D-B)|\kk+\gs_j|^2-M-\frac{A^2}{2B}\doteq\tilde{E}_v(\kk+\gs_j),\end{eqnarray} for $|\gs_j|>G_0$. Practically, with $M=0.1$ eV and $B=0.706$ $nm^2 \cdot$ eV, we can choose $G_0=2|\bb^M_1|$ with $|\bb^M_1|=0.63$ $nm^{-1}$, so that the dimensionless quantity $\frac{BG_0^2}{M}\approx 12\gg 1$. Under this approximation, the energy $\mathcal{E}_0$ of the bound state is determined by 
\begin{eqnarray}  \label{Eq: BHZ energy approximation 2}
&&\sum_{|\gs_j|< G_0}\frac{1}{\mathcal{E}_0-E_v(\gs_j)}+\frac{1}{\Omega_B}\int_{G_0}^{G_\Lambda} \frac{2\pi g}{\mathcal{E}_0-\tilde{E}_v(\gs_j)} dg =\frac{1} {\Delta_0}\nonumber\\
&&\sum_{|\gs_j|< G_0}\frac{1}{\mathcal{E}_0-E_v(\gs_j)}+\frac{\pi}{\Omega_B(B-D)}\ln \left(\frac{\mathcal{E}_0-\tilde{E}_v(G_\Lambda)}{\mathcal{E}_0-\tilde{E}_v(G_0)}\right)=\frac{1}{\Delta_0},
\end{eqnarray} 
where $g=|\gs_j|$ and $\Omega_B$ are the area of mBZ. The energy $\mathcal{E}_0$ of the bound state can be numerically solved from Eq.(\ref{Eq: BHZ energy approximation 2}) and compared with the solution from Eq.(\ref{Eq: BHZ E 0}) for different cut-off $G_\Lambda$ in Fig.\ref{Fig：BHZ delta TF}(a), which shows a good agreement for a larger $G_\lambda$. 
In the limit $G_\Lambda \rightarrow \infty$, the second term will diverge and thus be much larger than the first term in the left-hand side of Eq.(\ref{Eq: BHZ energy approximation 2}), so we can drop the first term and set $G_0$ as 0 to get a simplified analytical expression. We further approximate the $E_v(\gs_j)$ by $\tilde{E}_v(\gs_j)$, so the analytical expression for $\mathcal{E}_0$ can be obtained as
\begin{eqnarray}\label{Eq:BHZ delta e ana}
    \frac{\pi}{\Omega_B(B-D)}\ln \left(\frac{\mathcal{E}_0-\tilde{E}_v(G_\Lambda)}{\mathcal{E}_0-\tilde{E}_v(0)}\right)=\frac{1}{\Delta_0};\quad
    \mathcal{E}_0 = \tilde{E}_v(0)+\frac{ \tilde{E}_v(0)-\tilde{E}_v(G_\Lambda)}{-1+e^\frac{\Omega_B(B-D)}{\Delta_0\pi}}.
\end{eqnarray}

Now let's denote this bound state as $\ket{\tilde{f}}$ at $\kk=\bs{0}$ with eigen-energy $\mathcal{E}_0$.  Based on Eq.(\ref{Eq:wavefunction 1d3}), the eigen-wavefunction of the bound state $\ket{\tilde{f}}$ for the effective Hamiltonian $H^{eff}_f(\kk)$ in Eq.(\ref{eq_SM:Heff_fk1}) can be found as 
\begin{eqnarray}\label{Eq:wavefunction BHZ App0}
 \langle j | \tilde{f} \rangle=N_v\frac{1}{\mathcal{E}_0-E_v(\gs_j)},
\end{eqnarray}
similar to Eq.(\ref{Eq:wavefunction 1d2}), where $N_v$=$\sum_j\frac{1}{\mathcal{E}_0-E_v(\gs_j)}$ is a normalization factor and $\ket{j}$ is the unit-vector of the non-zero $\gs_j$ component of $H^{eff}_f(\kk)$. 
To project $H_{mC}$ in Eq.(\ref{eq_SM:HmC_HFT_form1}) to the THF model, we choose the basis wavefunction for the c-electron the same as that for $H_c(\kk)$, namely 
\begin{eqnarray} \label{Eq: wave function C}
    |c\rangle = (1, \bs{0}_{1 \times (2 N_g-1)})^T, 
\end{eqnarray}
and 
the basis wavefunction for f-electrons should be given by 
\begin{eqnarray}
    |f\rangle = (\bs{0}_{1 \times N_g}, \langle 1 | \tilde{f} \rangle, \langle 2 | \tilde{f} \rangle, \cdots, \langle N_g | \tilde{f} \rangle  )^T. 
\end{eqnarray}
On the basis of $|c\rangle$ and $|f\rangle$, we find the Hamiltonian $H_{mC}$ in Eq.(\ref{eq_SM:HmC_HFT_form1}) can be transformed into
\begin{eqnarray}\label{Eq:HAM_HF delta}
H^0_{HF}=\sum_{|\kk|<\Lambda_c}(E_c+\alpha k^2) c^{\dagger}_{\kk}c_{\kk}+\sum_{\RR}E_f f^{\dagger}_{\RR}f_{\RR}+\frac{\beta}{\sqrt{N}}\sum_{|\kk|<\Lambda_c,\RR}[e^{-i \kk \cdot \RR} (k_x +i k_y)c^{\dagger}_{\kk}f_{\RR}+h.c.],
\end{eqnarray}
where $E_c = M +\Delta_0$,  $E_f=\mathcal{E}_0$, $\alpha = B + D$ and 
\begin{eqnarray}\label{Eq:wavefunction BHZ App}
\beta = \left.\frac{\partial\bra{c}H_{mC}(\kk)\ket{f}}{\partial k_x}\right|_{\kk\rightarrow \bs{0}}=N_v\frac{\beta}{\mathcal{E}_0-E_v(0)}=N_v\frac{\beta}{\mathcal{E}_0+M}.
\end{eqnarray}

In the second equation of Eq.(\ref{Eq:wavefunction BHZ App}), we substitute Eq.(\ref{Eq:wavefunction BHZ App0}) and Eq.(\ref{Eq: wave function C}) into $\ket{f}$ and $\bra{c}$. When $E_f>E_c$, the band inversion happens and VB1 becomes topological. Using the parameters for our numerical calculations, we extract all the parameters of the THF model as $E_c=0.104$ eV, $E_f=0.127$ eV,$\alpha=1.238$ eV$\cdot nm^2$ and $\beta=0.120$ eV$\cdot nm$. In Fig.3(b) of the main text, we find an excellent consistency for the energy dispersion between the THF model (\ref{Eq:HAM_HF delta}) and the numerical solution solved from Eq.(\ref{Eq: BHZ E 0}).

\subsection{Other approximations for the $\delta$-function}
In this section, we would emphasize that the ideal topological flat band physics discussed in the above sections does not rely on the approximated form of the $\delta$-function-like Moir\'e potential. To substantiate this point, we consider alternative functional forms that approximate a $\delta$-function. Specifically, we consider (1) the Gaussian form, $f_G (\rr)=\frac{1}{\sqrt{2\pi \eta}} e^{-r^2/2\eta  }$, and (2) the bivariate Cauchy distribution  form \cite{molenberghs1997non}, $f_c(\rr)=\frac{\eta r}{(2\pi(\eta^2+r^2 )^(3⁄2) )}$, where $r=|\rr|$. In the momenta space, these two potentials are written as $f_G (\kk)=e^{-\eta/2 k^2 }$  and $f_c (\kk)=e^{-\eta k}$  with $k=|\kk|$. In the limit where $\eta$ goes to zero, both forms become $\delta$-function. The corresponding potential Hamiltonians under these two $\delta$-function-like functions are given by 
\begin{eqnarray}
    \Hat{H}_{m,G}  =\Delta_0 \sum_{\kk\alpha}\sum_{\gs\gs'}e^{-\eta/2 |\gs-\gs'|^2 }  a_{\alpha \kk \gs}^\dagger a_{\alpha \kk \gs'}; \quad \hat{H}_{m,C}  =\Delta_0 \sum_{\kk\alpha}\sum_{\gs\gs'}e^{-\eta|\gs-\gs'|} a_{\alpha \kk \gs}^\dagger a_{\alpha \kk \gs'}. \label{Eq.other two delta}    
\end{eqnarray}
                       
\begin{figure}
    \centering
    \includegraphics[width=0.7\linewidth]{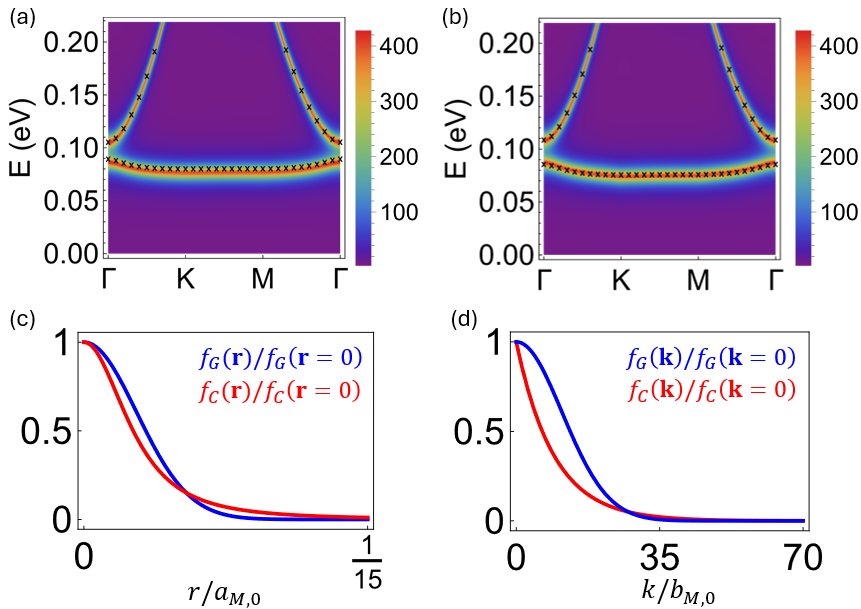}
    \caption{\justifying The crosses in (a) and (b) indicate the eigenenergies under a Gaussian $\delta$-function-like potential with $a_{M,0}/(\sqrt{2}\eta)=57.3$ and a bivariate Cauchy $\delta$-function-like potential with $a_{M,0}/\eta=66.23$,  where $a_{M,0}$ is the moiré lattice constant. The moiré potential strength is chosen as $\Delta_0$=6.3 meV for (a) and 10 meV for (b). The background colors in (a) and (b) show the spectral function $\mathcal{A}(\kk,E)$ gotten from our Green's function approach with equal moiré potential strength. The cutoff $G_\Lambda$ is chosen as $G_\Lambda a_{M,0}=64.5$ for (a) and $G_\Lambda a_{M,0}=33.2$ for (b).     (c) shows $f_C (\rr)$ and $f_G (\rr)$ forms in the real space, which are used for the calculation of band dispersions in (a) and (b). (d) shows the $f_C (\kk)$ and $f_G (\kk)$ forms in the momentum space.}    
    \label{fig:R2}
\end{figure}

We choose a small but finite $\eta$ in the above two expressions and diagonalize the Hamiltonians with a large enough momentum cut-off $G_\Lambda^0$, and the summations over $\gs$ and $\gs'$ in Eq.(\ref{Eq.other two delta}) are restricted within $G_\Lambda^0$ to obtain the energy spectrum, which are depicted as the black crosses in Fig.\ref{fig:R2}(a) and (b) for Gaussian and bivariate Cauchy $\delta$-function-like functions, respectively. The value of $G_\Lambda^0$ is chosen as $70|b^M_{1}|$. From the momentum-space distributions of these two potentials shown in Fig.\ref{fig:R2}(d), we find $f_c (G_\Lambda^0)$ and $f_G (G_\Lambda^0 )$ are close to zero for $G_\Lambda^0=70|b^M_{1}|$, which is the cutoff chosen for the calculations, so that the momenta outside the $G_\Lambda^0$ have little influence on the low-energy bands. In addition, the corresponding real-space distributions of these two potentials are shown in Fig.\ref{fig:R2}(c). In the real space, both potentials are peaked at $\rr=0$, and decay quickly with increasing $\rr$, thus approximating $\delta$-function. By comparing the black crosses in Fig.\ref{fig:R2} (a) and (b) with the red color lines, which show the peak of spectral function $\mathcal{A}(\kk,E)$ calculated from the approximate potential in Eq.(7) of main text, one can see different approximate functions give qualitatively similar band structures for both the lowest conduction band and the flat band of VB1. Thus, combined with the discussion about the first shell approximation to moir\'e potential in Sec.\ref{sec:first shell}, we conclude that our strategy for realizing topological flat bands is not sensitive to the approximate forms of the moir\'e potential. 

\section{Other candidates for Moir\'e heterostructures with type-II band alignment  }
\label{Sec_SM:2Dmaterial}


\subsection{ 2D materials heterostructures with the type-II band alignment}


In this section, we aim in searching for 2D material heterostructures with type-II band alignment and a narrow band gap. Our selection for two 2D materials, denoted as Material 1 and Material 2, to form a heterostructure is based on the following criteria:

(1) Two 2D materials should form type-II band alignment. Therefore, we extract the material information, including the conduction and valence band edge for individual materials with respect to vacuum and the lattice constants. These properties are obtained through DFT calculations using the Vienna Ab-initio Simulation Package (VASP) \cite{kresse1996efficiency}, with the HSE06 hybrid exchange-correlation functional for accurate determination of band edges \cite{heyd2003hybrid}.

(2) We search for materials with the conduction band minimum (CBM) of Material 1 and valence band maximum (VBM) of Material 2 at $\mathbf{\Gamma}$ in the atomic BZ.  We only focus on the materials with $\mathbf{\Gamma}$ valley, in which the moir\'e potential can be generated via interfacing Material 1 or 2 with another insulating layer to form moir\'e heterobilayer. 

(3) The irreps for CBM in Material 1 and VBM in Material 2 should be different. Particularly, we hope the angular momenta of the states at $\mathbf{\Gamma}$ between the conduction band from Material 1 and the valence band from Material 2 should be different by $\pm 1$, so that the linear-$k$ coupling will be the dominant term between two materials, allowing for the occurrence of the Chern-band model. This criterion gives a strong constraint on the possible material candidates. In the example of the group $p\bar{3}m1$ discussed below, we would expect the irreps of the CBM and VBM to belong to one 1D irrep $\mathbf{\Gamma}^{\pm}_{1,2}$ and one 2D irrep $\mathbf{\Gamma}^{\pm}_{3}$, respectively.

(4) The band gap between CBM in Material 1 and VBM in Material 2 should be smaller than 0.5 eV. As we discussed in main text, in order to achieve the THF limit, we require the moir\'e potential $\Delta_0$ to be larger than the atomic band gap $M$ in the moir\'e Chern-band model. The generated moir\'e potential has a limitation in its energy scale, normally around a few hundreds of meV. Thus, we require the atomic band gap of the whole heterostructure to be smaller than 0.5 eV. It should be noted that our search here is based on the DFT calculations without SOC due to time cost consideration. When including SOC, the type-II band gap can be further reduced, as shown in the example of Tl$_2$Se$_2$/Zn$_2$Te$_2$ heterostructure below. 

(5) The lattice mismatch between two materials should be smaller than $15\%$. A small lattice constant difference is expected to improve the sample quality in experiments. But this is \textit{not} an essential condition, and materials with a large lattice constant difference can also be possible. 

Based on these criteria, a class of possible 2D material hetero-structures for the groups $p\bar{3}m1$ is shown in the Tab.\ref{tab:type2 alignment pbar3m1}, in which we show Material 1 belonging to the material family of Tl$_2$(Se,Te,S)$_2$ with its conduction band at $\mathbf{\Gamma}$ belonging to the irrep $\mathbf{\Gamma}^{-}_2$ without SOC. Material 2 is selected according to the criteria discussed above. The irrep of Material 2 belongs to $\mathbf{\Gamma}^{\pm}_3$. The gaps listed in the Tab.\ref{tab:type2 alignment pbar3m1} describe the energy difference between the CBM from Material 1 and the VBM from Material 2 for individual compounds without SOC. When SOC is included, we anticipate that this gap will further be reduced since the SOC may split the valence band in Material 2 and thus increase the energy of the VBM. We emphasize that the energy gap in the real heterostructure will be different from the gap listed here since a charge transfer may exist between two layers and needs to be calculated directly from the DFT calculations.

\subsection{Chern-band model from 2D material heterostructures}
Below we will take the material combination of  Tl$_2$Se$_2$ and Zn$_2$Te$_2$ monolayers as an example to show the low-energy physics in the material list in Tab.\ref{tab:type2 alignment pbar3m1} that can be described by the Chern-band model in Eq.(\ref{Eq:Ham_Chernband}) when these two 2D materials form a heterostructure. Both Tl$_2$Se$_2$ and Zn$_2$Te$_2$ belong to the space group $P\bar{3}m1$ and  the corresponding wavevector group at $\mathbf{\Gamma}$ belongs to $D_{3d}$ group with the corresponding character table shown in Tab.\ref{Tab:charter table D3d c3v}(a). The conduction band of Tl$_2$Se$_2$ has a lower energy than that of Zn$_2$Te$_2$, while the valence band of Zn$_2$Te$_2$ has a higher energy than that of Tl$_2$Se$_2$. Thus, we would expect if these two materials are stacked 
  together  to form a heterostructure, the conduction band should come from Tl$_2$Se$_2$ while the valence band should originate from Zn$_2$Te$_2$.  Comparing these two 2D materials, we find the CBM of Tl$_2$Se$_2$ is around $200$ meV above the VBM of Zn$_2$Te$_2$ (without SOC), so we would expect type-II band alignment with a relatively small band gap for their heterostructure. Furthermore, the conduction band in Tl$_2$Se$_2$ belongs to the 1D irrep $\mathbf{\Gamma}^-_2$, which corresponds to the orbital angular momentum $L=0$, while the valence band in Zn$_2$Te$_2$ belongs to the 2D irrep $\mathbf{\Gamma}^-_3$, which corresponds to the orbital angular momentum $L = \pm 1$.

\begin{table}[htbp]
\centering
\renewcommand{\arraystretch}{1.2}
\resizebox{\textwidth}{!}{%
\begin{tabular}{cccccccc}
\toprule\toprule
\textbf{Material 1} & \textbf{Lattice const. (\AA)} & \textbf{CB edge (eV)} & \textbf{Material 2} & \textbf{Lattice const. (\AA)} & \textbf{VB edge (eV)} & \(\Delta\)\textbf{gap} & \textbf{Lattice mismatch (\%)} \\
\midrule
Tl$_2$Se$_2$& 4.229 & -5.433(\(\mathbf{\Gamma}_2^-\))  & Zn$_2$Te$_2$      & 4.33  & -5.642(\(\mathbf{\Gamma}_3^-\))  & 0.209 & 2.39 \\
Tl$_2$S$_2$ & 4.061 & -5.638(\(\mathbf{\Gamma}_2^-\))  & H$_2$Si$_2$      & 3.895 & -5.658(\(\mathbf{\Gamma}_3^+\))  & 0.02  & 4.09 \\
Tl$_2$Se$_2$& 4.229 & -5.433(\(\mathbf{\Gamma}_2^-\))  & Au$_2$I$_2$      & 4.444 & -5.801(\(\mathbf{\Gamma}_3^-\))  & 0.368 & 5.08 \\
Tl$_2$S$_2$ & 4.061 & -5.638(\(\mathbf{\Gamma}_2^-\))  & Zn$_2$Te$_2$      & 4.33  & -5.642(\(\mathbf{\Gamma}_3^-\))  & 0.004 & 6.62 \\
Tl$_2$Se$_2$& 4.229 & -5.433(\(\mathbf{\Gamma}_2^-\))  & H$_2$Si$_2$      & 3.895 & -5.658(\(\mathbf{\Gamma}_3^+\))  & 0.225 &7.9 \\
Tl$_2$S$_2$ & 4.061 & -5.638(\(\mathbf{\Gamma}_2^-\))  & Au$_2$I$_2$      & 4.444 & -5.801(\(\mathbf{\Gamma}_3^-\))  & 0.163 & 9.43 \\
Tl$_2$Te$_2$& 4.52  & -4.965(\(\mathbf{\Gamma}_2^-\))  & Ge$_2$H$_2$      & 4.077 & -5.337(\(\mathbf{\Gamma}_3^+\))  & 0.372 & 9.8 \\
Tl$_2$Se$_2$& 4.229 & -5.433(\(\mathbf{\Gamma}_2^-\))  & Cd$_2$Te$_2$      & 4.67  & -5.918(\(\mathbf{\Gamma}_3^-\))  & 0.485 & 10.43 \\
Tl$_2$S$_2$ & 4.061 & -5.638(\(\mathbf{\Gamma}_2^-\))  & Ag$_2$I$_2$      & 4.577 & -6.116(\(\mathbf{\Gamma}_3^-\))  & 0.478 & 12.71 \\
Tl$_2$S$_2$ & 4.061 & -5.638(\(\mathbf{\Gamma}_2^-\))  & Cd$_2$Te$_2$      & 4.67  & -5.918(\(\mathbf{\Gamma}_3^-\))  & 0.28  & 15 \\
\bottomrule\bottomrule
\end{tabular}
}
\caption{\justifying Candidate materials with the type-II band alignment in the space group $p\bar{3}m1$. The first to the third columns show the material providing the conduction band (CB), the corresponding lattice constant and the CB edge. The CB edge is at $\mathbf{\Gamma}$ point and the $\mathbf{\Gamma}_i$ in the  parentheses is the irrep of CB for the $D_{3d}$ point group. The fourth to the sixth columns show the material providing the valence band (VB), the corresponding lattice constant and the VB edge. The VB edge is at $\mathbf{\Gamma}$ point and the $\mathbf{\Gamma}_i$ in the  parentheses is the irrep of VB for the $D_{3d}$ group. The last two columns show the energy difference between CBM of material 1 and VBM of material 2 and the lattice mismatch between them.} 
\label{tab:type2 alignment pbar3m1}
\end{table}

\begin{table}[h]

\centering
\begin{subtable}{0.5\textwidth}
\centering
\renewcommand{\arraystretch}{1.2}
\begin{tabular}{|c|c|c|c|c|c|c|c|c|c|c|c|c|}
\hline
$D_{3d}$ & $\mathbf{\Gamma}_1^+$ & $\mathbf{\Gamma}_2^+$ & $\mathbf{\Gamma}_3^+$ & $\mathbf{\Gamma}_1^-$ & $\mathbf{\Gamma}_2^-$ & $\mathbf{\Gamma}_3^-$ & $\bar{\mathbf{\Gamma}}_4^+$ & $\bar{\mathbf{\Gamma}}_5^+$ & $\bar{\mathbf{\Gamma}}_6^+$ & $\bar{\mathbf{\Gamma}}_4^-$ & $\bar{\mathbf{\Gamma}}_5^-$ & $\bar{\mathbf{\Gamma}}_6^-$ \\
\hline
$E$& 1 & 1 & 2 & 1 & 1 & 2 & 2 & 1 & 1 & 2 & 1 & 1 \\
$2C_3$      & 1 & 1 & -1 & 1 & 1 & -1 & 1 & -1 & -1 & 1 & -1 & -1 \\
$3C_2$& 1 & -1 & 0 & 1 & -1 & 0 & 0 & $i$ & $-i$ & 0 & $i$ & $-i$ \\
$I$ & 1 & 1 & 2 & -1 & -1 & -2 & 2 & 1 & 1 & -2 & -1 & -1 \\
$2IC_3$& 1 & 1 & -1 & -1 & -1 & 1 & 1 & -1 & -1 & -1 & 1 & 1 \\
$3IC_2$& 1 & -1 & 0 & -1 & 1 & 0 & 0 & $i$ & $-i$ & 0 & $-i$ & $i$ \\
\hline
\end{tabular}
\caption{\centering }
\end{subtable} \hfill
\begin{subtable}{0.49\textwidth}
\centering

\renewcommand{\arraystretch}{1.2}
\begin{tabular}{|c|c|c|c|c|c|c|}
\hline
$C_{3v}$ & $\mathbf{\Gamma}_1$ & $\mathbf{\Gamma}_2$ & $\mathbf{\Gamma}_3$ & $\bar{\mathbf{\Gamma}}_4$ & $\bar{\mathbf{\Gamma}}_5$ & $\bar{\mathbf{\Gamma}}_6$ \\
\hline
$E$ & 1 & 1 & 2 & 2 & 1 & 1 \\
$2C_3$ & 1 & 1 & $-1$ & 1 & $-1$ & $-1$ \\
$3IC_2$ & 1 & $-1$ & 0 & 0 & $i$ & $-i$ \\
\hline
\end{tabular}

\caption{\centering   }

\end{subtable}
\hfill
\caption { \justifying Character tables for the point groups $D_{3d}$ and $C_{3v}$ in (a) and (b), respectively. The irreps with a bar denote double (spinful) representations, while those without a bar correspond to single (spinless) representations.}
\label{Tab:charter table D3d c3v}
\end{table}



Spin-orbit coupling (SOC) has not been included  in Tab.\ref{tab:type2 alignment pbar3m1} (a) . Since SOC in Zn$_2$Te$_2$ is expected to split the bands with orbital angular momentum $L = \pm 1$ into the $J_{z}=\pm 1/2$ bands and $J_{z} = \pm 3/2$ bands where $J_{z}$ is the z-directional total angular momentum that includes both orbital angular momentum  and spin. For the Tl$_2$Se$_2$/Zn$_2$Te$_2$ heterostructure, the inversion symmetry in the $D_{3d}$ group is broken and the remaining symmetry operations form an $C_{3v}$ group. According to the compatibility relation, the $\mathbf{\Gamma}^-_2$ and $\mathbf{\Gamma}^-_3$ irreps of $D_{3d}$ group for the conduction of Tl$_2$Se$_2$ and valence band of Zn$_2$Te$_2$ correspond to the $\mathbf{\Gamma}_1$ and $\mathbf{\Gamma}_3$ irreps in $C_{3v}$ group, respectively, whose character table is shown in Tab.\ref{Tab:charter table D3d c3v} (b). Thus, we will below construct the effective Hamiltonian of the Tl$_2$Se$_2$/Zn$_2$Te$_2$ heterostructure based on the irreps of the $C_{3v}$ group. Without spin, we would expect a spinless three-band model with the 1D $\mathbf{\Gamma}_1$ irrep for the conduction band and the 2D $\mathbf{\Gamma}_3$ irrep for the valence band. Including spin and SOC will lead to a six-band model, which has a general form 
\begin{eqnarray}
    H_{3b,hetero}(\kk)=\sum^{3}_{i=0}\sum^{8}_{j=0}F^{ij}(\kk)\sigma_i\lambda_j,
\end{eqnarray}
where the expansion coefficients $F^{ij}(\kk)$ are the polynomials of $\kk$, $\sigma_0$ and $\sigma_{1,2,3}$ are the $2\times 2$ identity and Pauli matrices in the spin basis, while $\lambda_0$ and $\lambda_i (i=1,...,8)$ are the $3\times 3$ identity and the Gell-Mann matrices \cite{zee2016group} in the orbital basis with the $\mathbf{\Gamma}_1$ and $\mathbf{\Gamma}_3$ irreps.  
The Gell-Mann matrices are defined by
\begin{eqnarray}
&&\nonumber \lambda_1 = \begin{pmatrix}
0 & 1 & 0 \\
1 & 0 & 0 \\
0 & 0 & 0
\end{pmatrix}, \quad
\lambda_2 = \begin{pmatrix}
0 & -i & 0 \\
i & 0 & 0 \\
0 & 0 & 0
\end{pmatrix}, \quad
\lambda_3 = \begin{pmatrix}
1 & 0 & 0 \\
0 & -1 & 0 \\
0 & 0 & 0
\end{pmatrix},\\ \nonumber
&&\lambda_4 = \begin{pmatrix}
0 & 0 & 1 \\
0 & 0 & 0 \\
1 & 0 & 0
\end{pmatrix}, \quad
\lambda_5 = \begin{pmatrix}
0 & 0 & -i \\
0 & 0 & 0 \\
i & 0 & 0
\end{pmatrix}, \quad
\lambda_6 = \begin{pmatrix}
0 & 0 & 0 \\
0 & 0 & 1 \\
0 & 1 & 0
\end{pmatrix},\\ 
&&\lambda_7 = \begin{pmatrix}
0 & 0 & 0 \\
0 & 0 & -i \\
0 & i & 0
\end{pmatrix}, \quad
\lambda_8 = \frac{1}{\sqrt{3}} \begin{pmatrix}
1 & 0 & 0 \\
0 & 1 & 0 \\
0 & 0 & -2
\end{pmatrix},
\lambda_0 = \begin{pmatrix}
1 & 0 & 0 \\
0 & 1 & 0 \\
0 & 0 &1
\end{pmatrix}
\end{eqnarray}
on the orbital basis $\{|L=0\rangle, |L=1\rangle, |L=-1\rangle\}$, where the $L =  0$ state belongs the $\mathbf{\Gamma}_1$ irrep and is from the Tl$_2$Se$_2$ layer while the $L =  \pm 1 $ states belong to $\mathbf{\Gamma}_3$ state and are from the Zn$_2$Te$_2$ layer. The $C_{3v}$ group includes two generators, the three-fold rotation $C_{3z}$ along the z axis and the mirror-y symmetry $M_{y}$. Under the spin and orbital basis, the $\sigma$ and $\lambda$ matrices transform as
\begin{eqnarray}
  &&D(C^s_{3z}) \sigma_i D^{-1}(C^s_{3z})=\sum_j\mathcal{D}^{ji}(C^s_{3z}) \sigma_j;D(C^o_{3z}) \lambda_i D^{-1}(C^o_{3z})=\sum_j\mathcal{D}^{ji}(C^o_{3z}) \lambda_j,\\
 && D(M^s_{y}) \sigma_i D^{-1}(M^s_{y})=\sum_j\mathcal{D}^{ji}(M^s_{y}) \sigma_j; D(M^o_{y}) \lambda_i D^{-1}(M^o_{y})=\sum_j\mathcal{D}^{ji}(M^o_{y}) \lambda_j.
\end{eqnarray}
Here $D(C^s_{3z})$, $D(C^o_{3z})$ are representation matrices of spin and orbital parts of $C_{3z}$, respectively, and given by 
\begin{eqnarray}
    D(C^s_{3z})=\left (\begin{array}{ccc}
        e^{-i\frac{\pi}{3}} &0  \\
        0 &  e^{i\frac{\pi}{3}}
    \end{array}\right); D(C^o_{3z})=\left (\begin{array}{ccc}
        1 &0&0   \\
        0 &  e^{-i\frac{2\pi}{3}}&0  \\
        0 & 0 &e^{i\frac{2\pi}{3}}
    \end{array}\right), 
\end{eqnarray}
while $D(M^s_{y})$, $D(M^o_{y})$ are representation matrices of spin and orbital parts of $M_{y}$ respectively, 
\begin{eqnarray}
    D(M^s_{y})=\left (\begin{array}{ccc}
        0&1  \\
        -1 & 0
    \end{array}\right); D(M^o_{y})=\left (\begin{array}{ccc}
        1 &0&0   \\
        0 &  0&1  \\
        0 & 1 &0
    \end{array}\right).
\end{eqnarray}.


Using these representation matrices of $C_{3z}$ and $M_y$, we can compute the transformation matrices $\mathcal{D}$, from which we can classify the $\sigma$ and $\lambda$ matrices into the irrep table of the $C_{3v}$ group, as shown in Table \ref{TAB: symmetry C3v matrix}. To simplify the notation, we make the linear combinations of $\lambda$ matrices to introduce  $\Lambda$ matrices as
\begin{eqnarray}
&&\nonumber \Lambda_1 = \frac{1}{\sqrt{2}}(\lambda_1+\lambda_4)=\frac{1}{\sqrt{2}}\begin{pmatrix}
0 & 1 & 1 \\
1 & 0 & 0 \\
1 & 0 & 0
\end{pmatrix}, \quad
\Lambda_2 = \frac{1}{\sqrt{2}}(\lambda_2-\lambda_5)=\frac{1}{\sqrt{2}}\begin{pmatrix}
0 & -i & i \\
i & 0 & 0 \\
-i & 0 & 0
\end{pmatrix}, \\ \nonumber
&&\Lambda_3 = \frac{1}{\sqrt{2}}(\lambda_2+\lambda_5)=\frac{1}{\sqrt{2}}\begin{pmatrix}
0 & -i & -i \\
i & 0 & 0 \\
i & 0 & 0
\end{pmatrix},\quad
\Lambda_4  = \frac{1}{\sqrt{2}}(\lambda_4-\lambda_1)=\frac{1}{\sqrt{2}}\begin{pmatrix}
0 & -1 & 1 \\
-1 & 0 & 0 \\
1 & 0 & 0
\end{pmatrix}\\\nonumber
&&\Lambda_5  = \frac{1}{2}(\sqrt{3}\lambda_3+\lambda_8)=\frac{1}{\sqrt{3}}\begin{pmatrix}
2 & 0 & 0 \\
0 & -1 & 0 \\
0 & 0 & -1
\end{pmatrix}, \quad
\Lambda_6  = \frac{1}{2}(\lambda_3-\sqrt{3}\lambda_8)=\begin{pmatrix}
0 & 0 & 0 \\
0 & -1 & 0 \\
0 & 0 & 1
\end{pmatrix} \nonumber \\
&& \Lambda_7 = \lambda_6\quad \Lambda_8 = \lambda_7\quad \Lambda_0 = \lambda_0.\end{eqnarray}

In the Table \ref{TAB: symmetry C3v matrix}, the first column shows the irrep, the second and third columns show the construction of the representation matrices using $\Lambda$ and $\sigma$ matrices, respectively, and the last column shows the construction of representations using $k$-polynomials up to the $k^2$ order. The $\pm$ signs in parentheses after the representation matrices and $k$-polynomials show their parity under TR.

\begin{table}[htbp]
\centering
\renewcommand{\arraystretch}{1.5}
\begin{tabular}{|c|c|c|c|}
\hline
\(C_{3v}\) & \(\Lambda\) & \(\sigma\) & \(k\) \\ 
\hline
\(\mathbf{\Gamma}_1 (A_1)\) & \(\Lambda_0(+), \Lambda_5(+)\) & \(\sigma_0(+)\) & \(k_x^2 + k_y^2(+), \text{C}(+)\) \\
\hline 
\(\mathbf{\Gamma}_2 (A_2)\) & \(\Lambda_6(-)\) & \(\sigma_3(-)\) & higher order \\ 
\hline
\(\mathbf{\Gamma}_3 (E)\) & \((\Lambda_7,\Lambda_8)(+)(\Lambda_1,\Lambda_2)(+),(\Lambda_3,\Lambda_4)(-)\) & \((\sigma_2,\sigma_1)(-)\) & \((k_x,-k_y)(-)\), $(k^2_x-k^2_y, 2k_x k_y)(+)$ \\ 
\hline
\end{tabular}
\caption{\justifying  The classification of the representation matrices and $k$-polynomials for the $C_{3v}$ group. Here $\Lambda$ matrices are in orbital basis and $\sigma$ matrices are in the spin basis. The + or - in the bracket indicates how the matrices transform under TR (even or odd). The symbol $\text{C}$ in the column of $k$-polynomial is for a constant. \label{TAB: symmetry C3v matrix} }
\end{table}

Using the representation matrices and $k$-polynomials in Table \ref{TAB: symmetry C3v matrix}, we can construct all the symmetry-invariant terms around $\kk=0$ for the $C_{3v}$ group order by order in the momentum $\kk$. Here we only keep the terms up to the $\kk^2$ order for the spin-independent terms ($\sigma_0$) and the linear $\kk$ order for spin-orbit coupling terms, and obtain the effective Hamiltonian around $\mathbf{\Gamma}$ point as
\begin{eqnarray}\label{eq_SM:H_hetero_full}
&&H_{3b,hetero}(\kk)=\left[\frac{1}{3}(E_1 + A_1 k^2)+\frac{2}{3}(E_2 + A_2 k^2)\right] \sigma_0 \Lambda_0 +\left[\frac{1}{\sqrt{3}
}(E_1 + A_1 k^2)-\frac{1}{\sqrt{3}}(E_2 + A_2 k^2)\right]\sigma_0 \Lambda_5 \nonumber \\ 
&&+\sqrt{2} B_0 (k_x \sigma_0 \Lambda_3 -k_y \sigma_0 \Lambda_4)+\sqrt{2} B_1 \left[(k^2_x-k^2_y) \sigma_0 \Lambda_1 +2k_xk_y \sigma_0 \Lambda_2)\right] + B_2 \left[(k^2_x-k^2_y) \sigma_0 \Lambda_7 +2k_xk_y \sigma_0 \Lambda_8)\right] \nonumber \\ 
&&-\Delta_{SOC}\sigma_3 \Lambda_6+\sqrt{2}\Delta    _{Inter}(\sigma_2\Lambda_3+\sigma_1\Lambda_4)+R_2(\sigma_3\Lambda_7k_y+\sigma_3\Lambda_8k_x)+R_3(\sigma_3\Lambda_1k_y+\sigma_3\Lambda_2k_x)\nonumber\\
&&+\frac{1}{3}R_0 \left[\sigma_2(\Lambda_0+\sqrt{3}\Lambda_5)k_x-\sigma_1(\Lambda_0+\sqrt{3}\Lambda_5)k_y\right ]+\frac{2}{3}R_1\left [\sigma_2(\Lambda_0-\frac{\sqrt{3}}{2}\Lambda_5)k_x-\sigma_1(\Lambda_0-\frac{\sqrt{3}}{2}\Lambda_5)k_y\right ],
\end{eqnarray}
where $E_1, E_2, A_1, A_2, B_0, B_1, B_2$ are material dependent parameters independent of spin, $\Delta_{SOC}$ and $\Delta_{Inter}$ describe the $\kk$-independent SOC terms, $R_0$ and $R_1$ terms are Rashba SOC in TI$_2$Se$_2$ and Zn$_2$Te$_2$ layers, respectively, and the $R_2$ and $R_3$ terms are spin-dependent hopping terms connected to out-of-plane spin ($\sigma_3$). 

To project the above Hamiltonian to the Chern-band model, we will below ignore the interlayer SOC term $\Delta_{inter}(\sigma_2\Lambda_3+\sigma_1\lambda_4)$, $\kk$-dependent SOC terms and the off-diagonal $\kk^2$ terms ($B_1$ and $B_2$ terms), which are expected to be small, so the effective 6-bands model becomes block diagonal and can be written as 
\begin{eqnarray}
&&H_{3b,hetero}(\kk)=\left (\begin{array}{ccc}
        H_{hetero,\uparrow}(\kk)&0  \label{Eq:3b model 1}\\ 
        0 & H_{hetero,\downarrow}(\kk)
    \end{array}\right),\\
&&H_{3b,hetero,\uparrow}(\kk) =
\begin{pmatrix}
E_1 + A_1 k^2 & i B_0 (k_x + i k_y) & i B_0 (k_x - i k_y) \\
-i B_0 (k_x - i k_y) & E_2 + \Delta_{SOC} + A_2 k^2 & 0 \\
-i B_0 (k_x + i k_y) & 0 & E_2 - \Delta_{SOC} + A_2 k^2
\end{pmatrix}\\
&&H_{3b,hetero,\downarrow}(\kk) =
\begin{pmatrix}
E_1 + A_1 k^2 & i B_0 (k_x + i k_y) & i B_0 (k_x - i k_y) \\
-i B_0 (k_x - i k_y) & E_2 - \Delta_{SOC} + A_2 k^2 & 0 \\
-i B_0 (k_x + i k_y) & 0 & E_2 +\Delta_{SOC} + A_2 k^2
\end{pmatrix} \label{Eq:3b model 3},
\end{eqnarray}  
on the basis wavefunctions $\ket{\mathbf{\Gamma}_1,J_{z}=1/2,\uparrow}$, $ \ket{\mathbf{\Gamma}_3,J_{z}=3/2,\uparrow}$, $\ket{\mathbf{\Gamma}_3,J_{z}=-1/2,\uparrow}$ and $ \ket{\mathbf{\Gamma}_1,J_{z}=-1/2,\downarrow}$, $ \ket{\mathbf{\Gamma}_3,J_{z}=1/2,\downarrow}$, $ \ket{\mathbf{\Gamma}_3,J_{z}=-3/2,\downarrow}$ for spin up and down, respectively.
Here, $E_1>E_2$ so that for each spin block of the Hamiltonian $H_{hetero}$, the first component gives the conduction band with $L=0$ while the second and third components provide the valence bands with $L=\pm 1$. We focus on the spin up block $H_{hetero,\uparrow}$. If $\Delta_{SOC}>0$, we would expect the $\ket{\mathbf{\Gamma}_3,J_{z}=3/2,\uparrow}$ state has a higher energy than the $\ket{\mathbf{\Gamma}_3,J_{z}=-1/2,\uparrow}$ state and thus serves as the topmost valence band, so we project out the $\ket{\mathbf{\Gamma}_3,J_{z}=-1/2,\uparrow}$ state and the resulting Hamiltonian on the basis $\ket{\mathbf{\Gamma}_1,J_{z}=1/2,\uparrow}$,$\ket{\mathbf{\Gamma}_3,J_{z}=3/2,\uparrow}$ is given by
\begin{eqnarray}\label{Eq:hetro eff}
   H^{eff}_{hetero,\uparrow}(\kk) = \begin{pmatrix}
E_1 + A_1 k^2+\frac{B_0^2 \kk^2}{E_1-E_2+\Delta_{SOC}} & i B_0 (k_x + i k_y) \\
-i B_0 (k_x - i k_y) & E_2 + \Delta_{SOC} + A_2 k^2  
\end{pmatrix},
\end{eqnarray}
which is identical to the Chern-band model in Eq.(\ref{Eq:Ham_Chernband}) up to a unitary transformation. 

If $\Delta_{SOC}<0$, we can project out the $\ket{\mathbf{\Gamma}_3,J_{z}=3/2,\uparrow}$ state instead and the Hamiltonian on the basis $\ket{\mathbf{\Gamma}_1,J_{z}=1/2,\uparrow}$ and $\ket{\mathbf{\Gamma}_3,J_{z}=-1/2,\uparrow}$ is
\begin{eqnarray}\label{Eq:hetro eff2}
   H^{eff}_{hetero,\uparrow}(\kk) = \begin{pmatrix}
E_1 + A_1 k^2+\frac{B_0^2 \kk^2}{E_1-E_2-\Delta_{SOC}} & i B_0 (k_x -i k_y) \\
-i B_0 (k_x +i k_y) & E_2 - \Delta_{SOC} + A_2 k^2  
\end{pmatrix}, 
\end{eqnarray}
which still recovers the Chern-band model form Eq.(\ref{Eq:Ham_Chernband}) up to a unitary transformation.



\subsection{Tl$_2$Se$_2$/Zn$_2$Te$_2$ heterostructure with moir\'e potential}

In this section, we will discuss the calculations of the flat bands in the heterostructure consisting of Tl$_2$Se$_2$ and Zn$_2$Te$_2$ when there is moir\'e potential on either Tl$_2$Se$_2$ layer of Zn$_2$Te$_2$ layer. 

We first consider a heterostructure with the lattice constant of Tl$_2$Se$_2$ being relaxed to match that of Zn$_2$Te$_2$, so that no moir\'e pattern is formed between Tl$_2$Se$_2$ and Zn$_2$Te$_2$ layers, and we can directly perform the DFT calculation for the heterostructure (See Fig.4 in the main text). From the DFT calculations, we find the conduction band in the Tl$_2$Se$_2$/Zn$_2$Te$_2$ heterostructure shows a strong Rashba-type of spin splitting from the SOC (See Fig.4(c) in the main text) while the valence band has negligible spin splitting around $\kk = 0$. To take into account the Rashba-type of spin splitting, we need to consider $\kk$-dependent intra-layer SOC terms in Eq.(\ref{eq_SM:H_hetero_full}), namely the $R_0, R_1, R_2$ terms, while the inter-layer SOC terms, namely the  $\Delta_{Inter}$ and $R_3$ terms, are dropped. We project out the $\ket{\mathbf{\Gamma}_3,J_{z}=-1/2,\uparrow}$ and $\ket{\mathbf{\Gamma}_3,J_{z}=1/2,\downarrow}$ with the second order perturbation theory and reduce the effective 6-bands model to 4-bands model, which reads
\begin{eqnarray} \label{Eq:hetero}
    &&H_{hetero}(\kk)=\begin{pmatrix}
H^{\uparrow\uparrow}_{hetero}(\kk) & H^{\uparrow\downarrow}_{hetero}(\kk)  \\
(H^{\uparrow\downarrow}_{hetero}(\kk))^\dagger & H^{\downarrow\downarrow}_{hetero}(\kk) 
\end{pmatrix},
\end{eqnarray}
where $H^{\uparrow\uparrow(\downarrow\downarrow)}_{hetero}(\kk)$ is the spin up (down) block with the same form as Eq.(\ref{Eq:hetro eff}). Here we keep the terms up to $k^2$ order for the diagonal terms and $k$-linear order for the off-diagonal terms, so that the Hamiltonian can be rewritten as
\begin{eqnarray}\label{eq_SM:hetero_Ham_block_upup_downdown}
    &&H^{\uparrow\uparrow}_{hetero}(\kk) =
\begin{pmatrix}
E_1 + \tilde{A}_1 k^2 & i B_0 (k_x + i k_y)  \\
-i B_0 (k_x - i k_y) & \tilde{E}_2 + A_2 k^2 \\
\end{pmatrix},\nonumber\\
&&H^{\downarrow\downarrow}_{hetero}(\kk)=(H^{\uparrow\uparrow}_{hetero}(-\kk))^*
\end{eqnarray}
on the basis wavefunctions of 
$|\mathbf{\Gamma}_1,J_{z}=1/2,\uparrow\rangle |\mathbf{\Gamma}_3,J_{z}=3/2,\uparrow\rangle$ for  $H^{\uparrow\uparrow}_{hetero}(\kk)$ and $|\mathbf{\Gamma}_1,J_{z}=-1/2,\downarrow\rangle |\mathbf{\Gamma}_3,J_{z}=-3/2,\downarrow\rangle$ for  $H^{\downarrow\downarrow}_{hetero}(\kk)$, where $\tilde{A}_1=A_1 + \frac{B_0^2}{E_1 -E_2 + \Delta_{SOC}}$, $\tilde{E}_2=E_2+\Delta_{SOC}$ and  $E_1$, $A_2$, $B_0$ are the same as those in Eq.(\ref{Eq:3b model 1})-Eq.(\ref{Eq:3b model 3}). $H^{\uparrow\downarrow}_{hetero}(\kk)$ describes the lowest order $\kk$-dependent SOC with the Rashba-type of SOC form for the conduction band,  
\begin{eqnarray}\label{Eq:Rashba}
    &&H^{\uparrow\downarrow}_{hetero}(\kk) =
\begin{pmatrix}
-i R_0  k_- &0  \\
0 & 0 \\
\end{pmatrix}.
\end{eqnarray}
By choosing the parameters $E_2=-E_1=0.0649$eV, $A_1=0.9$ $a^2_0\cdot$eV, $A_2=-0.896$ $a^2_0\cdot$eV, $B_0=0.21$ $a_0\cdot$eV and $R_0=0.168$ $a_0\cdot$eV with $a_0=4.32\AA$ as the lattice constant of the heterostructure, the effective model can fit the band dispersions from the DFT calculations in the energy range from -0.1 eV to 0.1 eV, as shown in Fig.4(c) in the main text. We note a strong Rashba-type spin splitting existing in the conduction band, but not in the valence band. This is because of different total angular momentum between the conduction bands with $J_{z}=\pm1/2$ and the valence bands with $J_{z}=\pm3/2$. 

Next we will introduce the moir\'e potential into the Tl$_2$Se$_2$/Zn$_2$Te$_2$ hetero-structure, which can be generally induced by interfacing either Tl$_2$Se$_2$ or Zn$_2$Te$_2$ with another insulating layer. Due to the strong Rashba-type of SOC in the conduction band, but not in the valence band, the effect of introducing the moir\'e potential into the Tl$_2$Se$_2$ layer will be completely different from that in the Zn$_2$Te$_2$ layer.

\begin{figure}
\centering
\includegraphics[width=0.8\textwidth]{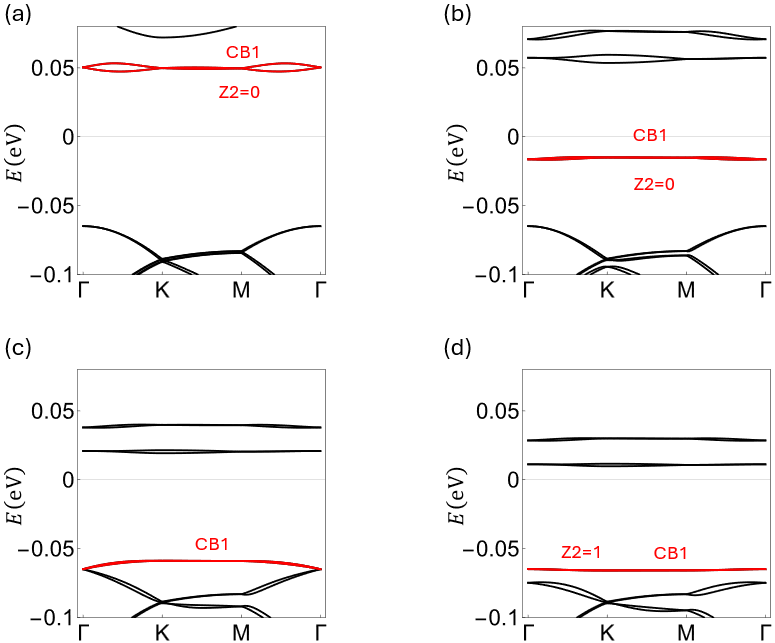}
\caption{\justifying \label{Fig: tlse2_potential} 
(a)–(d) show the band dispersions for the Tl$_2$Se$_2$/Zn$_2$Te$_2$ heterojunction for $\Delta_0 = -0.01$, $-0.03$, $-0.0445$, and $-0.048$ eV, respectively, with the moiré potential applied to the Tl$_2$Se$_2$ layer. We also choose the atomic gap $M$=0.0649 eV. }

\end{figure}

\begin{figure}
\centering
\includegraphics[width=1\textwidth]{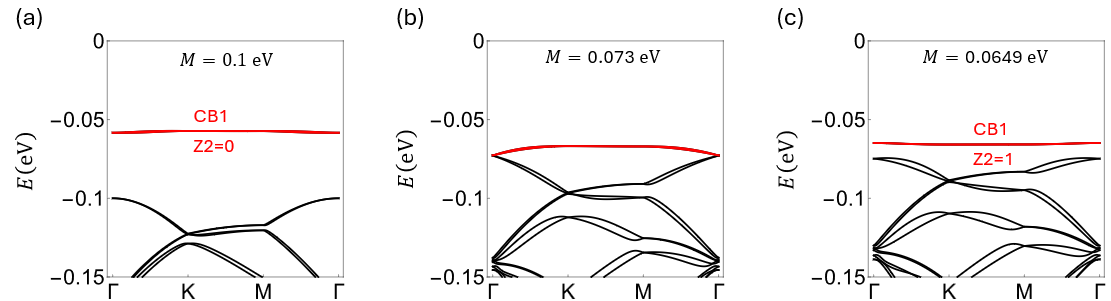}
\caption{\justifying \label{Fig: tlse2_M}
(a)–(d) show the band dispersions for the Tl$_2$Se$_2$/Zn$_2$Te$_2$ heterojunction for the atomic gap $M = 0.1$, $0.073$, and $0.0649$ eV, respectively, with the moiré potential applied to the Tl$_2$Se$_2$ layer. The moir\'e potential strength $\Delta_0$=0.0649 eV. }
\end{figure}

We first consider introducing moir\'e superlattice potential to the Tl$_2$Se$_2$ layer, and the whole system is described by the Hamiltonian 
\begin{eqnarray} \label{Eq:hetero moire}
   &&\hat{H}=\hat{H}_{hetero}+\hat{H}_{\text{m}},
\end{eqnarray}
where 
\begin{eqnarray} \label{Eq: hetero second}
&& \hat{H}_{hetero} = \sum_{\kk \alpha \beta}a_{\alpha \kk }^\dagger 
 [H_{hetero}(\kk)]_{\alpha \beta} a_{\beta \kk},
\end{eqnarray}
with $H_{hetero}(\kk)$ given in Eq.(\ref{Eq:hetero}) and the basis index $\alpha, \beta=1,2,3,4$, and the moir\'e potential at the Tl$_2$Se$_2$ layer is given by 
\begin{eqnarray}\label{Eq: hetero potential}
    \hat{H}_{\text{m}}  = \sum_{\alpha = 1, 3 }\int d^2r  a_{\alpha \rr }^\dagger H ^{\alpha \alpha}_ \text m (\rr) a_{\alpha \rr },
\end{eqnarray} 
and with the first shell approximation, we have $H_{\text{m}}^{\alpha\alpha}(\rr)=\sum_{\gs \in \GG^{\mathds{1}}_M} \Delta_0 e^{i \gs \cdot \rr}(\delta_{\alpha,1}+\delta_{\alpha,3})$. In the current basis, $\alpha = 1, 3$ is for conduction bands in Tl$_2$Se$_2$ layer and $\alpha = 2, 4$ is for valence bands in Zn$_2$Te$_2$ layer. Here, we take $\Delta_0$ as a tuning parameter, which is chosen to be a negative value $\Delta_0<0$ to introduce bound states for the conduction band, instead of the valence band.  The moir\'e lattice constant $a^M_0$ for the calculation is $ 24.75$ $nm$. The dispersions with different $\Delta_0$ are shown in Fig.\ref{Fig: tlse2_potential}(a)-(d). When $\Delta_0 = -0.01$ eV, the conduction bands CB1 originating from TI$_2$Se$_2$ are trivial and have a bandwidth of approximately $6$ meV, as shown in Fig.\ref{Fig: tlse2_potential}(a). This is due to the strong spin splitting that originates from strong Rashba-type of SOC for a small moir\'e potential strength. With increasing $\Delta_0$, we find the energy of CB1 is lowered and the spin splitting is also suppressed, so that the corresponding bandwidth decreases, as illustrated in Fig.\ref{Fig: tlse2_potential}(b) for $\Delta_0=-0.03$eV. For $\Delta_0\approx -0.0445$eV, we find the CB1 touches the valence band top at $\mathbf{\Gamma}$, leading to band inversion in Fig.\ref{Fig: tlse2_potential}(c). After the band inversion, the CB1 becomes topologically non-trivial while its bandwidth remains small, thus realizing topologically nontrivial flat bands depicted in Fig.\ref{Fig: tlse2_potential}(d). The Wannier function flow of CB1 in Fig.\ref{Fig: tlse2_potential}(d) is presented in Fig.4(e) of the main text. Similar to the calculation in Sec.\ref{sec:BHZ1}, the topology of VB1 can be tuned by the atomic gap $M$. The band dispersions with a fixed $\Delta_0$ and different $M$ values are shown in Fig.\ref{Fig: tlse2_M}. From Fig.\ref{Fig: tlse2_M}(a) to (c), as $M$ decreases, the CB1 evolves from trivial to topological bands, for a fixed $\Delta_0$. 

\begin{figure}
\centering
\includegraphics[width=1\textwidth]{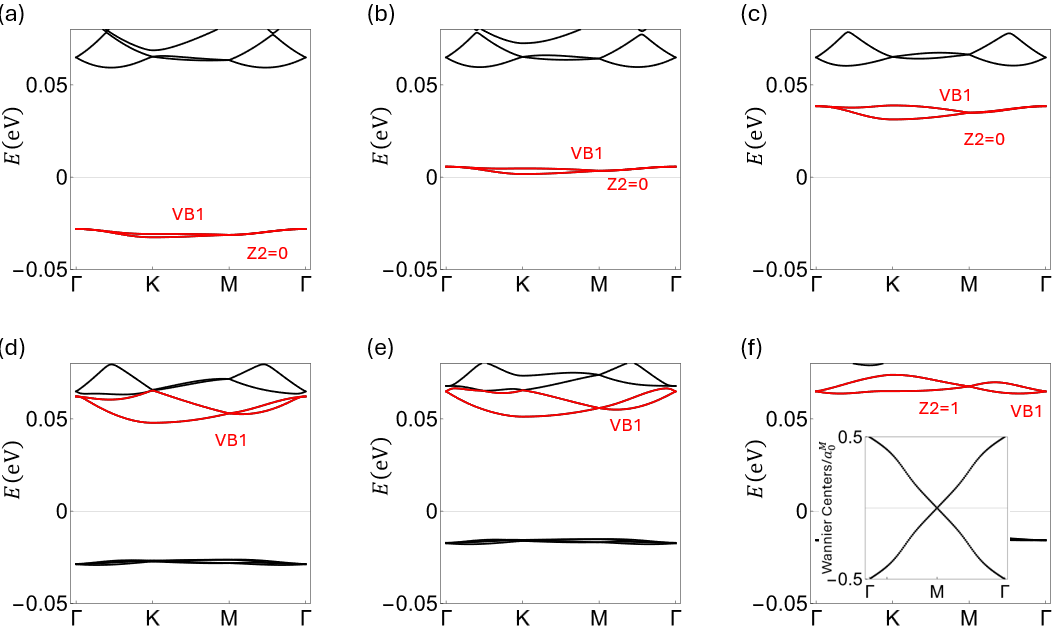}
\caption{\justifying \label{Fig: Znte2_potential} 
(a)-(e) show the band dispersions for the Tl$_2$Se$_2$/Zn$_2$Te$_2$ heterojunction with $\Delta_0$=0.02, 0.03, 0.04, 0.05 and 0.054 eV, respectively, with the moiré potential applied to the Zn$_2$Te$_2$ layer. The Rashba term is $R_0=0.168$ $a_0\cdot$eV  (f) shows the band dispersion for $\Delta_0 = 0.054$eV, with the Rashba term reduced to $R_0 = 0.05$ $a_0\cdot$eV. The inset in (f) shows the Wannier center flow of VB1.} 
\end{figure}

We next add the moir\'e potential to the Zn$_2$Te$_2$ layer, so that the moir\'e Hamiltonian $H_{\text{m}}^{\alpha\alpha}(\rr)$ in Eq.(\ref{Eq: hetero potential}) reads
\begin{eqnarray}
H_{\text{m}}^{\alpha\alpha}(\rr)=\sum_{\gs \in \GG^{\mathds{1}}_M} \Delta_0 e^{i \gs \cdot \rr}(\delta_{\alpha,2}+\delta_{\alpha,4}),
\end{eqnarray}
within the first shell approximation. We choose $\Delta_0 > 0$ to induce bound states for the valence band in Zn$_2$Te$_2$ layer. The band dispersions with different $\Delta_0$ are shown in Fig.\ref{Fig: Znte2_potential}(a)-(f). When $\Delta_0 = 0.02$ eV, the top valence mini-bands VB1 originating from the Zn$_2$Te$_2$ layer are trivial and have a bandwidth of approximately $5$ meV, as shown in Fig.\ref{Fig: Znte2_potential}(a). As $\Delta_0$ increases, VB1 are pushed upward in energy, as shown in Fig.\ref{Fig: Znte2_potential}(b) and (c) for $\Delta_0 = 0.03$eV and $0.04$eV, respectively. We also notice that when the VB1 approaches the conduction band bottom, the spin splitting of the VB1 becomes more pronounced, leading to an enhancement of bandwidth for VB1 in Fig.\ref{Fig: Znte2_potential}(b) and (c). At $\Delta_0 =0.05$ eV, the VB1 touch the conduction bands at the $\textbf{K}$ point in mBZ in Fig.\ref{Fig: Znte2_potential}(d), leading to a semi-metal phase for the VB1 in Fig.\ref{Fig: Znte2_potential}(e) for a larger $\Delta_0 = 0.054$eV. We slightly decrease the $R_0$ value of the Rashba SOC term, and find a $Z_2$ nontrivial insulator in Fig.\ref{Fig: Znte2_potential}(f), as demonstrated by the Wannier center flow of VB1 in the inset of Fig.\ref{Fig: Znte2_potential}(f). However, compared to the moir\'e potential at the Tl$_2$Se$_2$ layer, the bandwidth of VB1 is significantly larger for the moir\'e potential at the Zn$_2$Te$_2$ layer due to the strong Rashba SOC of the conduction band.



\subsection{Twisted Tl$_2$Se$_2$ homobilayer/Zn$_2$Te$_2$ monolayer heterostructure}\label{sec:twisted gamma}

\begin{figure}[h]
\centering
\includegraphics[width=1\textwidth]{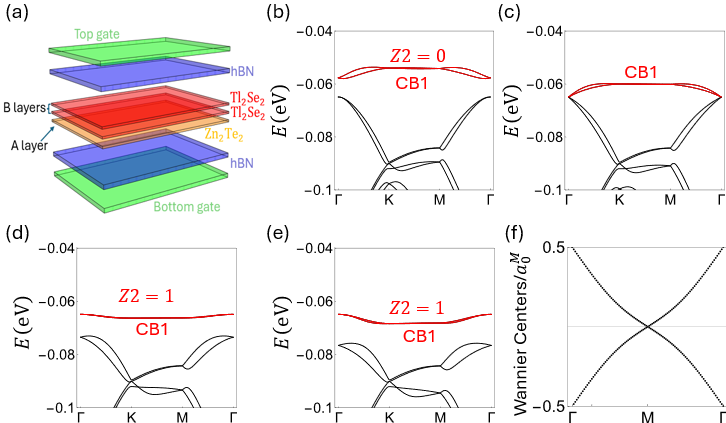}
\caption{\justifying \label{Fig: trilayer type2} 
(a) shows the schematics for the heterostructure with twisted Tl$_2$Se$_2$ bilayer and Zn$_2$Te$_2$.  In this setup, the twisted Tl$_2$Se$_2$ bilayer serves as  part B while the Zn$_2$Te$_2$ monolayer serves as the layer A for our general scenario in Fig. 1 of the main text. (b)-(e) show the band dispersions with $\Delta_0 = 0.019$eV, $0.0213$eV, $0.024$eV and $0.025$eV. (f) shows the Wannier center flow of CB1 highlighted in red in panel (d).} 
\end{figure}

In the above, we have discussed the role of moir\'e potential on the Tl$_2$Se$_2$ or Zn$_2$Te$_2$ layers. In this section, we will consider the heterostructure consisting of twisted Tl$_2$Se$_2$ homobilayer and Zn$_2$Te$_2$ monolayer, as schematically shown in Fig.\ref{Fig: trilayer type2}(a). For the twisted Tl$_2$Se$_2$ homobilayer with the twist angle around $1^\circ$, the primitive moir\'e reciprocal lattice vectors are $\bb^M_1=(2\pi,-\frac{2\pi}{\sqrt{3}})\frac{1}{a^M_0}$, $\bb^M_2=(0,\frac{4\pi}{\sqrt{3}})\frac{1}{a^M_0}$, where the moir\'e lattice constant is $a^M_0=24.75$ $nm$. 
The Hamiltonian for this hetero-structure takes the form
\begin{eqnarray} \label{Eq:trilayer}
    H = \begin{pmatrix}
H_{A}  & H_{AB} \\
H^\dagger_{AB} & H_B
\end{pmatrix},
\end{eqnarray}
where 
\begin{eqnarray}
    H_B = \begin{pmatrix}
H_{B0} + V_M(\rr) 1_{2\times 2}  & T_{M}(\rr) 1_{2\times 2}  \\
T^*_{M}(\rr) 1_{2\times 2} & H_{B0} + V_M(\rr) 1_{2\times 2}
\end{pmatrix},
\end{eqnarray}
with 
\begin{eqnarray}\label{Eq:trilayer HB0 HA}
&&H_{B0}=(E_1+A_1 k^2 + V_M(\rr))1_{2\times2}+\begin{pmatrix} \label{Eq;trilayer }
    0 &-i R k_-\\
    i R k_+ &0\\
\end{pmatrix};\nonumber \\
&&H_A=(E_2+A_2 k^2)1_{2\times2};\nonumber \\
&&H_{AB}=\begin{pmatrix}
    0 &0&-i B_0 k_- &0\\
    0 & 0 &0&-i B_0 k_+ 
\end{pmatrix}.
\end{eqnarray}
Here $H_A$ for the part A describes the Hamiltonian for the valence band in the Zn$_2$Te$_2$ layer on the basis wavefunctions $|\mathbf{\Gamma}_3,3/2,\uparrow\rangle$ and $|\mathbf{\Gamma}_3,-3/2,\downarrow\rangle$, while the Hamiltonian $H_B$ for the part B describes the Hamiltonian for the conduction bands of the twisted Tl$_2$Se$_2$ homobilayer on the basis wavefunctions $|\mathbf{\Gamma}_1,1/2,\uparrow,t\rangle$, $|\mathbf{\Gamma}_1,-1/2,\downarrow,t\rangle$, $|\mathbf{\Gamma}_1,1/2,\downarrow,b\rangle$ and $|\mathbf{\Gamma}_1,-1/2,\downarrow,b\rangle$, where $t$ and $b$ represent the top and bottom layers, respectively. The Hamiltonian $H_{AB}$ describes the tunneling between the parts A and B. In $H_B$, $T_M(\rr)$ and $V_M(\rr)$ describe the moir\'e tunneling between two Tl$_2$Se$_2$ layers and the moir\'e potential on each Tl$_2$Se$_2$ layer, respectively, and both should satisfy the symmetry of the moir\'e pattern in the twisted Tl$_2$Se$_2$ bilayer.
Thus, we can expand them as
\begin{eqnarray}
&& T_\text M (\rr) = \sum_{\gs \in \GG_M} T_\textbf{M}(\gs) e^{i \gs \cdot \rr}; \quad 
V_\text M (\rr) = \sum_{\gs \in \GG_M} V_\textbf{M}(\gs) e^{i \gs \cdot \rr}, 
\end{eqnarray}
similar to the expansion in Eq.(\ref{Eq:moirepotential_1}), where  $\GG_M = \{ n_1 \bb^M_1 + n_2 \bb^M_2 | n_1, n_2 \in \mathcal{Z}  \}$ includes all moir\'e reciprocal lattice vectors. It should be noted that since our effective model is expanded around $\mathbf{\Gamma}$ in ABZ, $\gs={\bf 0}$ is included in $\GG_M$, and $T_{\text M}(\gs={\bf 0})$ and $V_{\text M}(\gs={\bf 0})$ correspond to the uniform tunneling between two layers and uniform potential, respectively. The uniform potential $V_{\text M}(\gs={\bf 0})$ can be absorbed into the Fermi energy and thus we will drop $V_{\text M}(\gs={\bf 0})$ below. We denote $T_{\text M0}=T_{\text M}(\gs={\bf 0})$ and rewrite
\begin{eqnarray} \label{Eq:trilayer potential}
&& T_\text M (\rr) = T_{\text M0} + \tilde{T}_{\text M}(\rr) = T_{\text M0} + \sum_{\gs \neq {\bf 0}} T_\textbf{M}(\gs) e^{i \gs \cdot \rr}, 
\end{eqnarray}
where $\tilde{T}_{\text M}(\rr)$ includes the moir\'e tunneling other than $T_{\text M0}$.  We make the assumption that the inter-layer tunneling is dominated by the uniform part $T_{\text M0}$. This assumption has been justified in the previous studies on the moir\'e homobilayer systems with the low-energy atomic bands around ${\bf \Gamma}$ in the ABZ, as demonstrated in twisted TMD materials with  ${\bf \Gamma}$-valley \cite{angeli2021gamma,zhang2021electronic} and twisted ZrS$_2$ homobilayer\cite{claassen2022ultra}. We set the uniform tunneling term to be $T_{M0}=0.08$ eV and consider the first-shell potential approximation for the moir\'e tunneling and moir\'e potential, given by 
$T_\textbf{M}(\gs) = \Delta_0$ and $V_\textbf{M}(\gs) = V_0$ for $\gs\in \GG^{\mathds{1}}_M $ and $T_\textbf{M}(\gs) = V_\textbf{M}(\gs) = 0$ for $\gs\notin \GG^{\mathds{1}}_M $ and $\gs \neq {\bf 0}$. Below, we choose $V_0 = 0$ and $\Delta_0 \ll T_{M0}$ for the numerical calculations below, so that the uniform tunneling term dominates over the moir\'e tunneling and moir\'e potential.

The band dispersions with different values of $\Delta_0$ are shown in Fig.\ref{Fig: trilayer type2}(b)-(e). When $\Delta_0 = 0.019$ eV, the CB1 originating from the conduction band of twisted Tl$_2$Se$_2$ homobilayer reveals a spin splitting and is topologically trivial, shown in Fig.\ref{Fig: trilayer type2}(b). By increasing $\Delta_0$, the CB1 touches the valence band top and its spin splitting is reduced, shown in Fig.\ref{Fig: trilayer type2}(c). When $\Delta_0 = 0.024$ eV, the CB1 becomes topologically nontrivial and its spin splitting is negligible, thus realizing a topologically nontrivial flat band, which is demonstrated by the nontrivial winding of the Wannier center flow for CB1 in Fig.\ref{Fig: trilayer type2}(f). With further increasing $\Delta_0$, the CB1 becomes more dispersive while remains topologically nontrivial, and behaves similarly to a valence band with a negative effective mass around $\mathbf{\Gamma}$, as shown in Fig.\ref{Fig: trilayer type2}(e). From Fig.\ref{Fig: trilayer type2}(b)-(e), we find the evolution of the CB1 for twisted Tl$_2$Se$_2$ homobilayer case is quite similar to that for the case with adding moir\'e potential on Tl$_2$Se$_2$ layer in Fig.\ref{Fig: tlse2_potential}(a)-(d). 

Below we will provide an analytical understanding of the similarity between these two cases using the perturbation theory. Under the assumption $T_{\text M0} \gg |T_{\text M}(\gs\neq 0)|, |V_{\text M}(\gs\neq 0)|$, we can apply a unitary transformation 
\begin{eqnarray}
    U_0 = \frac{1}{\sqrt{2}}\begin{pmatrix}
-1_{2\times 2}  & 1_{2\times 2} \\
1_{2\times 2} & 1_{2\times 2}
\end{pmatrix}
\end{eqnarray}
to $H_B$ and the resulting Hamiltonian $H'_B$ is given by
\begin{eqnarray} \label{eq_SM:homobilayerHBtransform}
&&H'_B = U_0 H_B U^{-1}_0 \nonumber \\
&& = \begin{pmatrix}
H_{B0} + \left(- T_{\text M0} - \frac{1}{2} (\tilde{T}_{\text M}(\rr) + \tilde{T}_{\text M}^{\dagger}(\rr))  + V_M(\rr) \right) 1_{2\times 2}  & \frac{1}{2} ( - \tilde{T}_{\text M}(\rr) + \tilde{T}_{\text M}^{\dagger}(\rr)) 1_{2\times 2}  \\
\frac{1}{2} (\tilde{T}_{\text M}(\rr) - \tilde{T}_{\text M}^{\dagger}(\rr)) 1_{2\times 2}  & H_{B0} + \left(T_{\text M0} + \frac{1}{2} (\tilde{T}_{\text M}(\rr) + \tilde{T}_{\text M}^{\dagger}(\rr)) + V_M(\rr) \right) 1_{2\times 2}
\end{pmatrix}. \nonumber\\
\end{eqnarray}
$H'_B$ is written in the basis of antibonding and bonding states between two layers, where $\alpha = 1, 2$ (upper block) and  $\alpha = 3, 4$ (lower block) correspond to the antibonding and bonding states, respectively. Based on the assumption of $T_{\text M0} \gg |T_{\text M}(\gs\neq 0)|$, we can treat $\tilde{T}_{\text M}(\rr)$ as a perturbation, and neglect the off-diagonal terms in $H'_B$, since the energy difference between the diagonal terms in $H'_B$ is dominated by $2 |T_{M0}|$. The energy ordering of two bands for part B depends on the sign of $T_{\text M0}$. In the numerical calculations, $T_{M0}$ is chosen to be positive, so that the antibonding states have lower energy for $H'_B$ in Eq.(\ref{eq_SM:homobilayerHBtransform}). 
After the unitary transformation $U_0$, the full Hamiltonian becomes
\begin{eqnarray}
    H'= \begin{pmatrix}
H_{A}  & H'_{AB} \\
(H')^\dagger_{AB} & H'_B
\end{pmatrix},
\end{eqnarray}
where $H_A$ remains the same, $H'_B$ is given by Eq.(\ref{eq_SM:homobilayerHBtransform}) and $H'_{AB}$ after the unitary transformation is given by
\begin{eqnarray}
    H'_{AB}=\begin{pmatrix}
    -i \frac{B_0}{\sqrt{2}} k_- &0&-i \frac{B_0}{\sqrt{2}} k_- &0\\
    0 & -i \frac{B_0}{\sqrt{2}} k_+  &0&-i \frac{B_0}{\sqrt{2}} k_+ 
\end{pmatrix}.
\end{eqnarray}

For the whole moir\'e heterostructure, $H_A$ for the part A is from the Zn$_2$Te$_2$ layer and provides the dispersive valence band, while $H'_B$ for the part B is from the twisted Tl$_2$Se$_2$ homobilayer and provides the localized orbital from the conduction band, as shown in Fig. 4(e) of the main text. Thus, we need to pick out the lowest energy bands in $H'_B$. For $T_{M0} > 0$, the antibonding state (the upper block) in $H'_B$ (Eq. (\ref{eq_SM:homobilayerHBtransform})) has the lowest energy and thus we project out the bonding states (lower block) in $H'_B$ and only keep the antibonding states (upper block). The resulting four-band model Hamiltonian, which includes $H_A$ and the upper block in $H'_B$, reads
\begin{eqnarray}\label{eq_SM:Heff_homobilayer_twist}
    H'_{\text eff} = \begin{pmatrix}
H_{A}  &  H''_{AB} \\
 (H''_{AB})^\dagger & H_{B0} - T_{\text M0} 1_{2\times 2}  + \left( - \frac{1}{2} (\tilde{T}_{\text M}(\rr) + \tilde{T}_{\text M}^{\dagger}(\rr))  + V_M(\rr) \right) 1_{2\times 2} 
\end{pmatrix}, 
\end{eqnarray}
where
\begin{eqnarray}
    H''_{AB}=\begin{pmatrix}
    -i \frac{B_0}{\sqrt{2}} k_- &0\\
    0 & -i \frac{B_0}{\sqrt{2}} k_+  
\end{pmatrix},
\end{eqnarray}  
and $H_A$ and $H_{B0}$ take the same form as those in Eq.(\ref{Eq:trilayer HB0 HA}). 
In Eq.(\ref{eq_SM:Heff_homobilayer_twist}), one can see $- \frac{1}{2} (\tilde{T}_{\text M}(\rr) + \tilde{T}_{\text M}^{\dagger}(\rr))  + V_M(\rr)$ plays the role of moir\'e potential for the antibonding states of the conduction bands in the twisted Tl$_2$Se$_2$ homobilayer. Thus, the Hamiltonian (\ref{eq_SM:Heff_homobilayer_twist}) describes a dispersive valence band without any moir\'e potential and a conduction band with moir\'e potential given by $- \frac{1}{2} (\tilde{T}_{\text M}(\rr) + \tilde{T}_{\text M}^{\dagger}(\rr))  + V_M(\rr)$, which is equivalent to the Hamiltonian described in Eqs. (\ref{Eq:hetero})-(\ref{Eq: hetero potential}). This conclusion can be applied to other 2D ${\bf \Gamma}$-valley moir\'e materials under the assumption $T_{\text M0} \gg |T_{\text M}(\gs\neq 0)|$.



\bibliography{ref}